\documentclass[11pt]{article}

\usepackage[utf8]{inputenc} %
\usepackage[T1]{fontenc}    %
\usepackage[english]{babel}
\usepackage[super,comma,sort&compress]{natbib}
\usepackage{url}            %
\usepackage{authblk}
\usepackage{booktabs}       %
\usepackage{amsfonts}       %
\usepackage{nicefrac}       %
\usepackage{microtype}      %
\usepackage{xcolor}         %
\usepackage{graphicx}
\usepackage{array}
\usepackage{amsmath,bm}
\usepackage{subcaption}
\usepackage{wrapfig, blindtext}
\usepackage{fancyhdr}
\usepackage{lipsum}
\usepackage{relsize}
\usepackage[export]{adjustbox}
\usepackage{enumitem}       %
\usepackage{todonotes}
\usepackage{amssymb}
\usepackage[labelfont=bf]{caption}
\usepackage{titlesec}
\usepackage{afterpage}
\usepackage{numprint}
\usepackage{xhfill}
\usepackage{soul}
\usepackage{csquotes}
\usepackage[a4paper,
            bindingoffset=0.2in,
            left=1in,
            right=1in,
            top=1in,
            bottom=1in,
            footskip=.25in]{geometry}
\usepackage{changepage}

\newcommand{\figtitle}[1]{\noindent\xrfill[0.7ex]{0.75pt}\textsf{#1}\xrfill[0.7ex]{0.75pt}\vspace{0.5em}}
 \renewcommand*{\Affilfont}{\normalsize\normalfont}
\titlespacing*{\section}
{0pt}{5.5ex plus 1ex minus .2ex}{4.3ex plus .2ex}
\titlespacing*{\subsection}
{0pt}{5.5ex plus 1ex minus .2ex}{4.3ex plus .2ex}
\DeclareUnicodeCharacter{00B0}{ }

\DeclareCaptionFormat{myformat}{\textbf{#1#2 $|$} #3}
\newcommand{\themethod}{Phenformer}
\newcommand{\thehypothesis}{⟨sequence \textrightarrow{} cell context \textrightarrow{} expression \textrightarrow{} phenotype⟩}

\usepackage{color}
\usepackage{tcolorbox}
\usepackage{CJK}
\usepackage{adjustbox}
\tcbset{width=0.9\textwidth,boxrule=0pt,colback=red, arc=0pt, auto outer arc,left=0pt,right=0pt,boxsep=5pt}
\usepackage{hyperref}
\usepackage[capitalise]{cleveref}

\newcommand{\resultsubplotsize}{0.20}
\newcommand{\rowheadersize}{0.04}

\newcommand{\papertitle}{Multi-megabase scale genome interpretation with genetic language models}
\title{\textbf{\papertitle}}

\begin{document}
\author[1,2,a]{Frederik Träuble}
\author[1,a]{Lachlan Stuart}
\author[1,a]{Andreas Georgiou}
\author[3]{Pascal Notin}
\author[1]{Arash Mehrjou} 
\author[1]{Ron Schwessinger}
\author[1,4]{Mathieu Chevalley}
\author[1]{Kim Branson}
\author[2]{Bernhard Schölkopf}
\author[5]{Cornelia van Duijn}
\author[3]{Debora Marks}
\author[1,*]{Patrick Schwab}
\affil[1]{GSK plc, Zug, Switzerland}
\affil[2]{Max Planck Institute for Intelligent Systems \& ELLIS Institute, Tübingen, Germany}
\affil[3]{Harvard Medical School, Boston, United States}
\affil[4]{ETH Zurich, Switzerland}
\affil[5]{Nuffield Department of Population Health, University of Oxford, Oxford, United Kingdom}
\affil[a]{Joint first authors}
\affil[*]{Corresponding authors}

\date{}
\setcounter{Maxaffil}{0}
\renewcommand\Affilfont{\itshape\small}
\newcommand{\ManuscriptFormat}{nature} %

\maketitle
\thispagestyle{fancy}
\pagestyle{fancy}

\begin{adjustwidth*}{1.5cm}{1.5cm}
\section*{Abstract}
Understanding how molecular changes caused by genetic variation drive disease risk is crucial for deciphering disease mechanisms.
However, interpreting genome sequences is challenging because of the vast size of the human genome, and because its consequences manifest across a wide range of cells, tissues and scales - spanning from molecular to whole organism level.
Here, we present \themethod{}, a multi-scale genetic language model that learns to generate mechanistic hypotheses as to how differences in genome sequence lead to disease-relevant changes in expression across cell types and tissues directly from DNA sequences of up to 88 million base pairs. 
Using whole genome sequencing data from more than \numprint{150000} individuals, we show that \themethod{} generates mechanistic hypotheses about disease-relevant cell and tissue types that match literature better than existing state-of-the-art methods, while using only sequence data.
Furthermore, disease risk predictors enriched by \themethod{} show improved prediction performance and generalisation to diverse populations. 
Accurate multi-megabase scale interpretation of whole genomes without additional experimental data enables both a deeper understanding of molecular mechanisms involved in disease and improved disease risk prediction at the level of individuals. \end{adjustwidth*}
\newpage{}

\begin{figure*}[pt!] 
\vspace{-2.1em}
\centering	
\includegraphics[width=0.95\textwidth, valign=t]{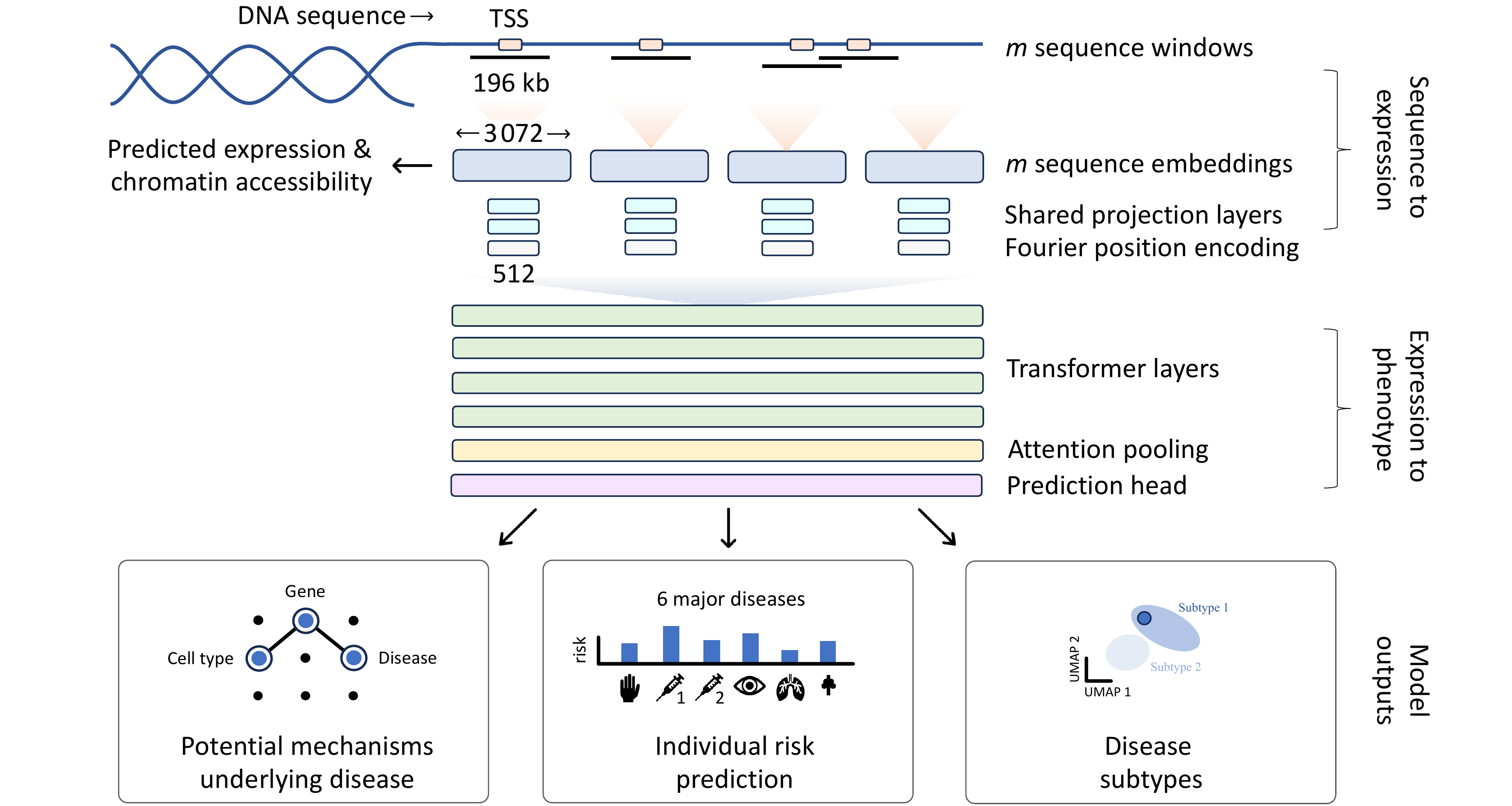}
\vspace{-0.4em}
\caption{\textbf{\themethod{} is a genetic language model that learns to connect individual genomes to changes in cell-type-specific expression to disease directly from sequence.} \themethod{} is an end-to-end multi-scale model that directly processes genomes following the information flow in molecular biology\cite{crick1970central} \thehypothesis{}. A variable number of $m$ windows of 196 kilobases (kb) centred around the transcription start site (TSS) of genes are first transformed by a sequence-to-expression backbone (Enformer\cite{avsec2021effective}) that was pretrained to predict expression and chromatin accessibility across a wide range of cell types. Tokens of sequence embeddings (\numprint{3072} dimensions per TSS) are then passed to an expression-to-phenotype core that consists of multiple transformer encoder layers\citep{vaswani2017attention} that later aggregate information across sequence embeddings using Pooling by Multihead Attention\cite{lee2019set} (PMA). A prediction head outputs individual risk predictions for the phenotype of interest. \themethod{} integrates up to 88 million base pairs - almost 3\% of an individual genome and an order of magnitude larger than the largest existing genetic language model\cite{nguyen2023hyenadna} - to highlight potential molecular mechanisms underlying diseases, predict disease risk, and identify disease subtypes.}
\label{fig:intro}
\vspace{-0.9em}
\end{figure*}

\section{Introduction} 
\label{sec:intro}
The advent of population-scale genetic sequencing\cite{chen2011china,sudlow2015uk,gaziano2016million,all2019all,kurki2023finngen} spurred by the dramatic drop in the cost of  sequencing\cite{bennett2005toward,mardis2017dna} has led to significant advances in human genetics, including a deeper understanding of the human genome and increased appreciation of the contribution of genetic variation to disease susceptibility \cite{macarthur2017new,thompson2022uk}. The wealth of data generated by population-scale genetic studies has enabled researchers to systematically associate specific genetic variations on the level of single nucleotide polymorphisms (SNPs) with diseases, shedding light on the genetic basis of numerous conditions\cite{macarthur2017new,cano2020gwas}, helping predict individual disease risk\cite{torkamani2018personal,lewis2021polygenic} and advancing the development of therapeutics targeted at disease-causing mechanisms\cite{nelson2015support,king2019drug,mehrjou2021genedisco,lyle2023discobax,minikel2023refining}.

Genomes are typically investigated on the population level in genome-wide association studies (GWAS) that attempt to relate SNPs to observed phenotypes using linear or logistic regression models\cite{uffelmann2021genome}. These studies identify variants statistically associated with disease that can be aggregated in sets of up to a few hundred SNPs into polygenic risk scores (PRS). PRS methods use an appropriate weighting function to achieve higher performance in predicting individual disease risk than those SNPs would have by themselves\cite{lewis2020polygenic}. However, a typical genome is reported to differ from the reference genome in around 20 million bases \cite{10002015global}, and existing methods that consider single to several hundred independent variants therefore fall far short of accounting for the entire variation in a single individual. Furthermore, existing methods do not consider the broader sequence context of variants, are dependent on ancestral linkage disequilibrium (LD) structures, prone to overfitting to the (typically European) populations they were derived from\cite{charles2014accounting,duncan2019analysis,ruan2022improving}, and do not by themselves provide further insights into the downstream functional effects of those variants on molecular processes\cite{gallagher2018post}. %

A major challenge in more comprehensively interpreting genetic variation is the sheer scale of the human genome with approximately three billion base pairs\cite{venter2010multiple}. New methods that aim to integrate more of the information contained in the genome than existing approaches necessitate large amounts of storage and compute, and scalable architectures. Researchers have attempted to improve modeling of gene-gene interactions by separately modeling non-linear effects using neural networks\cite{mccaw2022deepnull} and by utilizing non-linear models directly\cite{lopez2018single,elgart2022non,ghose2022genome}. However, these approaches are limited to modeling disease risk from SNPs rather than from sequence, which places SNPs into context. In related work that operates on sequence data, machine learning was used to predict pathogenicity of protein-coding missense variants from protein sequences\cite{brandes2023genome,rives2021biological}, multiple sequence alignments (MSA) of evolutionary data \cite{frazer2021disease}, and from protein sequences and predicted structures\cite{cheng2023accurate}. However, existing methods are limited to predicting the pathogenicity (benign or not benign) of the relatively small subset \cite{backman2021exome,dong2023annotating} of protein-coding missense variants. In addition, existing methods only consider variants for a single protein in isolation without considering the genetic context of an individual that may carry a particular variant, and do not link variation back to phenotypes. Other genetic language models include the nucleotide transformer \cite{dalla2024nucleotide}, Evo\cite{nguyen2024sequence} and HyenaDNA\cite{nguyen2023hyenadna} that are limited to respectively 6 kilobase (kb), 131 kb and up to 1 megabase (mb) sequence length (from more than 1000x to 88x smaller than presented here) and did not connect the genome sequence to organism-scale polygenic phenotypes. In another direction of research, previous studies used machine learning to model the relationship between genetic sequences to changes in gene expression across tissue and cell contexts\cite{kelley2016basset,kelley2018sequential,avsec2021effective,linder2023predicting} - without however linking changes in expression back to high-level phenotypes and diseases. More recently, Polygenic Transcriptome Risk Scores (PTRS) proposed predicting phenotypes based on the gene expression predicted for several cell types\cite{liang2022ptrs}. However, PTRS alone could not match the performance of state-of-the-art PRS.

\begin{figure*} 
\centering	
\vspace{-2.15em}

 \begin{subfigure}[b]{0.487\textwidth}
\figtitle{a) Cell type id compared to literature}
\includegraphics[width=\textwidth]{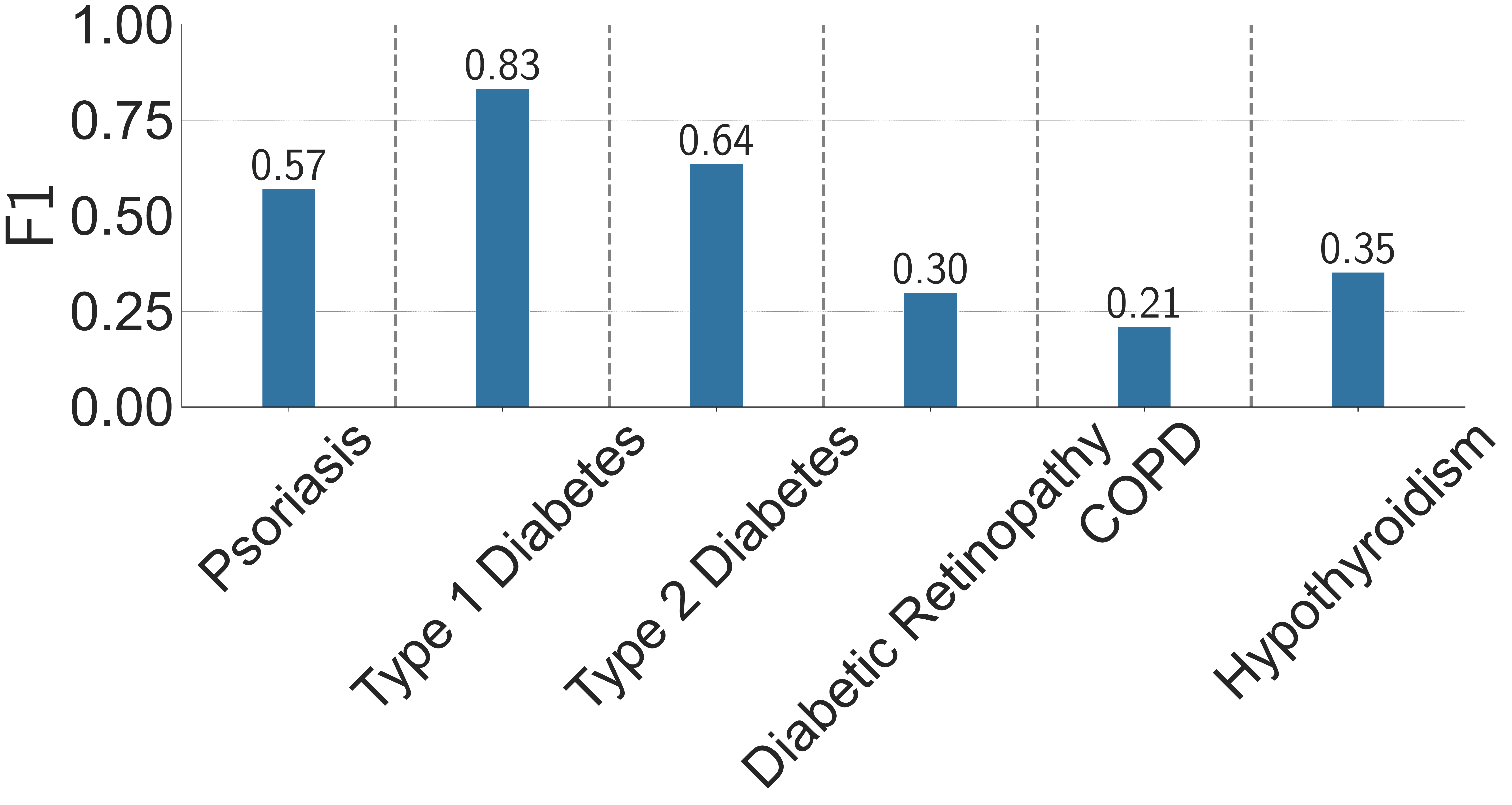}
  \end{subfigure}\hspace{4em}
 \begin{subfigure}[b]{0.377\textwidth}
\figtitle{b) Cell type id performance}
\includegraphics[width=\textwidth]{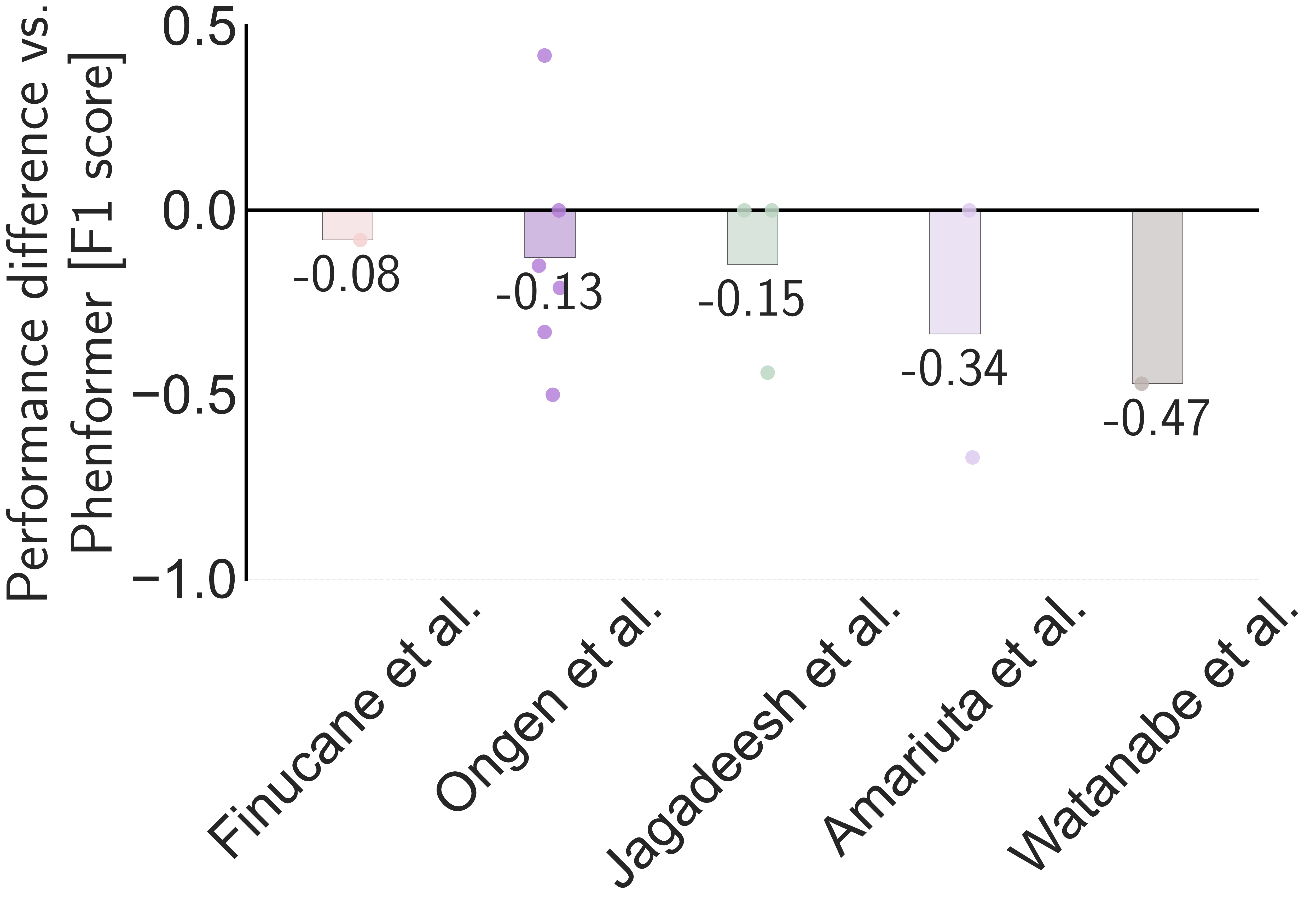}
  \end{subfigure}

\figtitle{c) Overview of identified cell and tissue type associations with disease}

 \begin{subfigure}[b]{\rowheadersize\textwidth}
    \rotatebox{90}{\hspace{3.75em}\textsf{\themethod{}}}
  \end{subfigure}
  \begin{subfigure}[t]{0.62\textwidth}\centering
  \includegraphics[width=\textwidth]{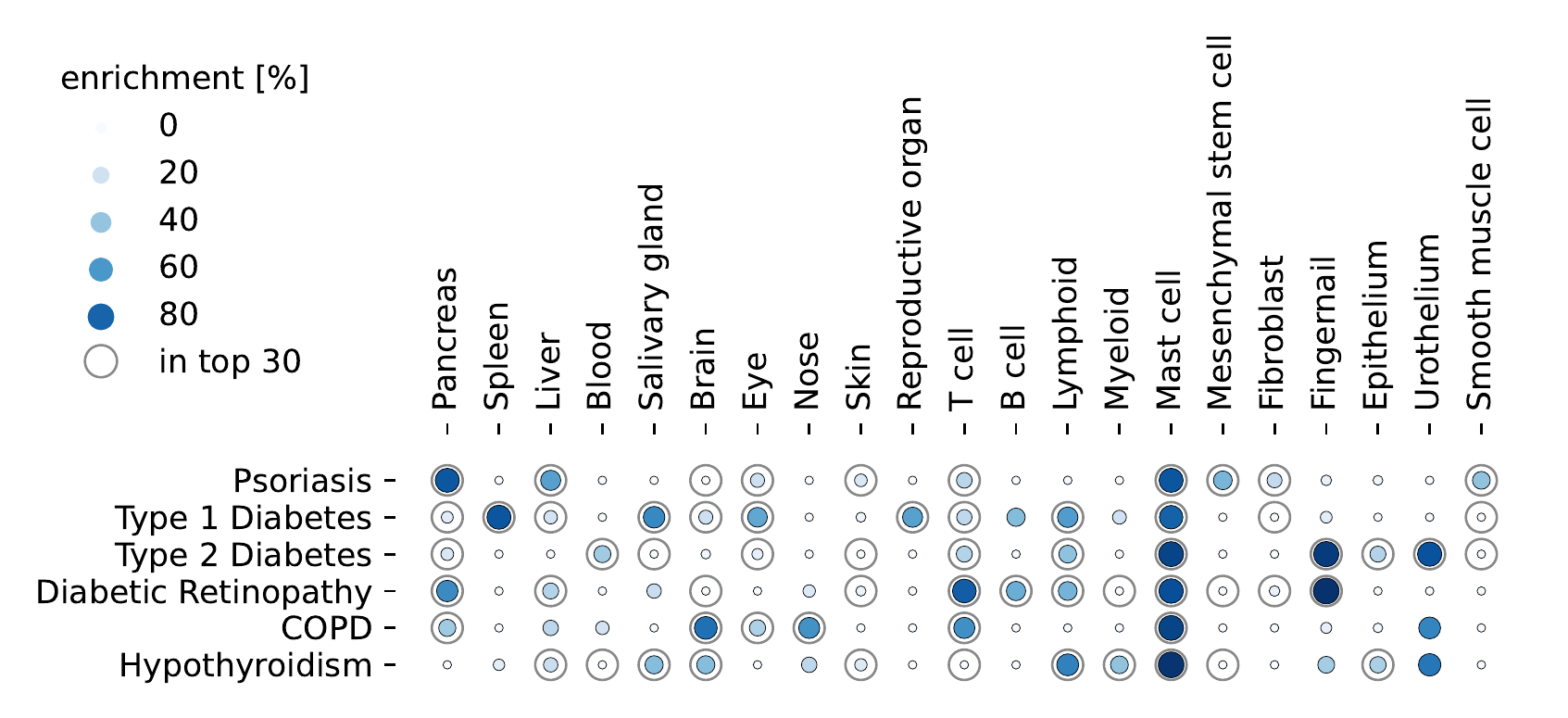}
\end{subfigure}\quad

 \begin{subfigure}[b]{\rowheadersize\textwidth}
    \rotatebox{90}{\hspace{4.5em}\textsf{Literature}}
  \end{subfigure}
  \begin{subfigure}[t]{0.62\textwidth}\centering
  \includegraphics[width=\textwidth]{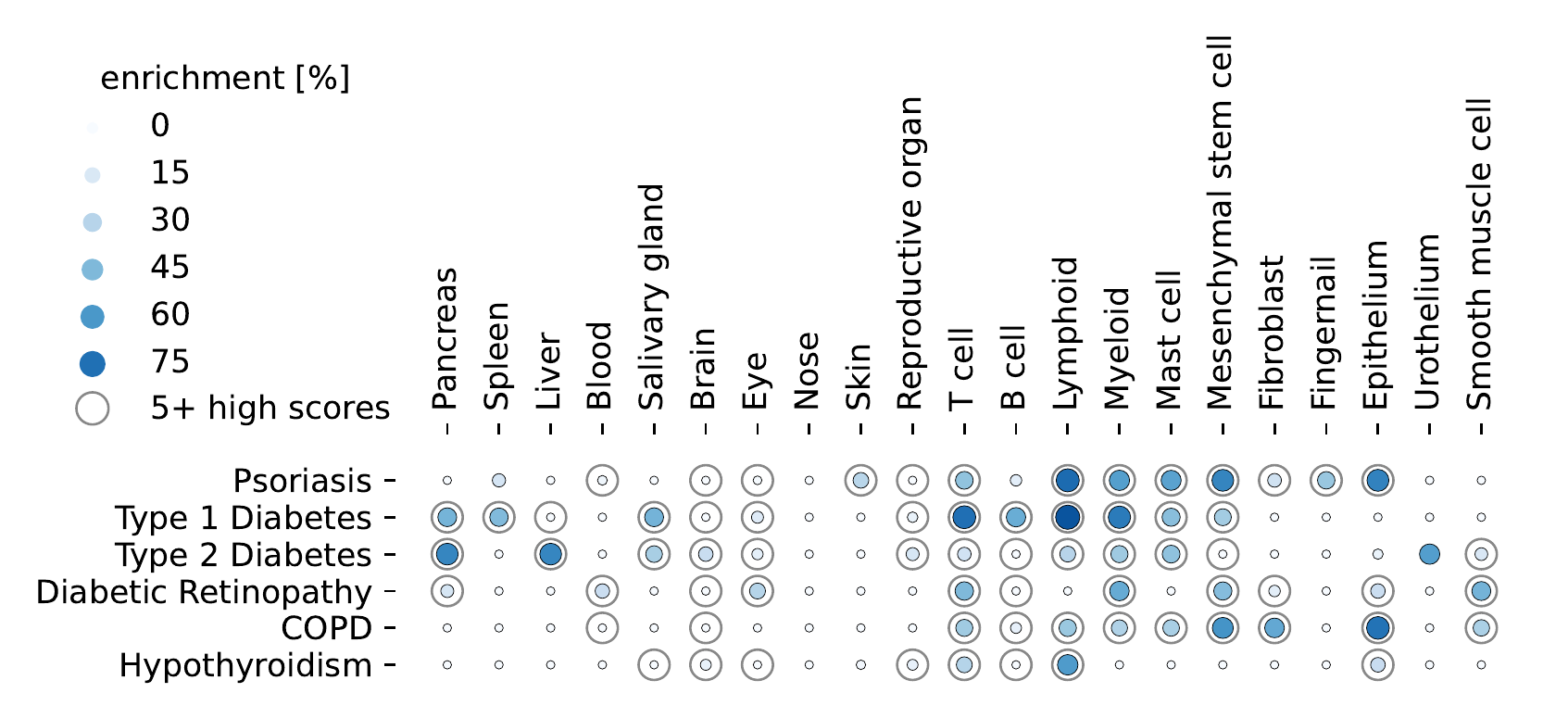}
\end{subfigure}
\caption{\textbf{\themethod{} identifies disease-associated cell and tissue types.} 
\textbf{a.} \themethod{} independently recovers cell and tissue type to disease associations previously reported in literature as measured via F1 score through enrichment (at least 5\% enrichment as a threshold for \themethod{}). \textbf{b.} We compared \themethod{} to state-of-the-art cell type identification methods that leverage genetic and/or single cell RNA sequencing (scRNAseq) data\cite{ongen2017estimating,finucane2018heritability,watanabe2019genetic,jagadeesh2022identifying,amariuta2023modeling} and found that \themethod{} more accurately identified the cell types reported in literature to be associated with disease by average F1 score (dots represent per-disease differences). For fairness, the comparison was conducted in pairwise fashion on the overlap of diseases and cell types for which predictions were available for both \themethod{} and the method being compared to.
\textbf{c.} An overview of categories of cell types highlighted by \themethod{} to be enriched in differential disease risk predictions (top) and - for comparison - an overview of the cell type-disease associations supported by scientific literature (bottom). Larger size circles indicate that more members of the respective category of cell type were ranked highly by \themethod{} (\Cref{fig:interpretation}) or scientific literature (see Section \enquote{\nameref{par:methods_literature}} for methodology), respectively. Grey circles indicate that at least one member of the cell type category was ranked in the top 30 most predicted differential cell types for a disease for \themethod{} or that 5 or more abstracts scoring highly for evidence of association between the cell type and disease were found in literature. Cell types were assigned to the most specific category shown, i.e. mast cells were not also part of the myeloid cells category.}
\label{fig:interpretation_quantification}
\end{figure*}

In this work, we present \themethod{} -- a first-of-its-kind deep-learning model that learns to predict disease risk end-to-end directly from individual genome sequences. The architecture of \themethod{} follows the direction of biological information flow from DNA sequence to expression\cite{crick1970central} to disease, and therefore permits rich \thehypothesis{} attributions that unlock a fine-grained in-silico understanding of {how} variants influence mechanisms underlying disease (\Cref{fig:intro}). Quantitatively, we show that candidate mechanisms independently predicted by \themethod{} are more enriched for those reported in scientific literature than those derived from state-of-the-art methods that require single-cell RNA sequencing data in addition to genetic information (\Cref{fig:interpretation_quantification}). Moreover, we qualitatively find that the variant-transcript-cell type-disease mechanisms highlighted by \themethod{} reflect clinically established disease pathologies that are to date molecularly poorly understood, including, for example, increased prevalence of non-alcoholic fatty liver disease (NAFLD) in psoriasis patients\cite{prussick2015nonalcoholic} and appendicitis\cite{tsai2008complicated,wei2016diabetes} complications in type 1 diabetes (\Cref{tab:qualitative_results}). In addition, we experimentally demonstrate that ensembles of PRS methods with \themethod{} significantly improve performance in predicting disease risk across diseases while achieving more robust performance across diverse human populations than base PRS methods alone. \themethod{} is a powerful method for whole sequence genome interpretation that promises to both improve our ability to annotate genetic variation with the putative mechanisms they influence as well as to predict disease risk on the level of individual genomes.

\section{Results}
\label{sec:results}

\paragraph{\themethod{}.} \themethod{} is a deep-learning model that predicts individual disease risk directly from whole genome sequences. \themethod{} input consists of $m=512$ windows of DNA sequence, each spanning 196 kilobases (kb) and centered on transcription start sites (TSS). This data is first processed in parallel across all windows into a sequence-to-expression backbone (frozen Enformer\cite{avsec2021effective}), which was pretrained to predict gene expression patterns across various cell types. The sequence embeddings are then passed to an expression-embedding-to-phenotype core that consists of multiple transformer layers that later aggregate information across sequence embeddings using Pooling by Multihead Attention\cite{lee2019set} (PMA). \themethod{} generates hypotheses for potential mechanisms underlying disease and individual per-disease risk predictions (\Cref{fig:intro}). We trained a separate \themethod{} model for each disease of interest. We used a subset corresponding to almost 88 million base pairs (almost 3\% of an individual's genome) of whole genome sequencing (WGS) data from \numprint{150076} individuals in the UK Biobank\cite{sudlow2015uk} and \numprint{12500} hours of graphics processing unit (GPU) compute time to train \themethod{} models on multiple major diseases to evaluate its interpretability and predictive performance relative to state-of-the-art methods.

\begin{figure*}[pt!] 
\vspace{-1.7em}
\centering	
\figtitle{a) Predictive performance (whole genome, mixed ancestry)}

\begin{subfigure}[b]{0.37\textwidth}
\includegraphics[width=\textwidth]{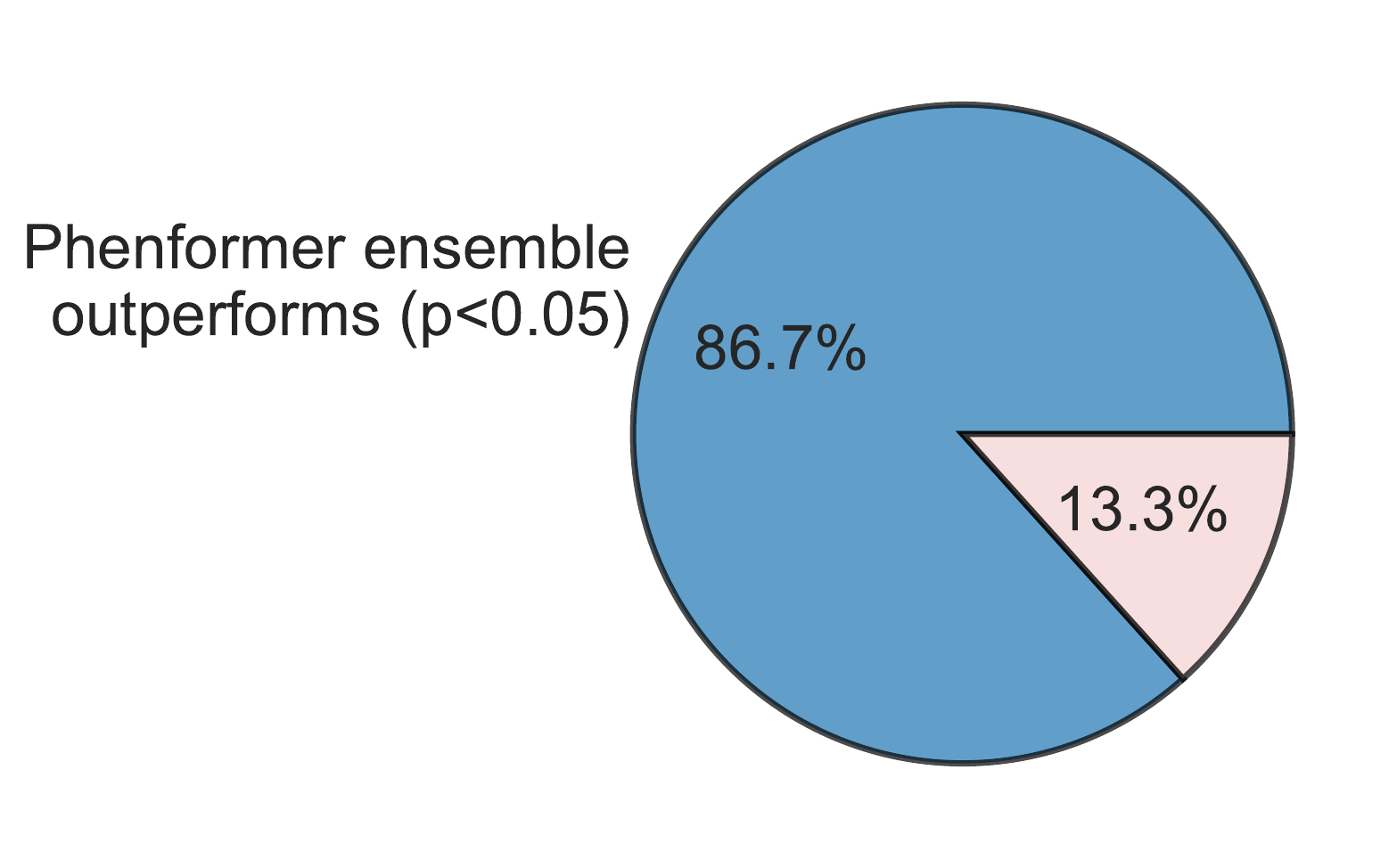}\vspace{1.35em}
  \end{subfigure} 
\begin{subfigure}[b]{0.45\textwidth}  
\includegraphics[width=\textwidth]{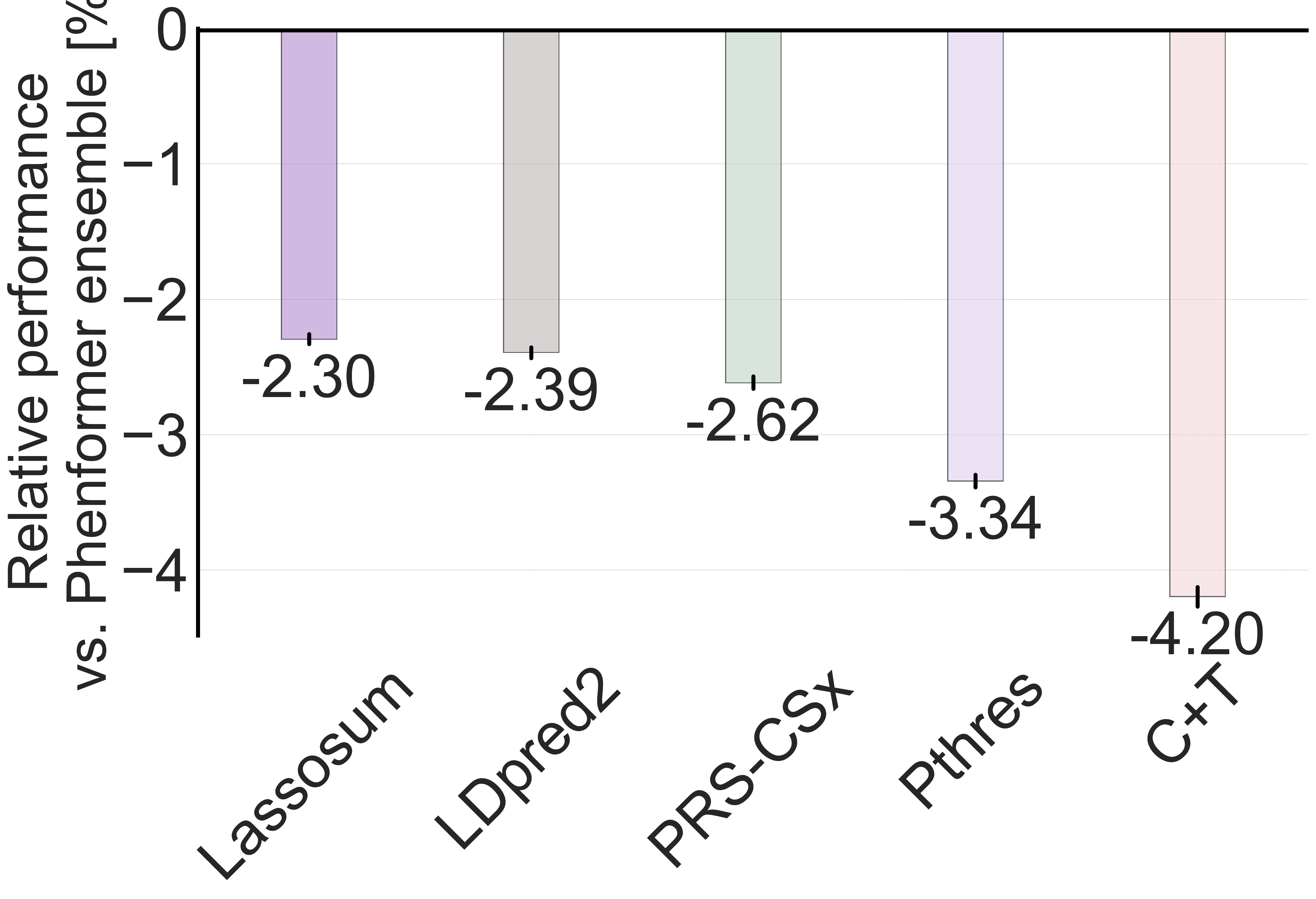}
  \end{subfigure} 
  
\figtitle{b) Predictive performance (whole genome, non-European ancestry)}

\begin{subfigure}[b]{0.37\textwidth}
\includegraphics[width=\textwidth]{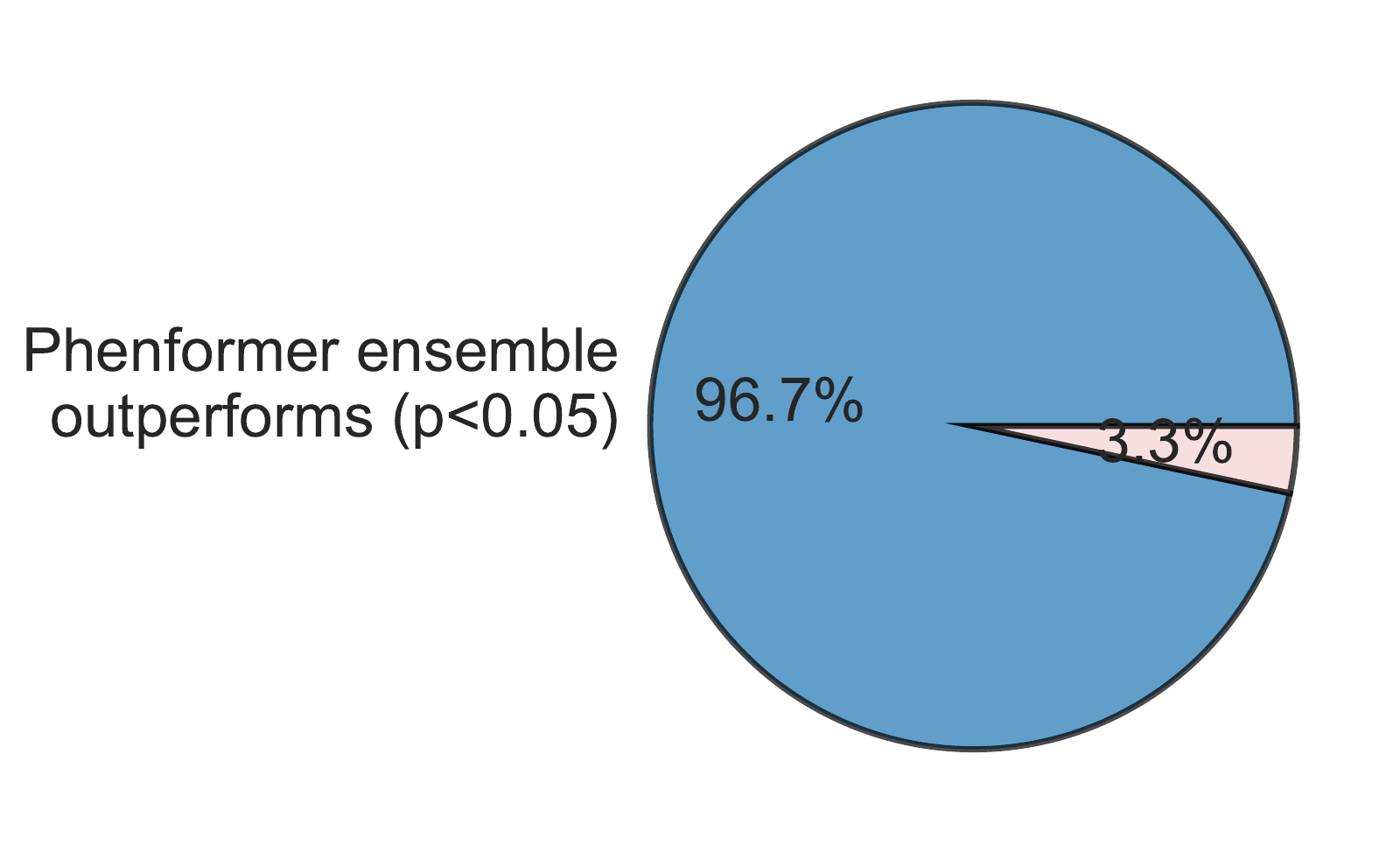}\vspace{1.35em}
  \end{subfigure} 
\begin{subfigure}[b]{0.45\textwidth}  
\includegraphics[width=\textwidth]{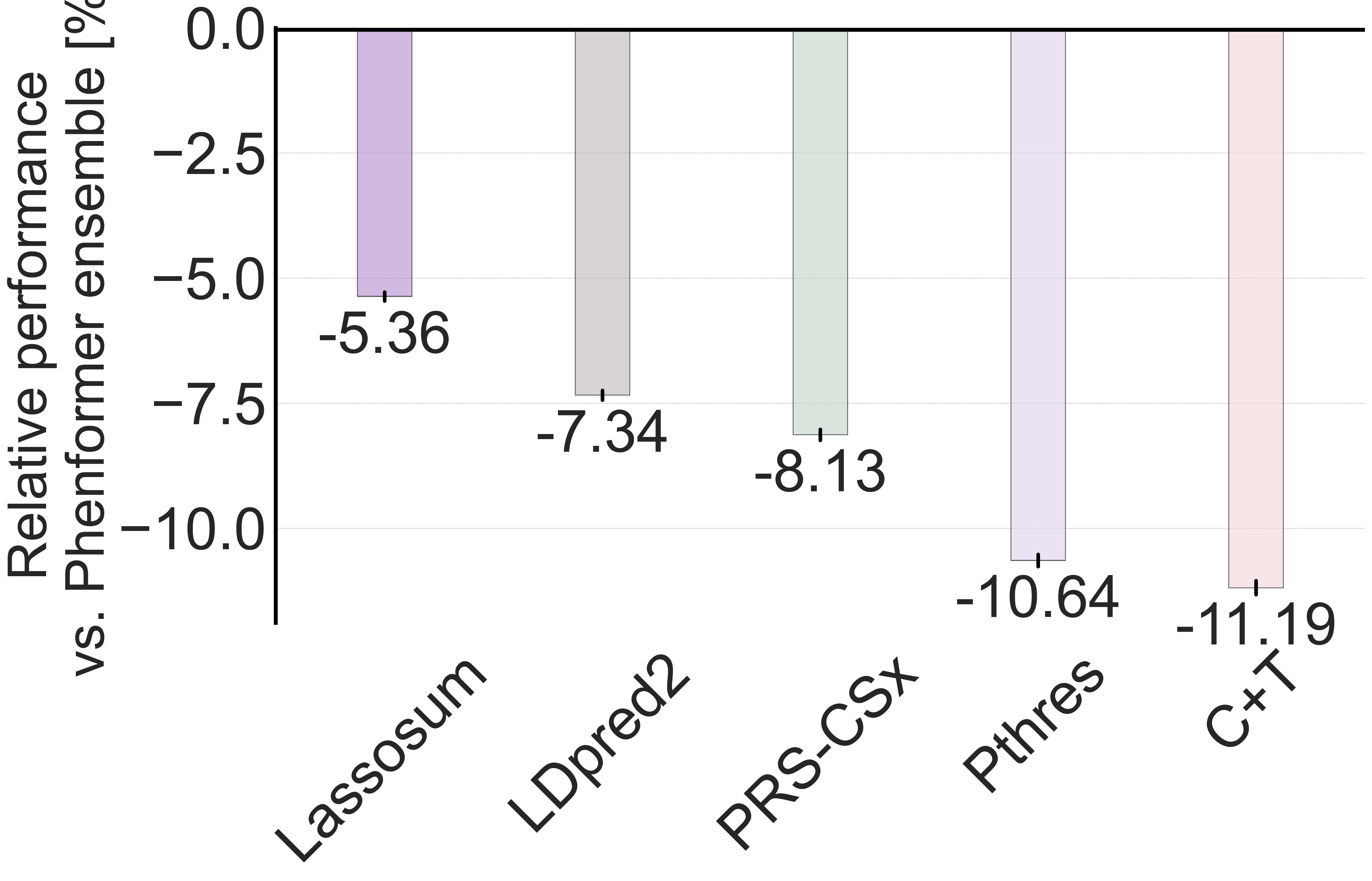}
  \end{subfigure} 
  
  \vspace{1em}
  
 \begin{subfigure}[b]{0.45\textwidth}
\figtitle{c) Performance (512 genes, mixed)}
\includegraphics[width=\textwidth]{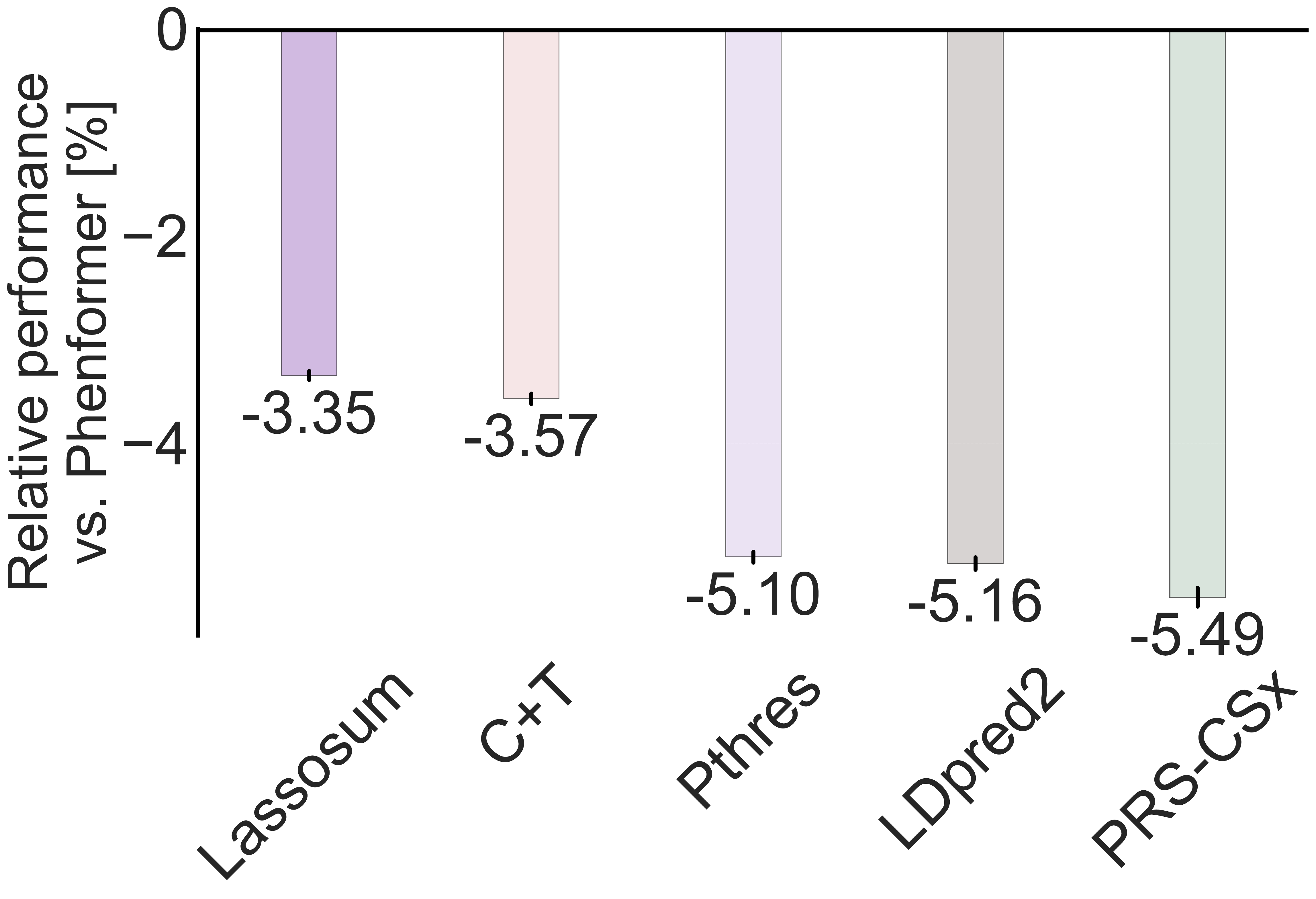}
  \end{subfigure}\hspace{2em}
 \begin{subfigure}[b]{0.455\textwidth}
\figtitle{d) Performance (512 genes, non-European)}
\includegraphics[width=\textwidth]{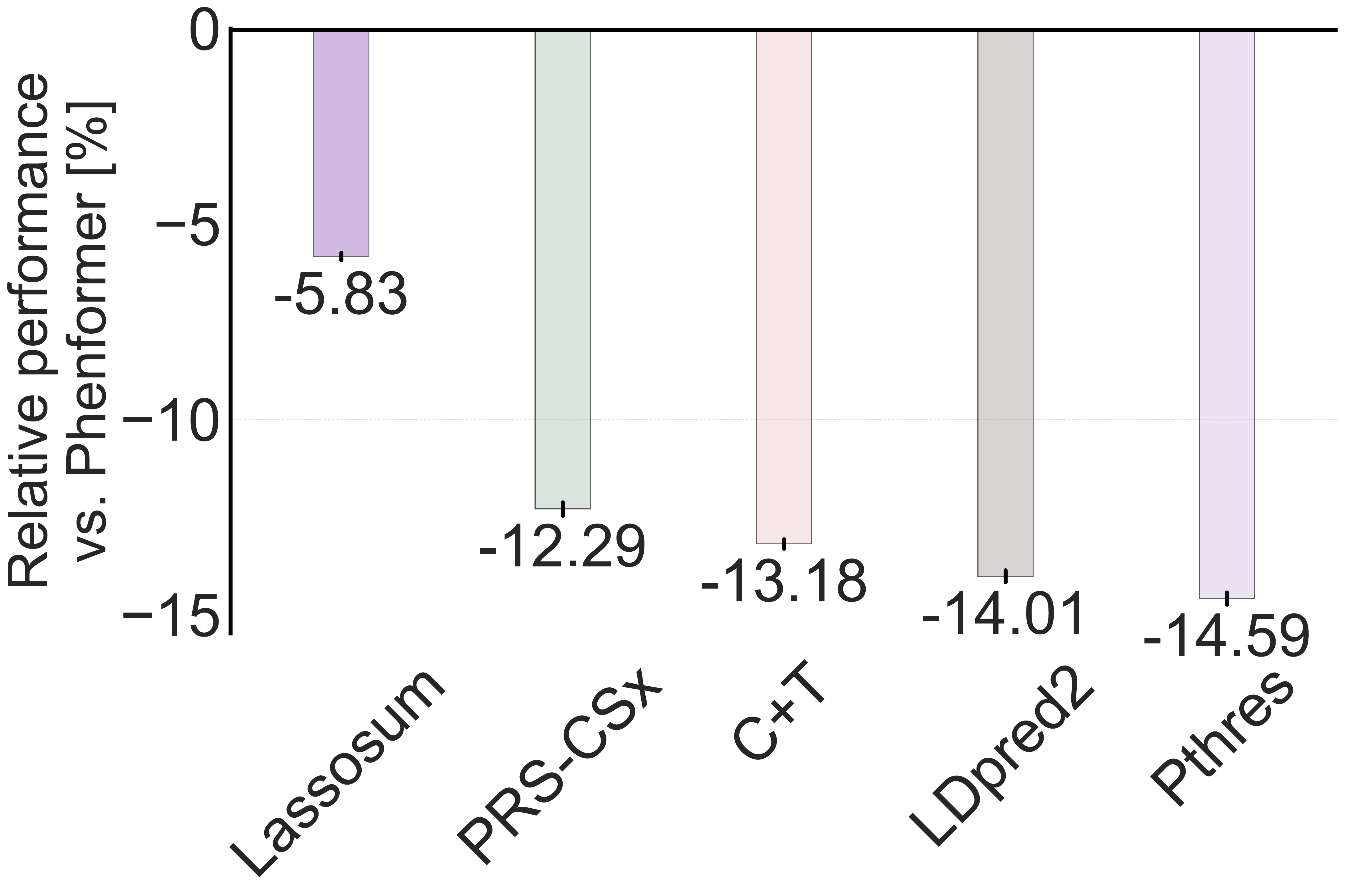}
  \end{subfigure} 
  
\caption{\textbf{\themethod{} improves prediction of disease risk from whole genomes.} We used ensembles of \themethod{} (trained on approximately 3\% of the whole genome) and state-of-the-art polygenic risk score (PRS) methods (Lassosum, LDpred2, PRS-CSx, Pthres, C+T) to improve risk prediction performance across 6 major diseases (psoriasis, type 1 diabetes, type 2 diabetes, diabetic retinopathy, chronic obstructive pulmonary disease [COPD] and hypothyroidism) on held-out test set individuals with \textbf{a.} mixed ancestry and with  \textbf{b.} non-European ancestry. We found that enhancing PRS methods with \themethod{} predictions significantly (p $\leq 0.05$; Mann-Whitney Wilcoxon test for superiority) improves disease risk prediction compared to predicting risk using only the ensemble partner for 86.7\% and 96.7\% of diseases and ensemble partners with average performance benefits across diseases of up to 4.2\% and 11.19\% higher area under the receiver operator curve (AUROC) in populations of mixed ancestry and non-European ancestry, respectively. When restricting the evaluation to the same subset of approximately 3\% of the genome sequence that \themethod{} was trained on (corresponding to sequence windows around 512 genes), \themethod{} achieves up to 5.49\% and 14.59\% higher prediction performance in terms of average AUROC across diseases for populations of \textbf{c.} mixed ancestry and with \textbf{d.} non-European ancestry, respectively. Uncertainty was evaluated using bootstrap resampling with \numprint{2000} samples.
}
\label{fig:aggregate_performance}
\end{figure*}

\paragraph{\themethod{} identifies disease associated molecular mechanisms from sequence.}
\themethod{} generates multiscale \thehypothesis{} hypotheses connecting disease mechanisms to phenotypes at the molecular level (see Section \enquote{\nameref{par:model_interpretation}} for methodology). The generated hypotheses provide a rich basis for further mechanistic evaluation of how genetic variation may give rise to a change in individual disease susceptibility. We first sought to validate whether hypotheses generated by \themethod{} are able to identify disease-associated cell types. We analysed \themethod{} predictions for all studied major diseases, identified categories of cell and tissue types enriched in \themethod{} hypotheses and evaluated their overlap with associations previously  reported in scientific literature (see Section \enquote{\nameref{par:methods_literature}} for methodology). We found that cell and tissue type hypotheses generated by \themethod{} directly from sequence more accurately reflected those reported in literature than state-of-the-art cell type identification methods that leverage genetic and additional data, such as single cell RNA sequencing (scRNAseq) data\cite{ongen2017estimating,finucane2018heritability,watanabe2019genetic,jagadeesh2022identifying,amariuta2023modeling} (\Cref{fig:interpretation_quantification}). 

\begin{figure*}[pt!] 
\centering	

\figtitle{Liver-associated psoriasis sequence windows}

  \begin{subfigure}[b]{\rowheadersize\textwidth}
    \rotatebox{90}{\hspace{6.0em}\textsf{SELENOW}}
  \end{subfigure}
  \begin{subfigure}[t]{0.43\textwidth}\centering
	\includegraphics[width=1.0\textwidth, valign=b]{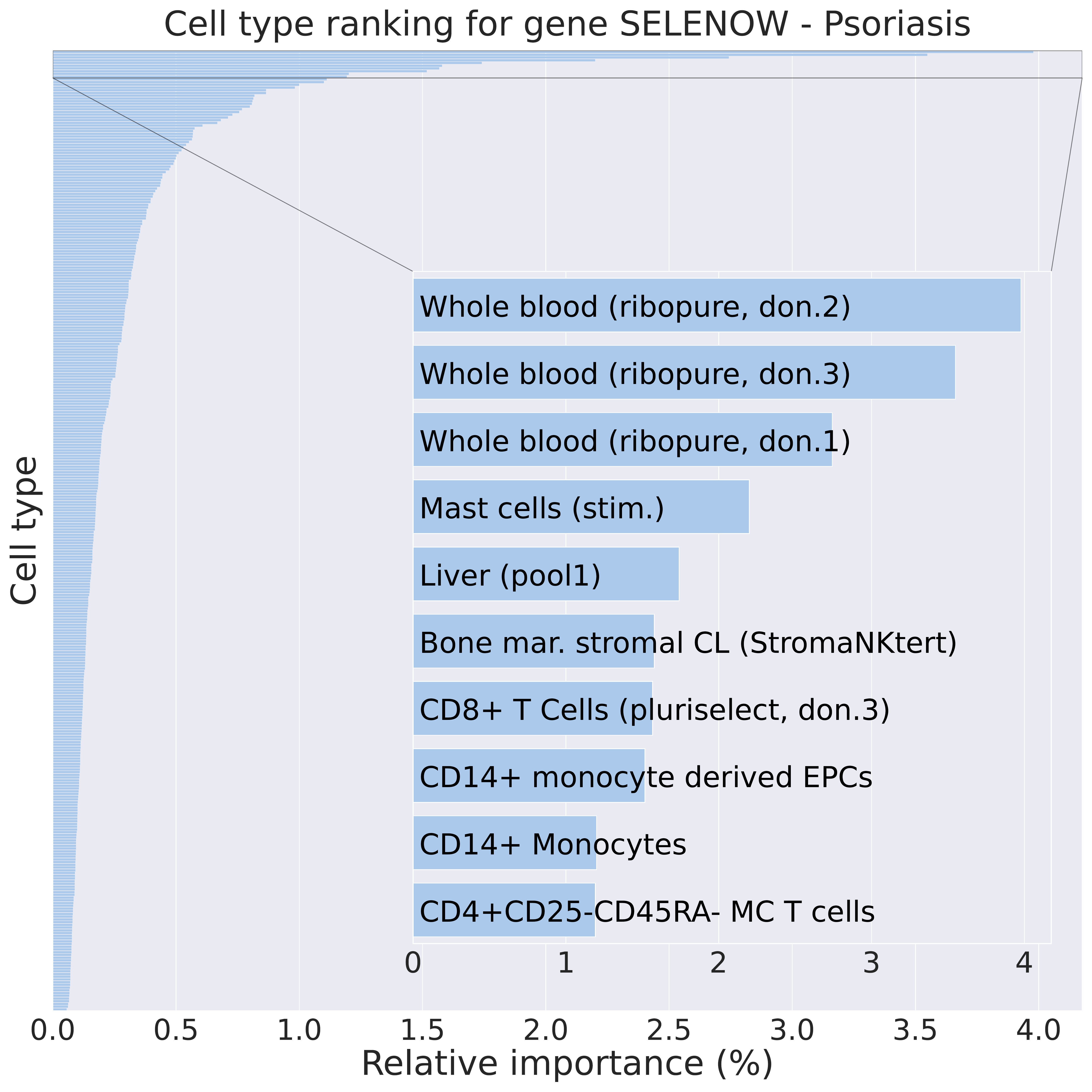}
  \end{subfigure}\hspace{1em}
    \begin{subfigure}[b]{\rowheadersize\textwidth}
    \rotatebox{90}{\hspace{7.5em}\textsf{SPX}}
  \end{subfigure}
  \begin{subfigure}[t]{0.43\textwidth}
\centering	
	\includegraphics[width=1.0\textwidth, valign=b]{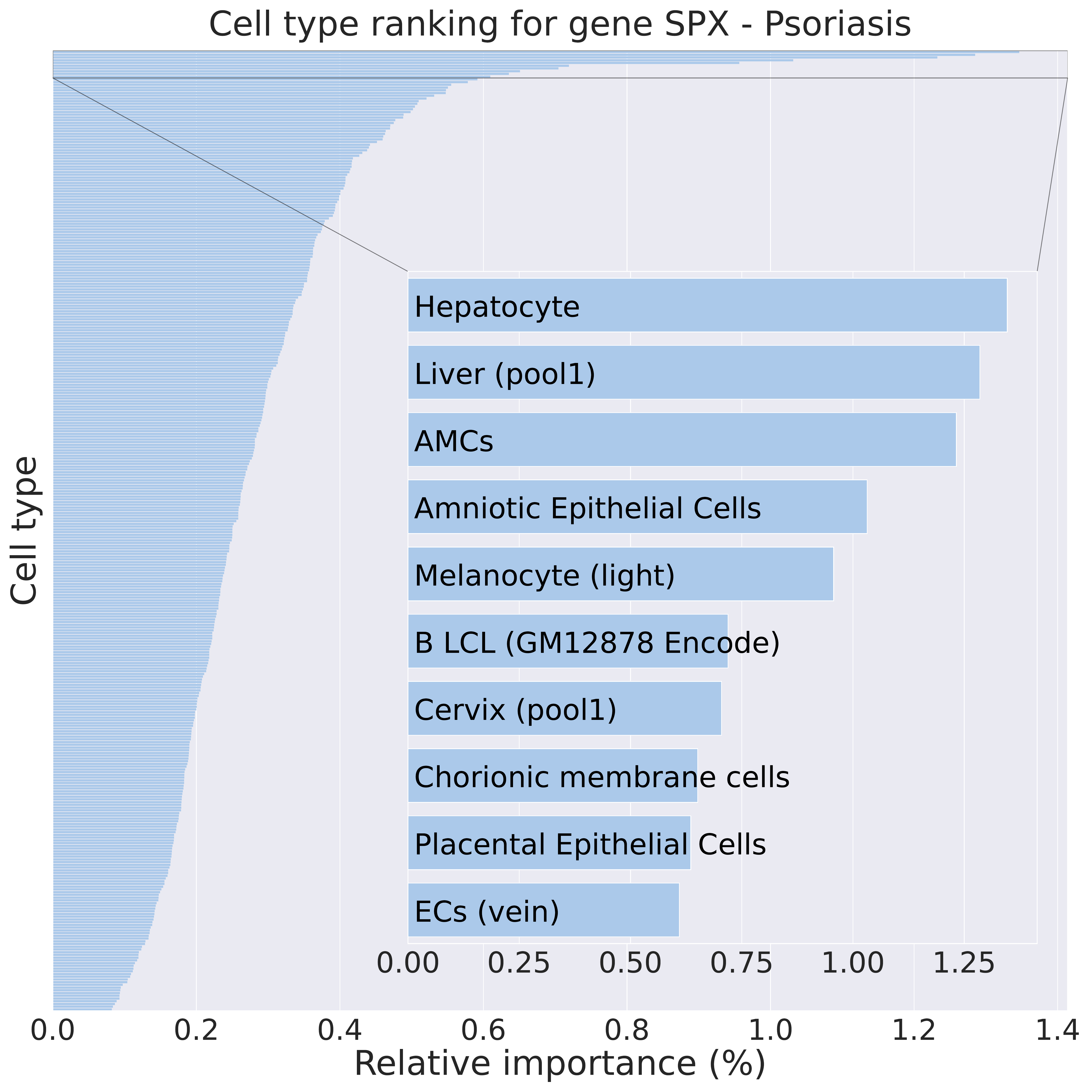}
  \end{subfigure}
\vspace{1em}

\figtitle{Small intestine-associated type 1 diabetes sequence windows}

  \begin{subfigure}[b]{\rowheadersize\textwidth}
    \rotatebox{90}{\hspace{6.8em}\textsf{CYP7A1}}
  \end{subfigure}
  \begin{subfigure}[t]{0.43\textwidth}\centering
	\includegraphics[width=1.0\textwidth, valign=b]{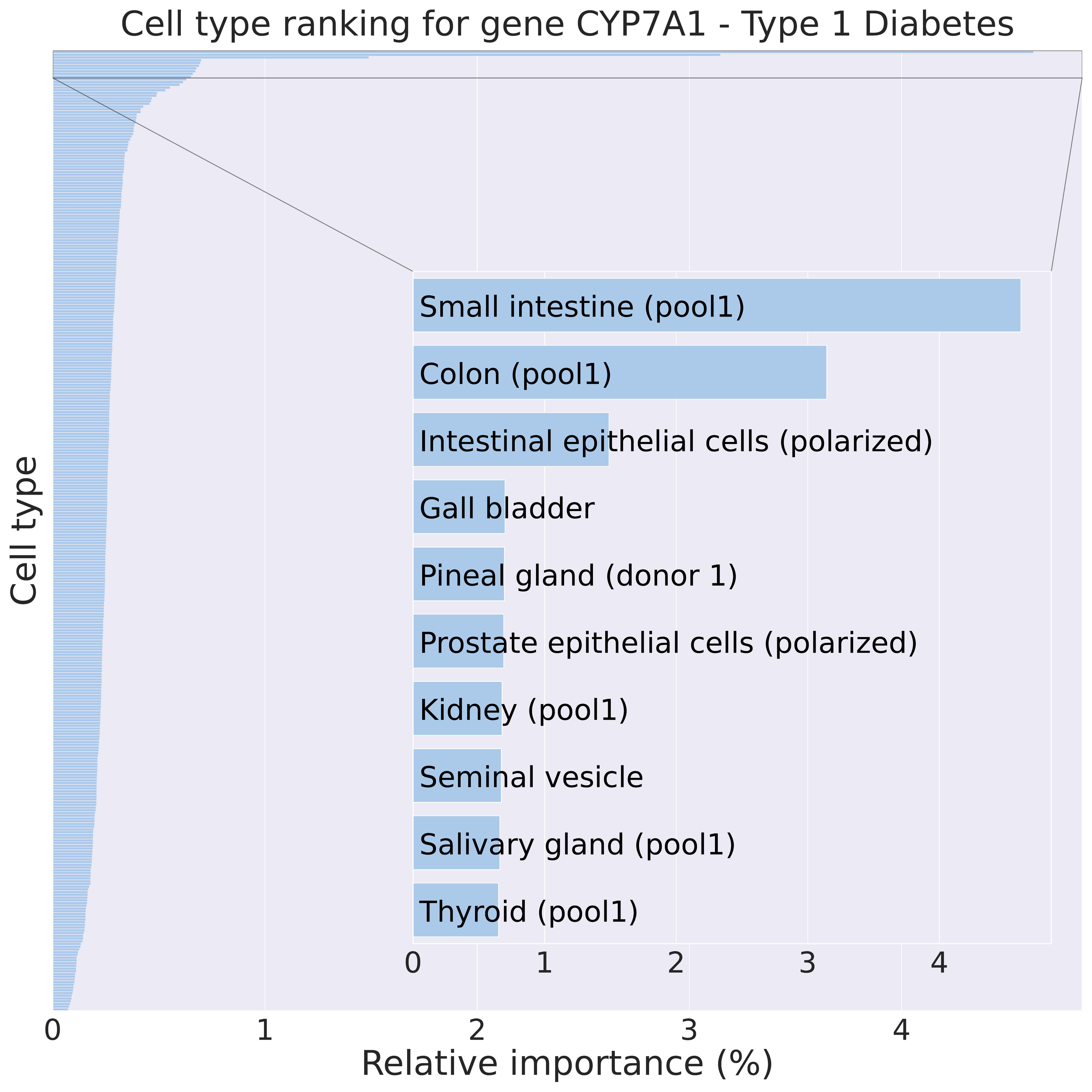}
  \end{subfigure}\hspace{1em}
    \begin{subfigure}[b]{\rowheadersize\textwidth}
    \rotatebox{90}{\hspace{7.0em}\textsf{GIMD1}}
  \end{subfigure}
  \begin{subfigure}[t]{0.43\textwidth}
\centering	
	\includegraphics[width=1.0\textwidth, valign=b]{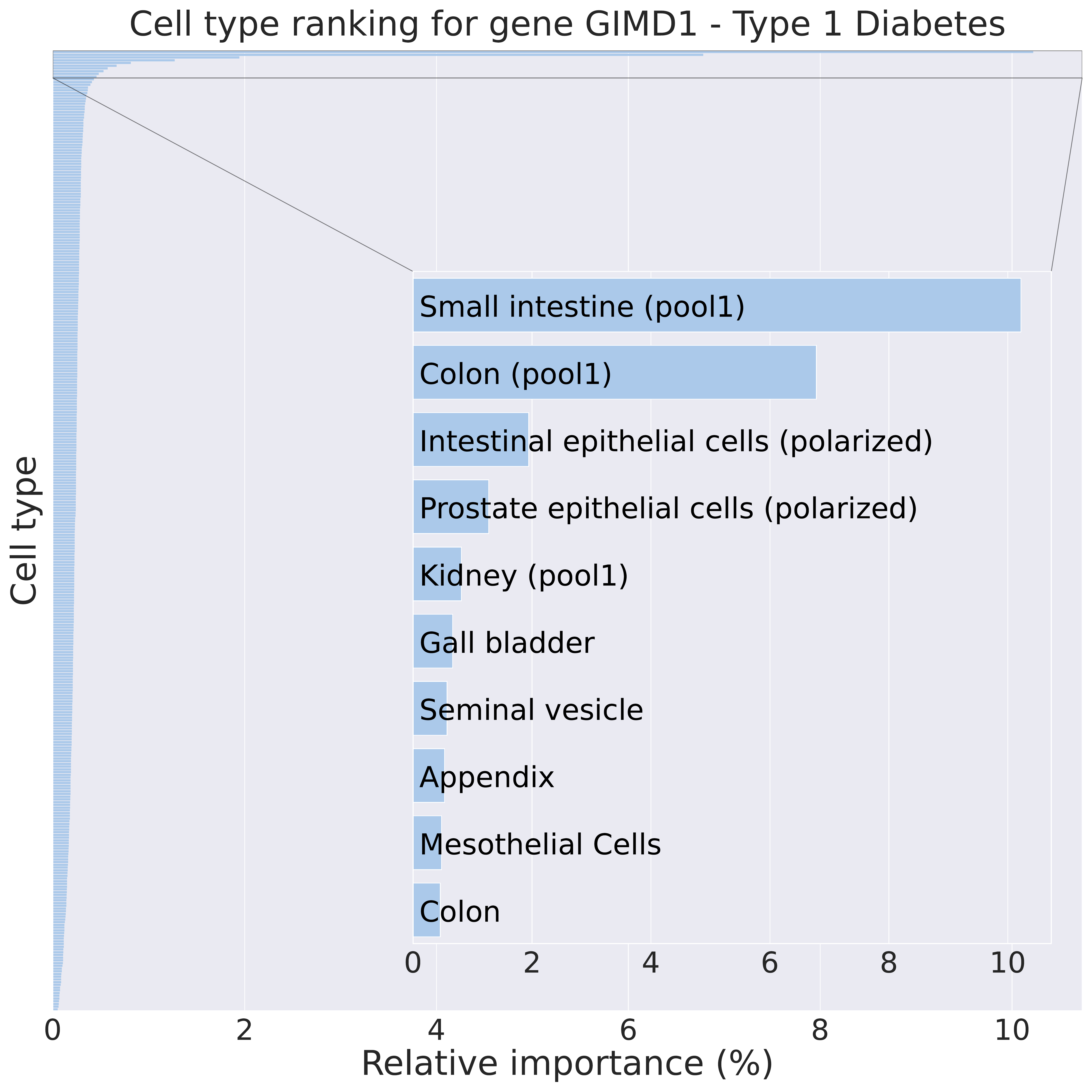}
  \end{subfigure}
  
\caption{\textbf{\themethod{} provides cell type rankings for sequence windows associated with the liver in psoriasis and the small intestine in T1D.} \themethod{} attributions highlight the sequence window around the TSSs of SELENOW (top left) and SPX (top right) as potentially relevant for differential expression changes in liver and hepatocyte cellular contexts in psoriasis-affected individuals (top row), and CYP7A1 (bottom left) and GIMD1 (bottom right) as potentially relevant in the small intestine in T1D-affected individuals (bottom row). We note that SELENOW (CRX, EHD2, NOP53, TPRX1, TPRX2), SPX (GOLT1B, GYS2, PYROXD1, RECQL), CYP7A1 (SDCBP, UBXN2B) and GIMD1 (AIMP1, TBCK) 196 kb sequence windows overlap with multiple other genes which may partially or fully explain the importance assigned to the respective sequence windows (see Section \enquote{\nameref{par:interpretation_meaning}} for additional guidance on interpretation). The ability of \themethod{} to highlight cell and tissue contexts of importance for particular gene sequence windows may provide hypotheses that may help substantiate known - but not yet molecularly understood - disease-associated pathologies, such as for example, increased frequency and severity of non-alcoholic fatty liver disease (NAFLD) in psoriasis patients\cite{prussick2015nonalcoholic} and changes in cholesterol synthesis and absorption markers in T1D patients \cite{semova2019type}.
}
\label{fig:liver_psoriasis}
\end{figure*}

Going one level deeper, we next analyzed the top predicted differential cell types and genes for specific diseases (\Cref{fig:interpretation}, \Cref{fig:interpretation_cont} and \Cref{fig:interpretation_cont2}). We observed that the attributions implicitly learnt by \themethod{} genetically substantiate several epidemiologically and clinically observed clinical disease pathologies of -- to the best of our knowledge -- to date unknown molecular mechanism, such as liver involvement and non-alcoholic fatty liver disease (NAFLD) in psoriasis patients\cite{prussick2015nonalcoholic}, appendicitis\cite{tsai2008complicated,wei2016diabetes} complications in T1D, and optic nerve involvement in COPD\cite{ozge2005cranial,mikaeili2015correlation} (\Cref{fig:liver_psoriasis}, tabular overview in \Cref{tab:qualitative_results}). We note that \themethod{} attributions are best understood as potential mechanistic hypotheses and not necessarily causal (see Section \enquote{\nameref{par:interpretation_meaning}} for further guidance on interpretation).%

\paragraph{\themethod{} improves disease risk prediction from sequence.} We evaluated the relative performance of ensembles of \themethod{} with state-of-the-art PRS methods - including p-value thresholding (Pthres), clumping and thresholding (C+T), lassosum\cite{mak2017polygenic}, LDPred2\cite{prive2020ldpred2}, and PRS-CSx\cite{ruan2022improving} - and compared to the PRS methods alone in terms of Area under the Receiver Operating Characteristic Curve (AUROC) at whole-genome level on the same held-out test set of individuals in predicting major diseases in both mixed and non-European ancestry populations. We found that enhancing existing PRS methods via ensembling with \themethod{} significantly (p $\leq 0.05$) outperforms base PRS methods alone in 86.7\% and 96.7\% combinations of disease and base PRS methods in mixed and non-European ancestry populations, respectively (\Cref{fig:aggregate_performance}a-b; more metrics in \Cref{fig:performance}). Additionally, we compared \themethod{} directly to state-of-the-art PRS methods on the subset of the genome that covers the sequence windows around the 512 genes that \themethod{} was trained on and found that \themethod{} achieves up to 5.49\% and 14.59\% higher prediction performance in terms of average AUROC across diseases for populations of mixed ancestry and with non-European ancestry, respectively (\Cref{fig:aggregate_performance}c-d). These results demonstrate that the comprehensive coverage of whole-genome sequence context provided by \themethod{} considerably enhances risk prediction performance while maintaining better robustness across diverse genetic populations.

\begin{figure*}[pt!] 
\centering	

\figtitle{Subtyping by molecular mechanisms}

\begin{subfigure}[b]{\rowheadersize\textwidth}
    \rotatebox{90}{\hspace{8.5em}\textsf{Psoriasis}}
  \end{subfigure}
  \begin{subfigure}[t]{0.775\textwidth}\centering
\includegraphics[width=\textwidth]{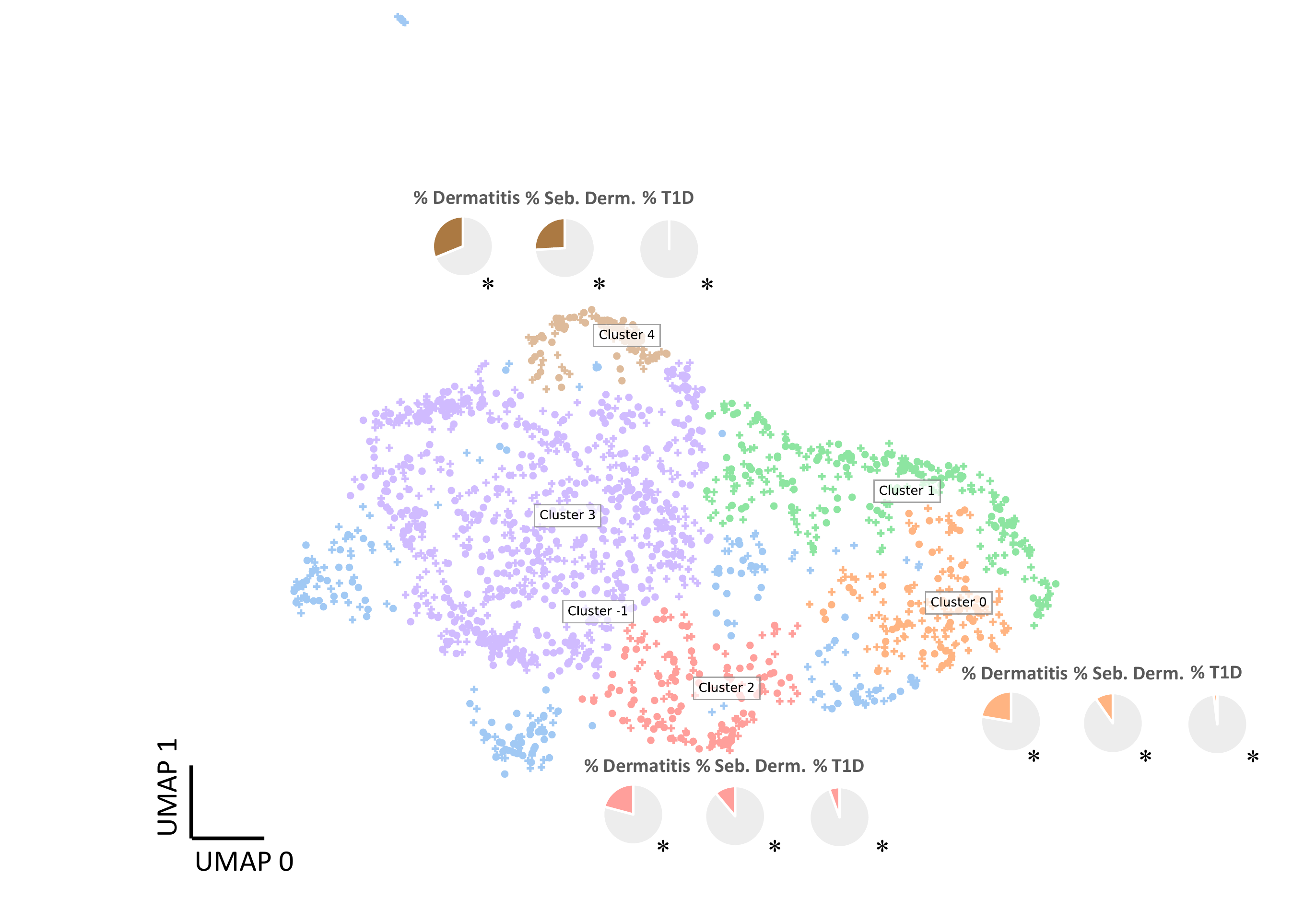} 
\end{subfigure}\quad

  \begin{subfigure}[b]{\rowheadersize\textwidth}
    \rotatebox{90}{\hspace{6.5em}\textsf{Diabetic Retinopathy}}
  \end{subfigure}
  \begin{subfigure}[t]{0.775\textwidth}\centering
\includegraphics[width=\textwidth]{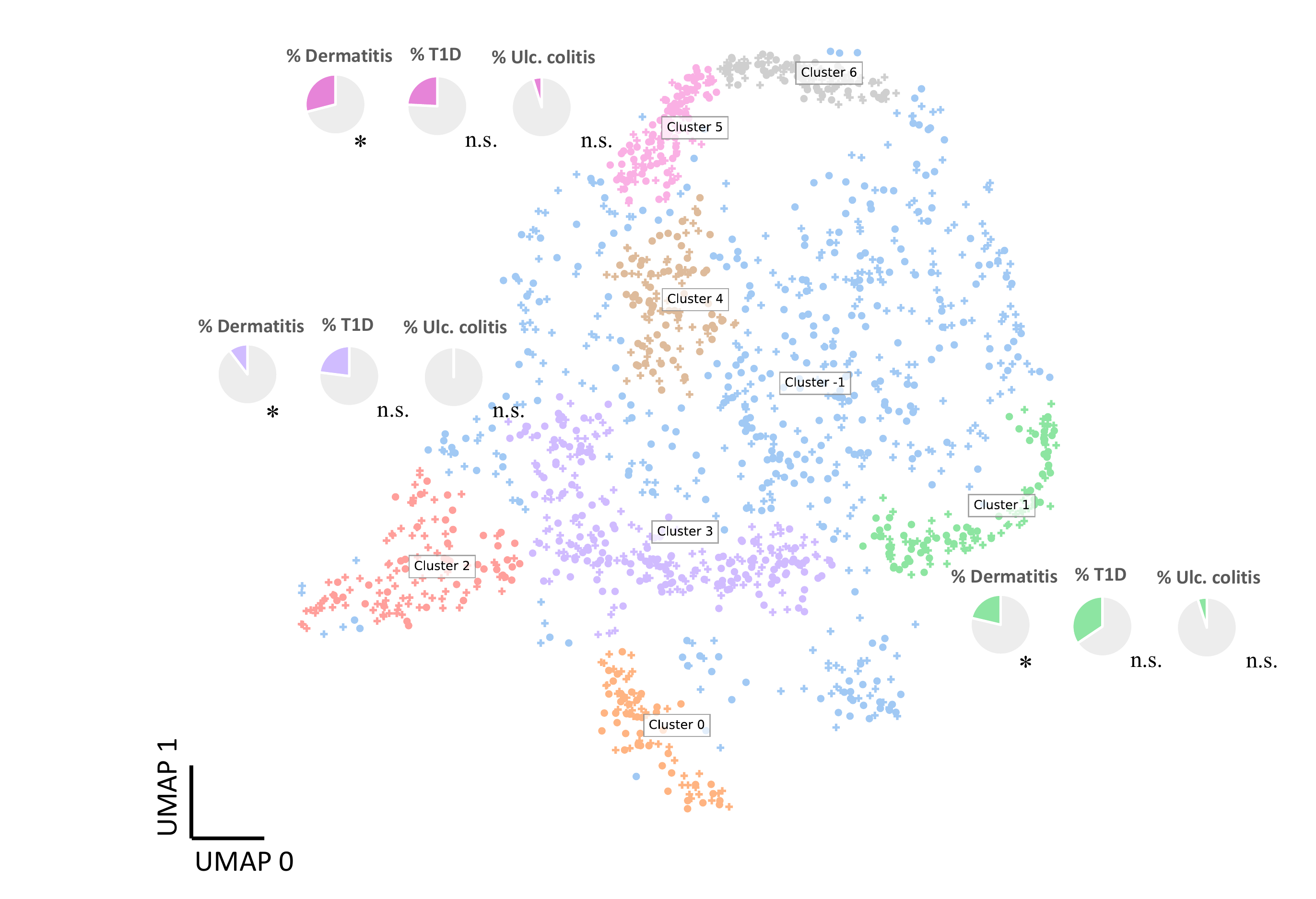} 
\end{subfigure}\quad

\caption{\textbf{\themethod{} embeddings enable grouping of individuals by their underlying differences in disease-related molecular mechanisms.} Latent space embeddings of \themethod{} can be used to subtype individuals according to their differences in molecular processes induced by genetic variation, enabling a fine-grained understanding of molecular subtypes in broader disease categories. Circles and plus (+) symbols represent diagnosed and an equal amount of reference undiagnosed individuals (not used for clustering), respectively. Using \themethod{} trained to predict psoriasis (top) and diabetic retinopathy (bottom; visualised using UMAP \cite{mcinnes2018umap}), we identified molecular subtypes (colors with associated cluster labels). Molecular subtypes were associated with differences in terms of co-morbidity rates (pie chart insets) among diagnosed cluster members (highlighted for clusters with the largest differences). We find statistically significant (* = p $\leq 0.05$; $\chi^2$ test) differences in dermatitis, seborrheic dermatitis and T1D comorbidity rates in psoriasis subtypes, and in dermatitis in diabetic retinopathy subtypes - suggesting differences in underlying molecular processes identified by the \themethod{} embeddings of individual genomes. Subtype differences in T1D ($p=0.0684$) and ulcerative colitis ($p=0.1374$) in diabetic retinopathy do not reach significance (n.s.).}
\label{fig:clusters}
\end{figure*}

\paragraph{\themethod{} highlights subtypes of disease putatively governed by different mechanisms.} In addition to identifying cell and tissue types that contribute to disease risk across a population, \themethod{} also enables the analysis of genetic variation on the level of subgroups and individuals. To demonstrate these capabilities, we visualise a latent space embedding of individuals based on their individual \themethod{} attributions (\Cref{fig:clusters} for psoriasis and diabetic retinopathy and \Cref{fig:clusters_more} for others). We found that \themethod{} trained to predict disease risk identified molecular clusters that were characterised by significant (p $\leq 0.05$) differential prevalence of disease-related co-morbidities, including for example a psoriasis subtype associated with higher dermatitis and seborrheic dermatitis risk (cluster 4) and a diabetic retinopathy subtype associated with higher dermatitis risk (cluster 5). The presence of differential co-morbidity risk by subtypes suggests that \themethod{} is able to stratify individuals by their differences in underlying molecular processes caused by genomic variation.

\section{Discussion}

We present \themethod{}, an end-to-end multi-scale deep learning model to associate individual genomes with disease phenotypes directly from DNA sequence. To the best of our knowledge, we demonstrate for the first time the computational and methodological feasibility of integrating an order of magnitude larger fraction of individual genomes in an end-to-end model connecting sequence, molecular mechanisms and disease susceptibility, demonstrating performance that exceeds that of existing state-of-the-art methods on the same data -- an achievement that was not previously known to be within reach of current technology. 

\themethod{} opens up new avenues for interpretation of how and where disease risk may be conferred through its latent space representations that are grounded in context-dependent gene expression and epigenetic features. Quantitatively, we found that the associations between cell and tissue types and diseases identified by \themethod{} are more enriched for those reported in literature than the associations reported by state-of-the-art methods that use genetic information in addition to requiring additional experimental data, such as single-cell RNA sequencing (scRNAseq) data. Additionally, we qualitatively found that the disease-sequence-expression-cell type relationships highlighted by \themethod{} provide genetic substantiation for clinically and epidemiologically observed, but, to our knowledge, not yet molecularly understood, disease-associated pathologies such as for example, increased frequency and severity of non-alcoholic fatty liver disease (NAFLD) in psoriasis patients\cite{prussick2015nonalcoholic} and heightened risk for appendicitis complications in T1D\cite{tsai2008complicated,wei2016diabetes}. These findings are notable because \themethod{} provides accurate and fine-grained mechanistic attributions on the level of individual genomes - which may in the future enable not only the prediction of risk but also which pathological changes and disease symptoms may be expected by an individual based on their genetic background.

In terms of predictive performance, \themethod{} is able to more comprehensively account for gene-gene interactions and rare variants than existing methods by incorporating a considerably larger fraction of the genetic sequence context into its predictions. Our experimental results further show that the integrative approach to modeling represented by \themethod{} leads to significant gains in predictive performance in quantifying individual disease susceptibility.  Additionally, we determined that \themethod{} improves transportability to diverse, non-European ancestries over using existing PRS methods alone. We hypothesize that this is an effect of the preconditioned gene-to-expression backbone of \themethod{} that helps combat the overfitting that is commonly observed when training on SNP data without sequence context\cite{torkamani2018personal}. \themethod{} may therefore potentially be an effective approach for individual genome interpretation that addresses the poor transportability of existing methods for quantifying genetic disease risk limits to more diverse cohorts \cite{gyawali2023improving}.

A limitation of the presented study is that - for computational reasons and the need for even larger-scale training data - a selected subset of 3\% of the entire genome sequence of individuals was available to \themethod{} for predictions and training. While 3\% of the genome is an order of magnitude more comprehensive coverage of individual genomes than existing methods, it is likely that the performance of \themethod{} could be improved by increasing the coverage of the genome sequence further. The incomplete sequence context may also be a challenge when interpreting the mechanistic hypotheses highlighted by \themethod{} since highly predictive variant-induced changes in risk may have been missed if they were outside of the sequence region available to \themethod{}. The selection of gene sequence regions for inclusion into predictions of \themethod{} is biased towards regions around the TSS of included genes due to the sequence-to-expression backbone utilized. Nonetheless, the experimental results show that - already at the training context size presented herein - \themethod{} considerably improves genome-wide risk prediction and interpretation. Furthermore, in a similar vein and also due to computational limitations, the presented study only included WGS data from \numprint{150076} individuals and therefore does not reach the same population sizes as some of the largest genetic studies conducted to date in up to \numprint{500000} to up to millions of individuals\cite{bellenguez2022new,yengo2022saturated}. We expect that future studies may expand the genome coverage and the size of training datasets of sequence-to-phenotype models as the limits of hardware and software shift and more WGS data is made publicly available. We note that the type of variants studied in this work is constrained to SNPs - although insertions and deletions (indels) could technically be processed by \themethod{} in sequence. Furthermore, while the experimental evidence supporting end-to-end sequence-to-phenotype models via a sequence-to-expression backbone is encouraging, it is also clear that there are several areas for potential future methodological improvements. For example, the sequence-to-expression backbone of \themethod{} has not been trained specifically for variant-induced effect prediction and has in related work been demonstrated to therefore perform not particularly well at this task \cite{sasse2023far}. Although \themethod{} uses the derived embeddings from the sequence-to-expression backbone as tokens (which also reflect chromatin predictions and may be more robust than using the expression predictions themselves), a suboptimal sequence-to-expression backbone may not fully capture the elements of the sequence context that are relevant for disease risk prediction and therefore could potentially be reducing the overall predictive performance of \themethod{} - presenting an avenue for future improvements. Finally, like any predictive tool for genetic susceptibility, \themethod{} must be scrutinized through an ethical lens. The potential to influence decisions based on genomic predispositions raises concerns about data bias\cite{chauhan2024sampleselectionbiasmachine}, the risk of misinterpretation and misuse, broader social implications such as genetic discrimination, and potential unintended biases within the data can inadvertently lead to inequitable healthcare outcomes. While we presented evidence on the potential robustness of the performance of \themethod{} in individuals of non-European background, the dataset considered in this study is known to be biased towards healthy volunteers\cite{fry2017comparison,schoeler2023participation} and it is paramount to critically assess the predictions of \themethod{} in diverse scenarios and ensure that they align with society's expectations towards ethical and responsible healthcare.

\themethod{} is a powerful approach to sequence-based genome interpretation that enables both a deeper understanding of molecular mechanisms involved in disease and improved disease risk prediction on the level of individuals. As such, \themethod{} considerably improves our ability to model whole genome sequences across biological scales and could therefore in the future be used to better understand and interpret individual genomes, including how genetic variation gives rise to differences in bio-molecular processes and how these differences contribute to disease risk.

\renewcommand{\figurename}{Supplementary Figure}
\renewcommand{\thetable}{S\arabic{table}}
\renewcommand{\thefigure}{S\arabic{figure}}
\setcounter{figure}{0}

\section{Materials and methods} 
\label{sec:methods}

\subsection{\themethod{} -- Neural Architecture}

\paragraph{Step 1: Tokenization via Enformer embeddings.} In a first step, we infer sequence embeddings from a pretrained and frozen sequence-to-expression model. Here, the Enformer model\cite{avsec2021effective}, a state-of-the-art gene expression model, is being leveraged for this task. Enformer was trained to predict gene expression from input sequence windows of \numprint{1536} tokens each comprising \numprint{128} base pairs. Specifically, we extract the \numprint{3072}-dimensional embedding vector which is being passed to predict the \numprint{5313}-dimensional human gene expression and epigenetic output (and \numprint{1643}-dimensional mouse output) from the token centred around the transcription start site (TSS) from the two Enformer prediction heads. In total, the model is given $m$ sequence embeddings per individual from $m$ distinct raw sequence windows chosen (details in the \enquote{Data} paragraph below). The rationale of using these sequence embeddings is to capture the genetic variation of \numprint{196} kilobases (kb) long sequence windows centred around the TSS in compressed lower-dimensional tractable vectors that serve as tokens for the subsequent self-attention blocks.  In cases where a gene has multiple TSSs inside the \numprint{196} kb window, the embedding vectors of the tokens containing TSSs are averaged into a single embedding vector. 

\paragraph{Step 2: Process via \themethod{} backbone.} In the second step, the $m$ sequence embeddings serve as the input tokens of the \themethod{} backbone. First, the sequence embeddings are further down-projected using a shared two-layer Multi-Layer Perceptron (MLP) to $d_\text{model}=512$ dimensional embeddings. The size of hidden layers is 512 units. We add a 512-dimensional positional Fourier encodings to each down-projected token, which provides the necessary information about the genetic location of each vector. The $m$ 512-dimensional embeddings are further processed by four Transformer encoder layers\citep{vaswani2017attention}. A Transformer encoder layer is a neural network that seeks to learn a rich representation of its input. It comprises two main sub-components: (1) a multi-head self-attention mechanism and (2) a position-wise feed-forward network that facilitates the modeling of interactions between long-ranged variations in the genome of an individual. The position-wise feed-forward network transforms the output of the attention mechanism. Residual connections and layer normalization are applied around each sub-layer, facilitating deeper stacking of these layers and aiding in the model's convergence during training. We use 8 heads, pre-layer normalization, 0.2 as dropout rate, \numprint{2048} as the dimension of the feedforward network model as well as a variant of the gated linear unit (GLU) activation function, GEGLU\cite{shazeer2020glu}, throughout the model.

\paragraph{Step 3: Predicting disease risk.} Finally, the $m$ processed embeddings from the Transformer layers are pooled into a single representation. We use Pooling by Multihead Attention (PMA)\cite{lee2019set}. The PMA module incorporates one learnable query vector and the resulting pooled embedding is finally passed to a 2-layer (256, 128) MLP head which outputs a single-disease logit. Optionally, additional information such as age, sex and HLA type of the individual can be introduced to the model at the PMA layer, or adjusted for after model training.

\subsection{Data, Training and Evaluation Pipeline}

\paragraph{Data.}
\themethod{} was trained on the whole genome sequencing data of \numprint{150119} individuals with disease annotations\cite{kuan2019atlas} in the UK Biobank\cite{halldorsson2022ukb150k}. The dataset included the \numprint{150076} individuals who had both WGS and disease annotations available. These individuals formed the basis of a 60\%-20\%-20\% train-validation-test set split (\numprint{90046}, \numprint{30015} and \numprint{30015} individuals respectively), stratified across the 294 available disease labels using iterative stratification\cite{sechidis2011stratification,szymanski2017stratification}. For all experiments, we keep the validation and test set fixed. We studied the following 6 major diseases: Psoriasis, Type 1 Diabetes, Type 2 Diabetes, Diabetic Retinopathy, Hypothyroidism, and COPD. Diseased to control individual ratios were strongly imbalanced (\Cref{tab:disease frequencies}).
Training \themethod{} models requires large amounts of data and compute, especially when dealing with large numbers of tokens. We found training on the full set of \numprint{21725} TSS-centred windows corresponding to all annotated genes to be not feasible since the computational resources needed to train and infer grow quadratically in the number of input tokens passed due to the use of an attention mechanism in the model. We therefore focused on a subset of 512 genes, selected for their putative relevance in immune disorders. This dataset of 512 sequence windows centred on TSS of the selected genes corresponds to roughly 88 million base pairs or 3\% of an individual genome. However, we note that \themethod{} is not intrinsically limited in the size of gene sets it can process, and, with further progress in computation, processing even larger context windows may become feasible in the future.

\paragraph{Disease annotations.} The disease annotations for the diseases included in this study were based on validated phenotypes following the methodology described in \citet{kuan2019atlas} and integrating primary care records, hospital episode statistics, cancer and death registries, and UK Biobank health questionnaires including self-reported illnesses. We encoded individual disease status as either presence or absence of the disease annotation, and \themethod{} was trained to classify disease status based on the input whole genome sequencing data. 

\paragraph{Gene set selection.} Since including the full \numprint{21725} annotated genes was not technically feasible, we aimed to build an informed gene subset consisting of 512 immune-associated genes that we subsequently used for training and evaluation across all target diseases included in this study. For this gene subset, we aimed to include the genes with the most significant variance in expression between immune disease-affected individuals and controls. We therefore employed a heuristic approach: we first selected a cohort of 100 diseased individuals and 50 healthy controls for reference, and then utilized the Enformer to predict changes in Cap analysis of gene expression (CAGE) across \numprint{21725} annotated genes, omitting those on the Y chromosome. We chose psoriasis as the reference immune-disease due to its high prevalence in the UK Biobank population. We assessed the predicted CAGE changes using two metrics: log2-fold change and absolute change, setting a threshold of 0.5 for both. Through this approach, we identified a set of 206 genes by comparing the median gene expression of the immune-disease group against the median of the control group. These genes were those that exceeded our threshold criteria in the median-vs-median comparison. Subsequently, we expanded the gene selection by including genes that met the threshold in at least 50 of the 100 immune disease-affected participants. This step was based on a control median vs. individual comparison, which added 405 additional unique genes to the gene set. The final combined set comprised 611 genes, from which we selected the top 512 based on the magnitude of their relative log2-fold change. One gene (ZNF835) that was selected had to be excluded from \themethod{} input due to a numerical instability in generating Enformer embeddings. 

The selected gene set of 511 includes: ZG16B, EPN3, NIPAL4, STEAP4, BAIAP2L2, FCN1, USP7, EPS8L3, ENSG00000261147, SPDYC, NAB1, TPBG, TTLL13, KRT24, IL18, FAM110D, F12, ACER3, CAPN9, C1QA, LIPI, LGR6, SLC25A21, S100P, PLCD4, NRG4, SLC25A18, CD3D, C3AR1, DENND2D, ENSG00000285868, SLC1A7, CLPS, PRRG2, MMP7, SULT2A1, PRSS3, ASNSD1, ADH1C, OLIG2, SAA1, MCF2L, MSTN, SLC28A2, AHSP, NUAK2, TJP3, ENSG00000285188, ENSG00000284797, PDGFRB, PRSS58, PCDHA13, NTSR2, FAT2, LDB3, CRISP2, PPEF2, LILRB1, H2AC1, VIL1, SMIM31, IGSF23, CA12, C1QB, KIR2DL3, CHADL, PDE1B, MYH11, PDIA2, CHST13, DYNLT5, ZSCAN1, CCR2, ARHGEF10, TRIM31, SMPDL3B, LHFPL5, ARRDC5, LEFTY1, HLA-DQB1, ASDURF, CYP7A1, CLCNKA, SLC6A19, TTC29, DMBT1, PPBP, MS4A2, CHI3L1, HLA-DQA2, ANKRD2, KLHDC8B, MYH8, SSUH2, PXDNL, TOMM34, GMNC, CES1, MED25, MMP13, SLC26A1, C4orf17, FTCD, PPIC, OPTN, CALHM6, NCCRP1, REG3G, FAM163A, RCCD1, NNMT, TMEM176B, OGN, SERPINB13, HLA-F, APOL4, CCDC80, ZNF483, ENSG00000286165, CD53, COX7B2, CST4, INHBE, FABP2, FBL, TPSD1, SLC1A1, ABHD8, CD40, SLC27A6, CXCR6, SFRP2, RNF112, RNF207, MOB2, ZSCAN5B, C1GALT1C1L, AFAP1, AHI1, HSH2D, CD96, CNTLN, VWDE, INPP4B, STAC, WSCD2, FOXS1, F2RL1, PRSS54, SAA2-SAA4, PHLDB2, PDZD7, PRH2, PGM5, MPC1, ZNF165, RRH, GAS2, ADAM32, THY1, ANKRD34C, ELP3, SLC7A9, PRSS21, ZNF208, PTPRE, CDSN, MS4A6E, MYBPHL, ZNF91, HDC, CPLANE2, RRN3, CCDC148, RAB17, CYP4F12, CDK15, SBK2, KCNK13, POU2F3, PHETA2, RILPL2, USP37, WDR6, IFIT3, PKD1L1, IL19, DEPDC7, DPP10, N4BP1, NME6, SLC2A8, PSORS1C1, CCDC163, ABO, ZNF268, ENSG00000268870, SV2C, RHBDD1, SOSTDC1, ZSCAN12, ANKRA2, SESN3, HMGCS2, IKBKE, GABRP, PCSK1, AGT, HEPHL1, LRR1, POM121L12, ITIH4, MUSTN1, ASPA, B3GNT3, TRPV3, PGR, TRPC6, LRTM1, MMP8, DNASE1L3, SLC35F2, IGFBP2, ADAMTS9, CTXN3, UTS2, ALOX12, ZNF708, SLC2A7, SLC2A5, IHH, PTGES, GPX2, SPEGNB, NXPE1, GSAP, NUDT19, KCNE4, JAML, AKR1C2, TUBAL3, ALOX15B, SLC19A3, ENSG00000285635, STEAP2, MYMK, RNF222, COL5A1, KRTDAP, PDPN, GLP2R, UCMA, GAS7, MYH2, MYH3, PROX2, TRAT1, HSPB7, PON1, NBPF1, UGT1A9, MFAP2, UGT1A3, LRRC74A, LCN12, TMEM63C, COL6A3, ZNF80, ZFP30, PLA2G2E, PLA2G2A, UPK1B, PATE1, PDSS1, GPR35, AGXT, LYZL1, LYZL2, FGF1, FBXO27, ALDH3A1, SLC47A2, MUC3A, LGALS9, SERPINA6, SCGB3A2, SLC52A3, SERPINA11, SERPINA4, SERPINA5, IL22RA1, SERPINA3, C14orf132, MSMB, UROC1, CPXM1, ZNF488, EXTL1, CNKSR1, CDHR3, ADAM33, TMEM273, GALNT8, SLC26A3, HAVCR1, CCL11, FABP6, CD177, ETHE1, LAPTM5, GABRA6, DZIP1L, TINAGL1, LMOD2, CFAP61, A2M, CLEC2B, CLEC12A, CLEC1B, ADAMTS14, APOC2, CST5, TRPM1, PRR4, PRH1, TM4SF4, DEFB119, GSDMA, CSF3, DUSP29, STRA8, SUCNR1, SFTPA2, KRT12, KRT23, HPCAL4, SELENOW, ELSPBP1, ADIRF, SPTSSB, EDN2, PRSS1, PRSS2, SPX, KRT19, PLA2G4E, F13A1, CNTNAP2, LBP, SYCP2L, MCF2L2, RARRES2, GIMAP4, JPH2, MUC19, TMEM176A, AOC1, WFDC12, SLC27A2, PDZK1IP1, ADIPOQ, GFAP, MMP9, SIGLEC6, SIGLEC5, SIGLEC14, FPR1, ATP13A5, MYZAP, GCOM1, LRRC15, DEFB1, FAM43A, FAIM2, FAM151A, CYP17A1, C2CD4A, ZNF391, BCAS1, L1TD1, VSTM1, OSCAR, LILRB2, BLK, LILRA2, LILRB4, HLA-G, ITGA11, SLC18A2, MUC21, C6orf15, PSORS1C2, MUCL1, SFTPC, PHYHIP, MCCD1, SPATA1, LPAR3, MCOLN2, SAMSN1, DUXA, CLCA2, FANK1, CLCA1, CYYR1, ZNF772, ZNF419, FGFBP2, CRABP1, HLA-DRA, HLA-DRB1, HLA-DQB2, HLA-DOA, HLA-DPB1, HLA-DPA1, SOD3, ENSG00000288681, SNTG2, AVIL, CRACR2B, CD300H, CLIC6, MUC2, MUC5B, CHRNA9, AGBL1, CLPSL1, LSP1, IFNG, PSRC1, IL22, INS-IGF2, ETV7, PLIN1, GSTM4, GABRA4, PI16, CNGA1, FAM3B, UMODL1, TFF3, TFF2, UBASH3A, UNC5CL, ERVH48-1, TREML4, OLFML3, EPHA5, HBE1, CENPC, TMPRSS11A, KRTAP12-3, SULF1, TRIM22, COL6A2, TRPA1, AMTN, HAL, CXCL1, CXCL5, EREG, SYCP3, ODAPH, CA2, GJA5, CNGB3, TYMS, EMILIN2, CMKLR1, DLGAP1, ABCG5, ABCG8, ABCC8, LHCGR, ENSG00000279956, SPP1, MRGPRX3, SAA2, PTPN5, MRGPRX1, LCE2A, SLC6A5, GAS2L1, ADH4, SPRR3, DAPP1, CGA, SPRR2B, TMEM233, DSG1, DSG4, DSG3, CCN3, ENPP2, TCN2, GIMD1, ELF5, APIP, CD44, COQ3, CLEC4F, ATP6V1B1, SLC14A1, TMC5, GPRC5B, SYNPO2, APOL1, ACTG2, LIPG, QRFPR, CYTH4. 

\begin{table}[]
\vspace{-1.5em}
    \centering
    \begin{tabular}{l|rrr}
         Disease\hspace{9.4em} & \hspace{9.1em}Train & Validation & Test   \\
         \midrule
         Psoriasis & \numprint{2729} (\numprint{87317}) & \numprint{910} (\numprint{29105}) & \numprint{909} (\numprint{29106}) \\ 
         Type 1 Diabetes & \numprint{863} (\numprint{89183}) & \numprint{287} (\numprint{29728}) & \numprint{280} (\numprint{29735}) \\ 
         Type 2 Diabetes & \numprint{7526} (\numprint{82520}) & \numprint{2488} (\numprint{27527}) & \numprint{2468} (\numprint{27547}) \\ 
         Diabetic Retinopathy &\numprint{2291} (\numprint{87755}) & \numprint{756} (\numprint{29259}) & \numprint{800} (\numprint{29215}) \\ 
         COPD & \numprint{4152} (\numprint{85894}) & \numprint{1385} (\numprint{28630}) & \numprint{1384} (\numprint{28631}) \\
         Hypothyroidism & \numprint{7005} (\numprint{83041}) & \numprint{2297} (\numprint{27718}) & \numprint{2315} (\numprint{27700}) \\ 
         \bottomrule
    \end{tabular}
    \caption{\textbf{Distributions of diagnosed and undiagnosed individuals.} Counts of individuals with disease diagnoses in train-validation-test set for each of the major disease investigated. Counts for controls (no diagnosis of the respective type) are shown in parenthesis.}
    \label{tab:disease frequencies}
\end{table}

\paragraph{Training.} \themethod{} models were trained using \texttt{pytorch}\cite{paszke2019pytorch} on distributed Nvidia A100 DGX and Tesla V100 HGX environments with a total batch size of 512. We minimized the cross entropy loss using the Lion\cite{chen2023lion} optimizer with parameters $\beta_1=0.95$, $\beta_2=0.98$ with disease-frequency weights to counteract the imbalance in the dataset.\cite{loshchilov2017decoupled} The models were being trained for a total of \numprint{35175} steps (200 epochs). We further applied a learning rate schedule that increases linearly for the first \numprint{350 000} samples (\numprint{684} steps) from 0 to $3e^{-6}$. Additionally, a weight decay factor of $0.01$ was applied. Once trained, models were selected based on the best AUROC validation set performance. To combat overfitting, we added normally-distributed noise to each training sample, scaling the noise for each feature by \numprint{10} to \numprint{40}\% of the feature's range, then optionally further scaling each sample's noise magnitude by a unit log-normal distribution such that the model saw a mixture of low-noise and high-noise samples. We observed greater amounts of noise delayed or prevented overfitting, but often with the trade-off of reduced accuracy. Additionally, the amount of noise and whether to apply per-sample noise scaling was tuned separately for each disease, requiring 4 models to be trained per disease to find the best parameters. For the ensemble models, we incorporate logistic regression using as inputs a baseline polygenic risk score and the probability predictions from the Phenformer model. We conduct a grid search to optimize hyperparameters, examining both L1 and L2 regularization strategies, as well as varying the inverse regularization strength parameter $C$ over a range from $1 \times 10^{-7}$ to $1 \times 10^{7}$. Stratified 20-fold cross-validation is employed to determine the best-performing hyperparameters.

\paragraph{Baselines.}
As a baseline comparison, we conduct a Genome Wide Association Study (GWAS) and derived corresponding Polygenic Risk Score (PRS) models\cite{hayes2013overview, lewis2020polygenic} using the exact same genomic information as provided to \themethod{}. This entails considering all the Single Nucleotide Polymorphisms (SNPs) within the sequence windows of the genes under investigation.
To ensure a fair comparison and mitigate potential biases, we correct for ancestral bias which is a well-known issue in GWAS, especially in highly imbalanced datasets. We leverage the \texttt{hail} package\cite{Hail}, which provides robust methods for controlling for population stratification and ancestral bias and \texttt{regenie}\cite{mbatchou2021computationally} for performing whole genome regression modeing. We compute the top 10 principal components of the genotypes using \texttt{plink}\cite{purcell2007plink} and use them as covariates in the GWAS model. We train five baseline PRS models: p-value thresholding (Pthres), clumping and thresholding (C+T), lassosum\cite{mak2017polygenic}, LDPred2\cite{prive2020ldpred2}, and PRS-CSx\cite{ruan2022improving}. For C+T, we use \texttt{plink} to perform clumping with a distance threshold of 250 kb and LD threshold of $0.1$. For lassosum, we leverage the \texttt{bigsnpr}\cite{prive2018efficient} and \texttt{lassosum} R packages to compute LD blocks and to perform the rest of the training and evaluation, respectively. Additionally, the \texttt{bigsnpr} R package is used for the LDPred2 baseline due to its convenient availability within the package. The baseline PRS-CSx is a multi-discovery method that utilizes multiple GWAS summary statistics as input. To support this, we identify 5 distinct superpopulation clusters within our data. This is achieved by applying HDBSCAN\cite{campello2013density} clustering from \texttt{scikit-learn}\cite{pedregosa2011scikit} to the principal components, derived from a subset of SNPs. These SNPs are selected by first filtering for the most common variants and subsequently pruning variants that are highly correlated using \texttt{plink}'s LD-based variant pruner. For each population cluster, we compute LD blocks using the \texttt{bigsnpr} R package and we use \texttt{regenie} to perform GWAS and compute summary statistics suitable for PRS-CSx. For all baselines, we use the same train-validation-test split we use for our \themethod{} model. We use the validation set to optimise the p-value threshold for the p-value thresholding and C+T PRS as well as $\lambda$ and $s$ parameters for lassosum PRS.
By conducting the GWAS and training the PRS models using the same genomic information and correcting for ancestral bias, we established a comparable reference to assess the relative predictive power and accuracy of \themethod{}.

\paragraph{Evaluation.} We evaluate the performance of our model using two metrics: the Area Under the Receiver Operating Curve (AUROC) and Area under der Precision Recall Curve (AUC-PR). To ensure a comprehensive evaluation, we conduct experiments on three test sets: (1) the full 20\% test set, (2) a subset of the previous test set consisting exclusively of individuals with white European ancestry, and (3) a complementary test set comprising individuals who are non-white and non-European.
The use of these test sets allows for a fair and transparent evaluation, particularly in addressing the ancestral bias issue that often arises when dealing with highly imbalanced datasets in GWAS and PRS scores. By including test sets (2) and (3), we aim to shed light on any potential biases or discrepancies in the performance of \themethod{} across different ancestral groups.
For each test set, we calculate both, the AUROC and AUC-PR. The AUROC provides a measure of the ability of \themethod{} to discriminate between positive and negative instances, considering the full range of classification thresholds. Higher AUROC indicates superior performance in distinguishing between diseased and control individuals.
The AUC-PR offers a different perspective on model performance more focused on the positive class. In situations where the classes are imbalanced, with many more negative instances than positive, AUC-PR becomes especially important. It evaluates the trade-off between precision (the fraction of true positive predictions among all positive predictions) and recall (the fraction of true positive predictions among all actual positive instances). A higher AUC-PR value suggests that the model is adept at identifying true positives without incurring many false positives, making it a crucial metric for assessing the model's capability in scenarios with fewer positive instances.
Through this comprehensive evaluation, we aim to assess the performance of our model in a rigorous and transparent manner, accounting for potential biases and challenges associated with imbalanced GWAS datasets. We used the Mann-Whitney Wilcoxon test to assess statistical significance.

\begin{figure*}[pt!] 
\centering	

\figtitle{Sequence region and cell and tissue type attributions for risk predictions}

  \begin{subfigure}[b]{\rowheadersize\textwidth}
    \rotatebox{90}{\hspace{7.5em}\textsf{Psoriasis}}
  \end{subfigure}
  \begin{subfigure}[t]{0.47\textwidth}\centering
	\includegraphics[width=1.0\textwidth, valign=b]{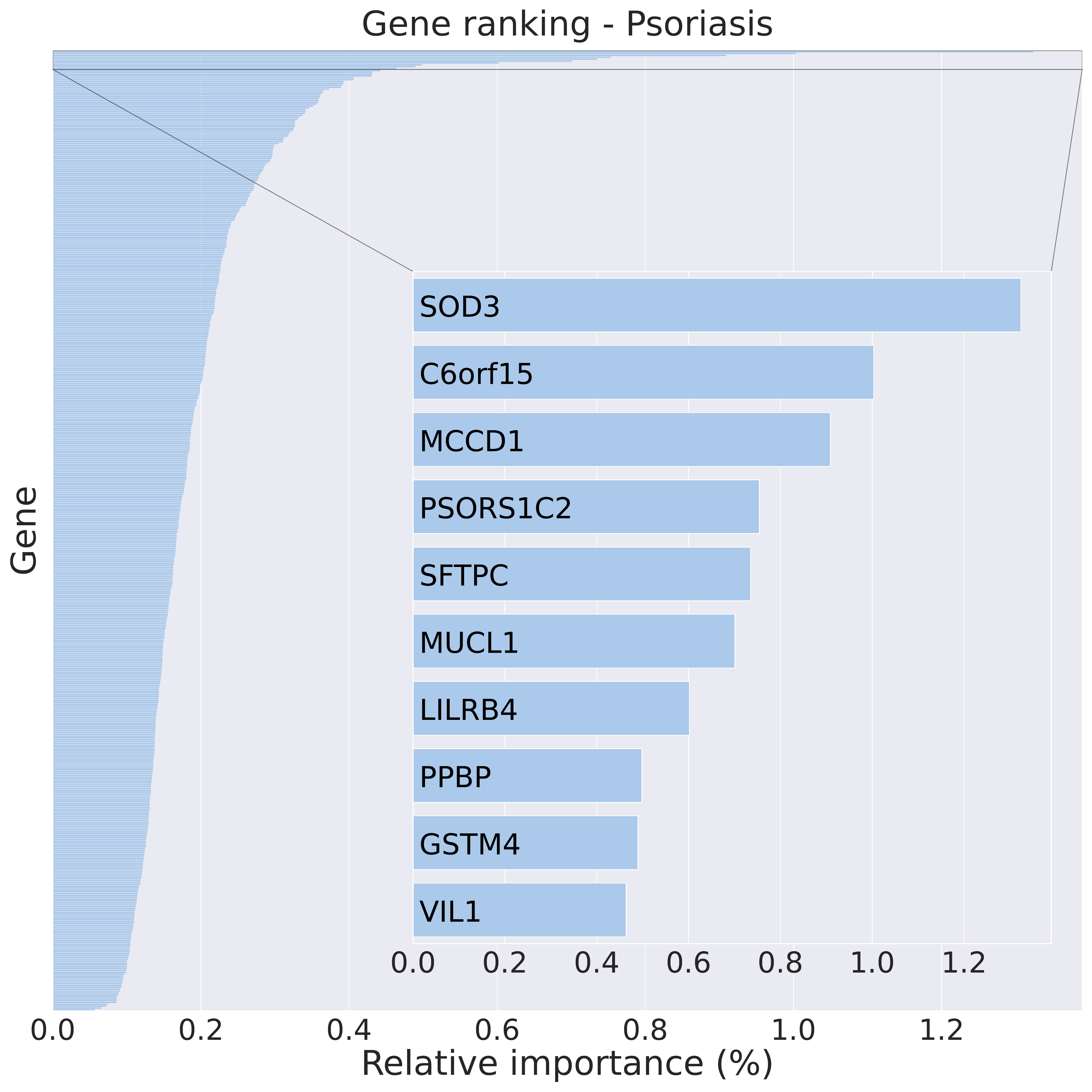}
  \end{subfigure}
  \begin{subfigure}[t]{0.47\textwidth}
\centering	
	\includegraphics[width=1.0\textwidth, valign=b]{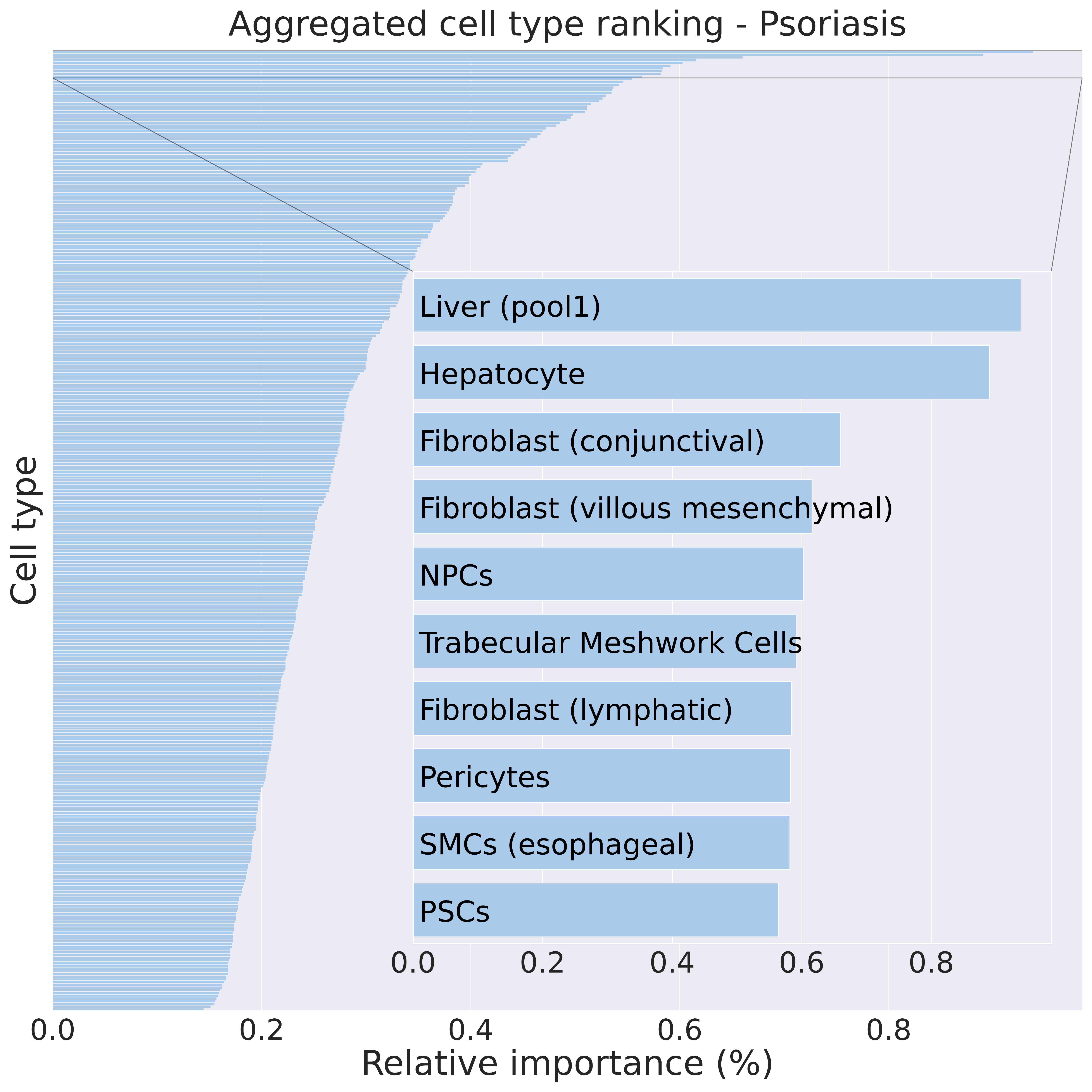}
  \end{subfigure}	
  
  \begin{subfigure}[b]{\rowheadersize\textwidth}
    \rotatebox{90}{\hspace{5.5em}\textsf{Type 1 Diabetes}}
  \end{subfigure}
  \begin{subfigure}[t]{0.47\textwidth}\centering
	\includegraphics[width=1.0\textwidth, valign=b]{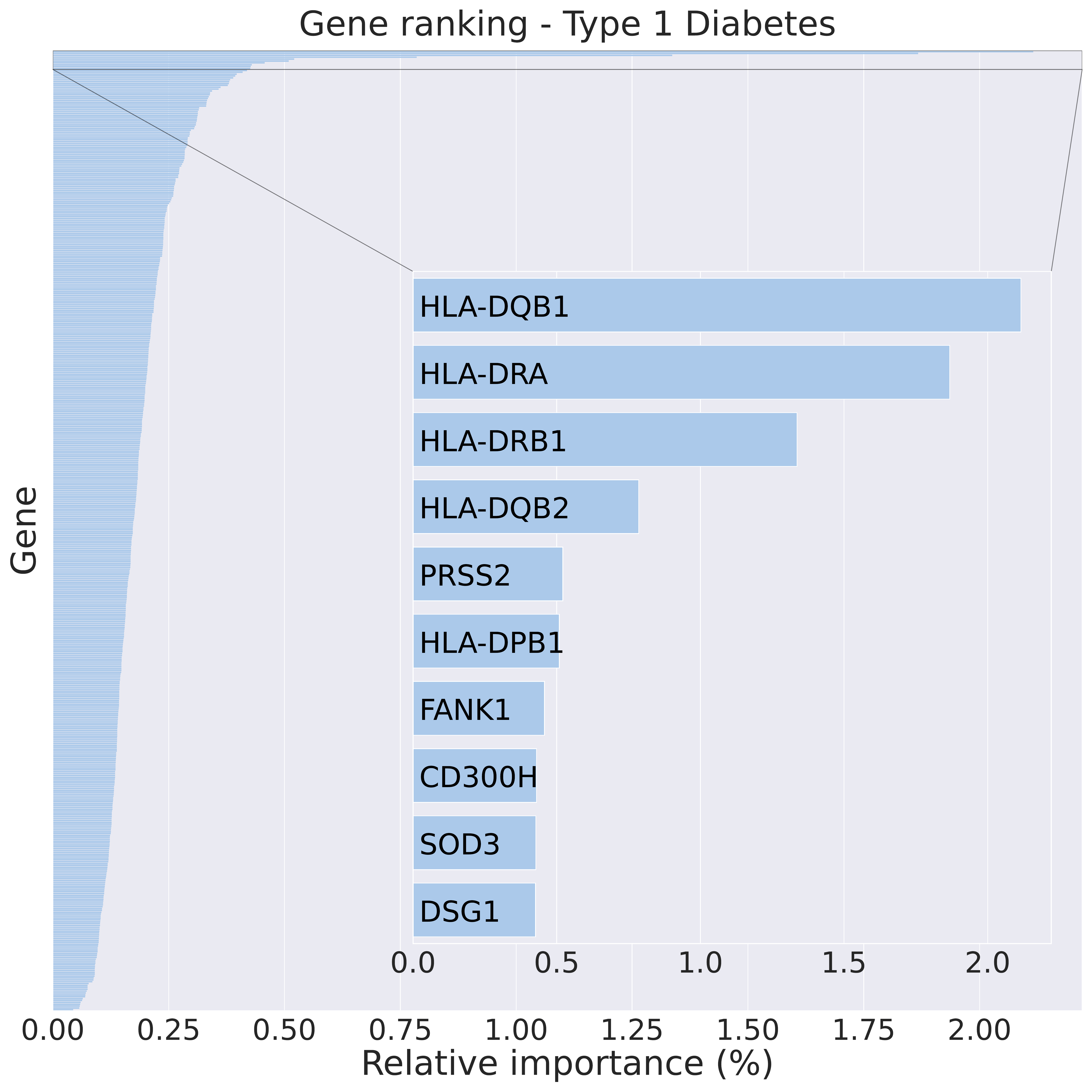}
  \end{subfigure}
  \begin{subfigure}[t]{0.47\textwidth}
\centering	
	\includegraphics[width=1.0\textwidth, valign=b]{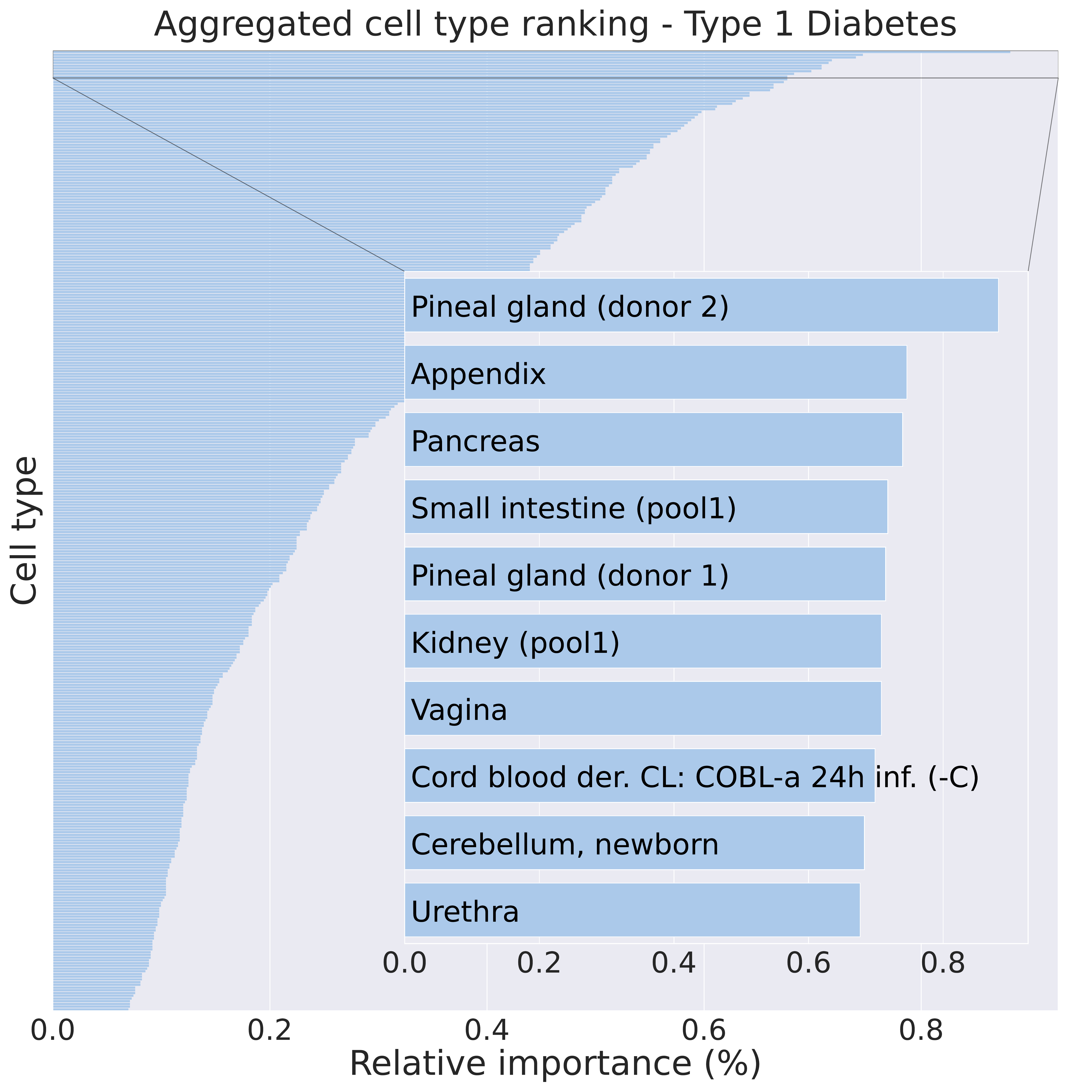}
  \end{subfigure}
  
\caption{\textbf{\themethod{} implicitly attributes cell types and sequence windows associated with predicted risk.} Internal computations of \themethod{} implicitly enable the attribution as to what changes in the transcripts of which sequence region (left column) and cell and tissue types (right column) differentiate individuals that are predicted to go on to develop a disease (top row: psoriasis, bottom: type 1 diabetes) compared to those that are not. Relative importance (\%) of sequence windows and cell types towards risk predictions of \themethod{} were derived using the saliency method\cite{simonyan2013deep}. Intriguingly, \themethod{} identifies liver and hepatocytes as the tissue and cell type contexts with the largest changes aggregated across all transcripts in individuals genetically susceptible to psoriasis. This provides a - to our knowledge not previously reported - genetic basis for the clinical observation of increased frequency and severity of non-alcoholic fatty liver disease (NAFLD) in psoriasis patients\cite{van2014psoriasis,prussick2015nonalcoholic} (top right). Similarly, in type 1 diabetes, we find evidence for the involvement of the appendix in gene expression changes induced by genetic variation which is substantiated by the epidemiological observation of increased risk of appendicitis complications in type 1 diabetes\cite{tsai2008complicated,wei2016diabetes}. Attributions for the other diseases are presented in \Cref{fig:interpretation_cont} and \Cref{fig:interpretation_cont2}. We note that sequence windows (referred to by the respective TSS-donating gene identifier) may encapsulate overlapping genes and gene products and are therefore not necessarily uniquely linked to a single gene region.
}
\label{fig:interpretation}
\end{figure*}

\paragraph{Model interpretation.}\label{par:model_interpretation} In order to calculate sequence window and cell importances, we use the saliency\cite{simonyan2013deep} attribution method from the package \texttt{captum}\cite{kokhlikyan2020captum}.  At first we compute the gradients with respect to the normalized input embeddings.  Following this, we sum the absolute values of these gradients across the embedding dimension, and aggregate by taking the mean over the sample dimension of only the true positive samples, which results in the sequence window importance ranking. The choice to use only true positive samples was intentionally made to reveal the features that the model depends on when it accurately predicts the positive class. For the cell importance ranking in \Cref{fig:interpretation}, we first perturb the normalized input embeddings by performing a single step of gradient descent with the computed gradients. Intuitively, we want to know how the predictions of the sequence-to-expression-embedding head change as we change the input embeddings guided by the gradients of \themethod{}. We de-normalize both the original and the perturbed embeddings and use the Enformer head to compute the values of the CAGE tracks for each gene for both embeddings. We aggregate by summing across the sequence window dimension and subsequently averaging across all true positive samples. We quantify the change between the tracks computed from the two embeddings by calculating the absolute log fold change. Finally, we filtered the output to exclude cancer cell line related output tracks since they are unlikely relevant for the included diseases to obtain the cell type rankings.

\begin{table}[h!]
\vspace{3em}
    \centering
    \begin{tabular}{>{\raggedright\arraybackslash}p{8em}|>{\raggedright\arraybackslash}p{15em}|>{\raggedright\arraybackslash}p{14em}}
         Hypothesis & \themethod{} findings & Supporting evidence   \\
         \midrule
         Liver-involvement in psoriasis & \themethod{} highlights the liver and hepatocytes as some of the most differentially affected cellular contexts in individuals genetically predisposed for psoriasis (\Cref{fig:interpretation}). SELENOW (liver and whole blood) and SPX (liver and hepatocytes) were highlighted as sequence windows most associated with changes in liver and/or hepatocytes (\Cref{fig:liver_psoriasis}).& Psoriasis patients are 1.5 to 3 fold more likely to have non-alcoholic fatty liver disease (NAFLD) after adjusting for common NAFLD risk factors \cite{van2014psoriasis,prussick2015nonalcoholic}. Reportedly, NALFD is also more frequently severe in psoriasis patients\cite{miele2009prevalence,abedini2015patients}. Serum selenium has been reported to be associated with NAFLD status\cite{thuluvath1992selenium}. SPX has been shown to mitigate hepatic steatosis in vitro and in vivo \cite{wang2022treatment}.\\
         \midrule
         Appendicitis in T1D & \themethod{} identifies the appendix as top ranking for differentially affected cellular contexts in T1D (\Cref{fig:interpretation}). No single gene-centred sequence window was enriched for differential changes in the appendix, and the importance was shared across multiple windows. & T1D has been reported to be associated with higher risk for acute appendicitis\cite{tsai2008complicated,wei2016diabetes}. \\
         \midrule
         Small intestine in T1D & \themethod{} hypotheses show the small intestine as a top ranking context in T1D (\Cref{fig:interpretation}). CYP7A1 and GIMD1 are top ranked gene windows enriched in their differential effects in the small intestine (\Cref{fig:liver_psoriasis}). & In mice, CYP7A1 (involved in bile acid synthesis\cite{pandak2001effects}) has been found to potentially exacerbate metabolic disorders\cite{li2012glucose}. T1D has been reported to be associated with changes in cholesterol synthesis and absorption markers\cite{semova2019type}. \\
         \midrule
         Optic nerve complications in COPD & \themethod{} surfaces the optic nerve as a potentially most differential cellular context in COPD (\Cref{fig:interpretation_cont2}). Importance is shared among implicated multiple gene sequence windows but notably include the OPTN-centred window. & The optic nerve has been implicated in COPD through visual evoked potential (VEP) abnormalities \cite{ozge2005cranial,mikaeili2015correlation}. Women with COPD are reportedly at higher risk of open angle glaucoma\cite{lee2022increased,wandell2022systemic}. Variants in OPTN have been connected to open-angle glaucoma\cite{sears2019mendelian,shiga2024identification}.\\
         \bottomrule
    \end{tabular}
    \caption{\textbf{Selected potential mechanistic hypotheses identified by \themethod{}.} We interpreted attributions provided by \themethod{} trained to predict 6 major diseases, and identified several potential hypotheses that connect disease pathologies to underlying mechanisms. Several \themethod{}-derived findings are substantiated by previous studies (rightmost column). Although some of the indicated findings are clinically and epidemiologically supported, they - to our knowledge - to date lack a potential mechanistic explanation.}
    \label{tab:qualitative_results}
\end{table}

\paragraph{Interpretation of \themethod{} cell and gene expression attributions.}\label{par:interpretation_meaning} It is important to note that the ⟨sequence \textrightarrow{} cell context \textrightarrow{} expression \textrightarrow{} phenotype⟩ paths highlighted by \themethod{} are not necessarily causal paths for a given disease. Conclusively establishing causal relationships in general requires controlled perturbation experiments in relevant systems\cite{chevalley2022causalbench,chevalley2024intersort}. \themethod{} explanations are best interpreted as highlighting a potential path through which genetic variation could give rise to expression in specific cellular contexts that is different between diseased and control individuals - pointing to increased risk. It is possible that this risk is not realised in practice even though such differences can be predicted for certain cell types. As an illustrative example, this can be the case because the highlighted cell type, state or tissue context is not present in the individual for which \themethod{} produced predictions. For example, the sequence-to-expression backbone used includes output tracks for cell types associated with newborns (that will not be present in adults), reproductive organs associated with a specific sex (that will not be present in the opposite sex) and for cell lines used in cancer research (that may not be representative of non-cancerous cells). Genetic variation of an individual may well lead to predictable differences in those not-realised cell types that can be used by \themethod{} to differentiate between diseased and control individuals, but they are not causal. In addition, sequence windows highlighted by \themethod{} as important warrant further interpretation as the region covered by the 196 kb sequence window frequently partially or fully overlap with multiple other genes beyond the TSS-associated one. For example, in the case of HLA-DQB2, there are 8 other overlapping genes in the 196 kb window (ENSG00000250264, HLA-DOB, HLA-DQA2, HLA-DQB1, PSMB8, PSMB9, TAP1, TAP2). Disambiguating the sequence subregion responsible for the importance of a sequence window requires an attribution analysis at the sequence level as the predictive signal may stem from an overlapping gene region.

\paragraph{Transportability of \themethod{}.} We hypothesise the increased transportability of \themethod{} is connected to its utilisation of a sequence-to-expression backbone that acts as a bottleneck for risk predictions. \themethod{} cannot rely on SNP-level variation - which is prone to correlation patterns defined by ancestry - directly to make differential risk predictions and can instead only leverage variation that leads to differential expression predictions in cellular contexts that were included in the pretraining dataset of its sequence-to-expression backbone. Through this mechanism, \themethod{} limits reliance on non-generalisable predictors that are only correlated with differential signal relevant for disease risk prediction, but do not lead to corresponding gene expression or chromatin accessibility changes. This inductive bias ensures \themethod{} predictions are more likely portable to diverse ancestral backgrounds.

\begin{figure*} 
\centering	

\figtitle{Cell type enrichment in disease}

\includegraphics[width=0.35\textwidth]{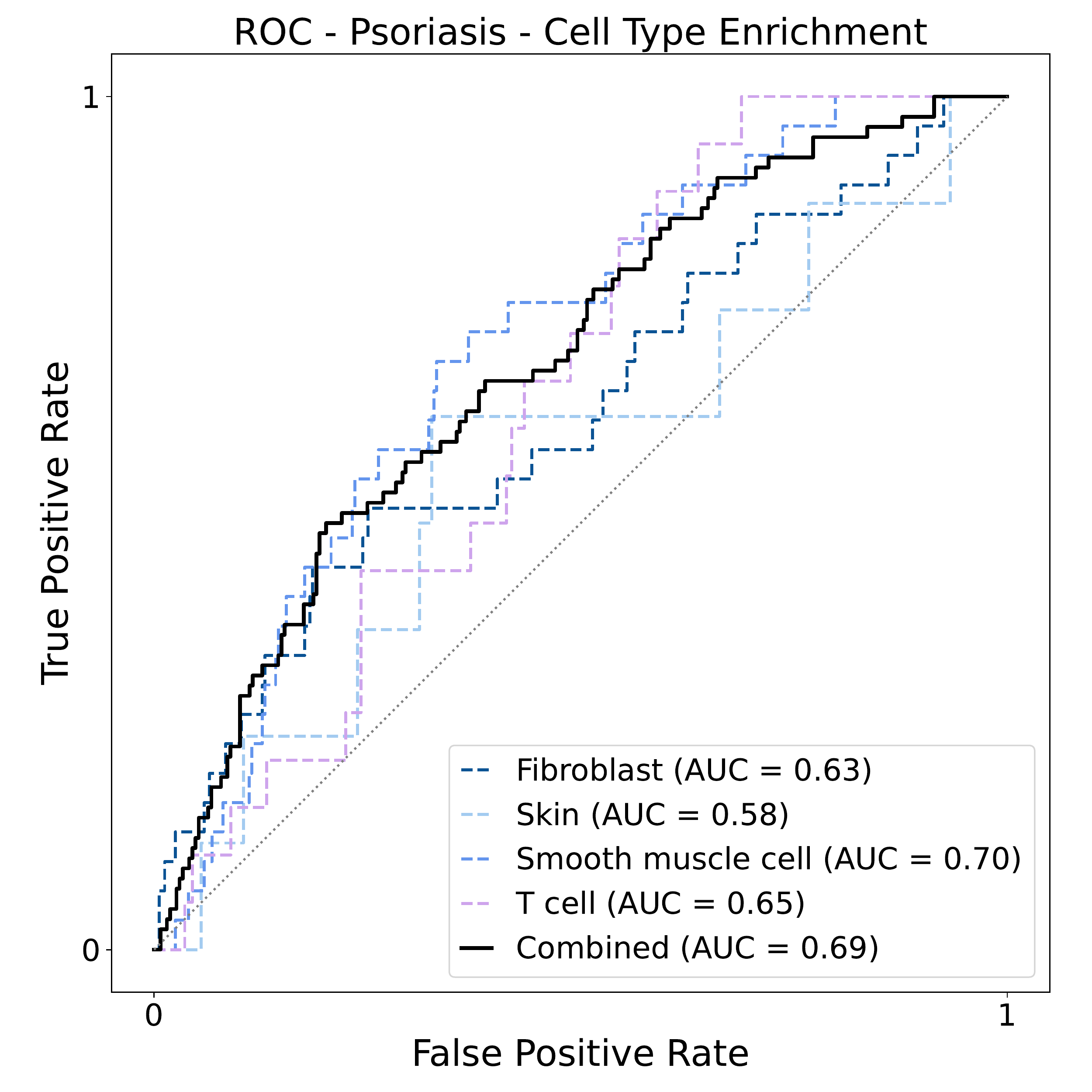}
\includegraphics[width=0.35\textwidth]{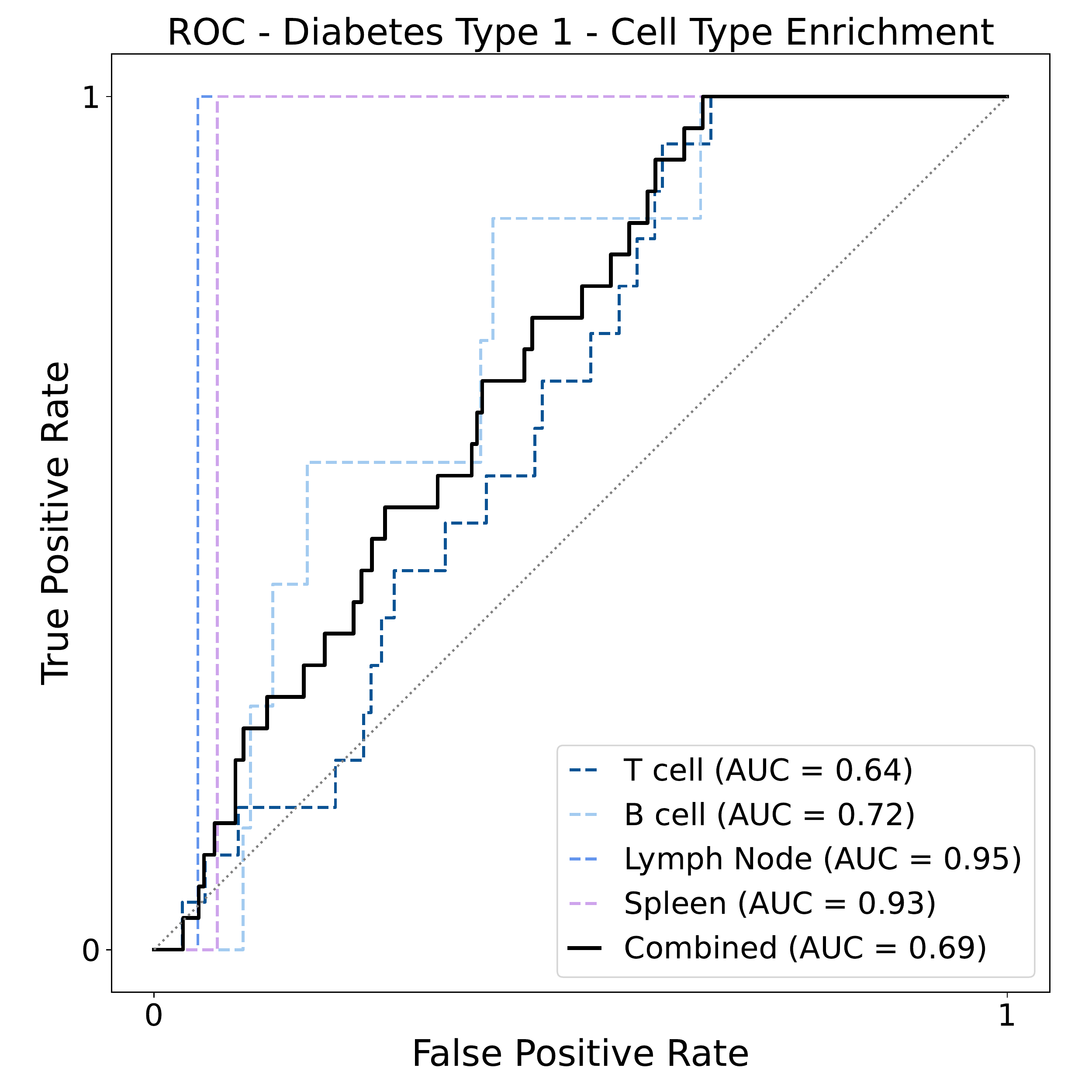}
\includegraphics[width=0.35\textwidth]{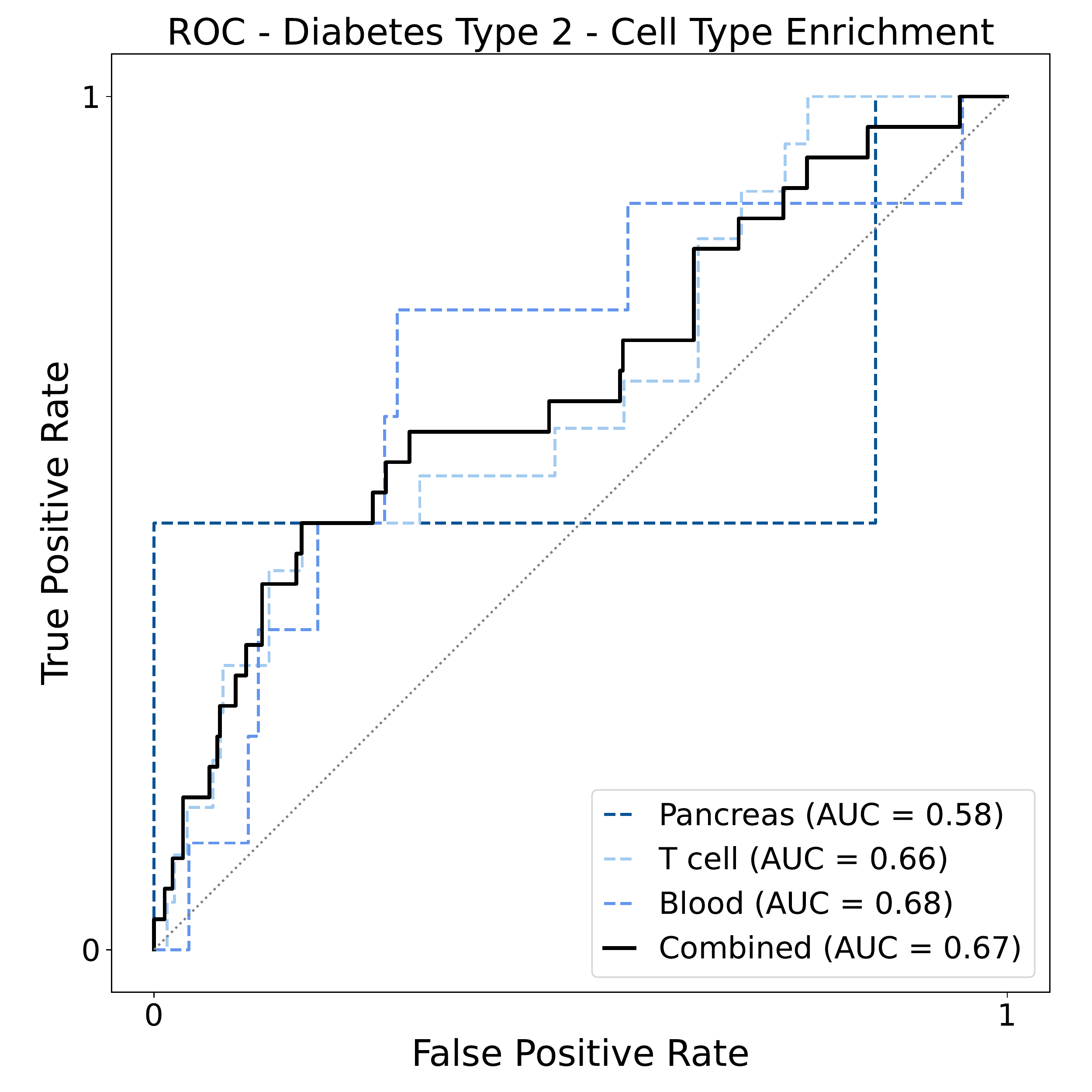}
\includegraphics[width=0.35\textwidth]{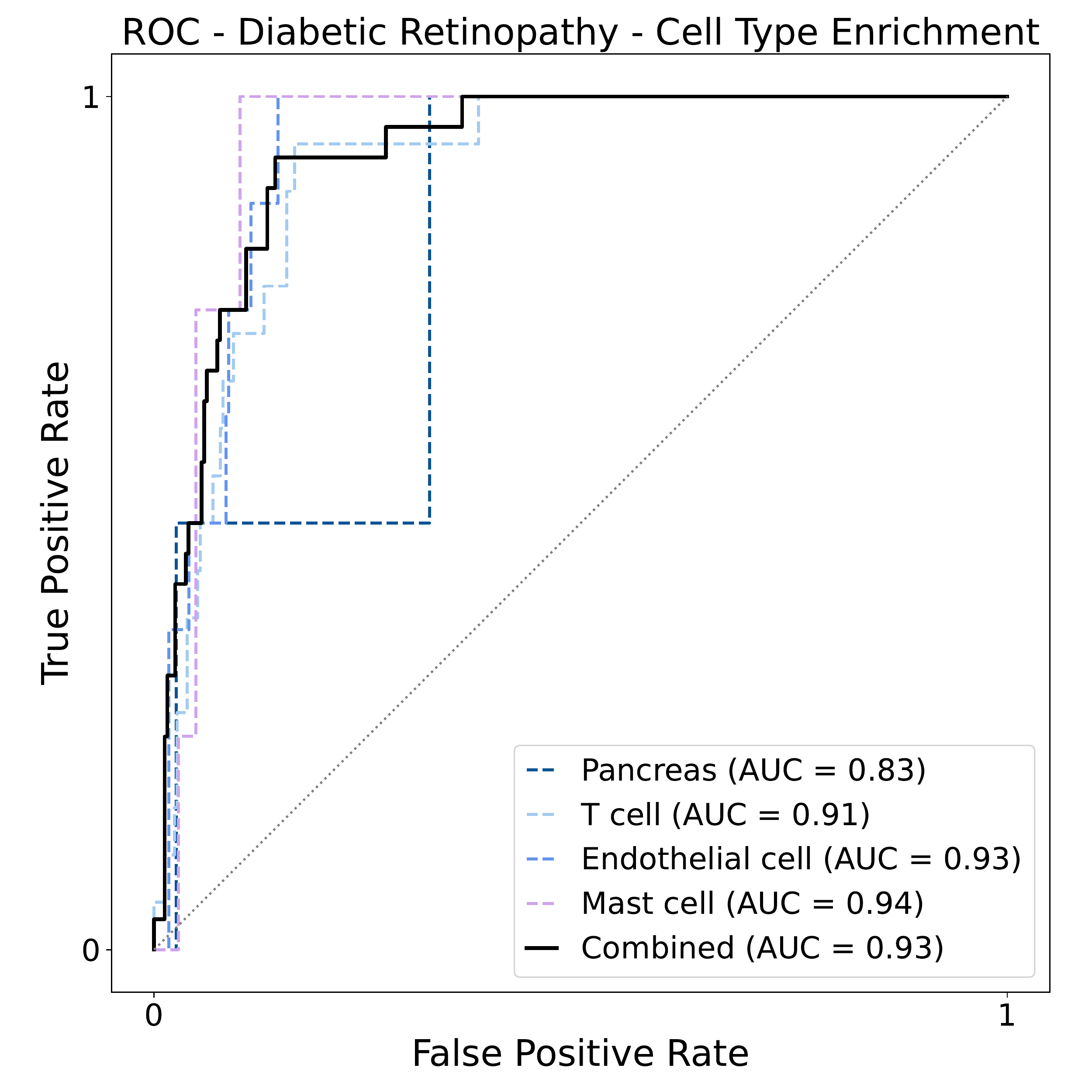}
\includegraphics[width=0.35\textwidth]{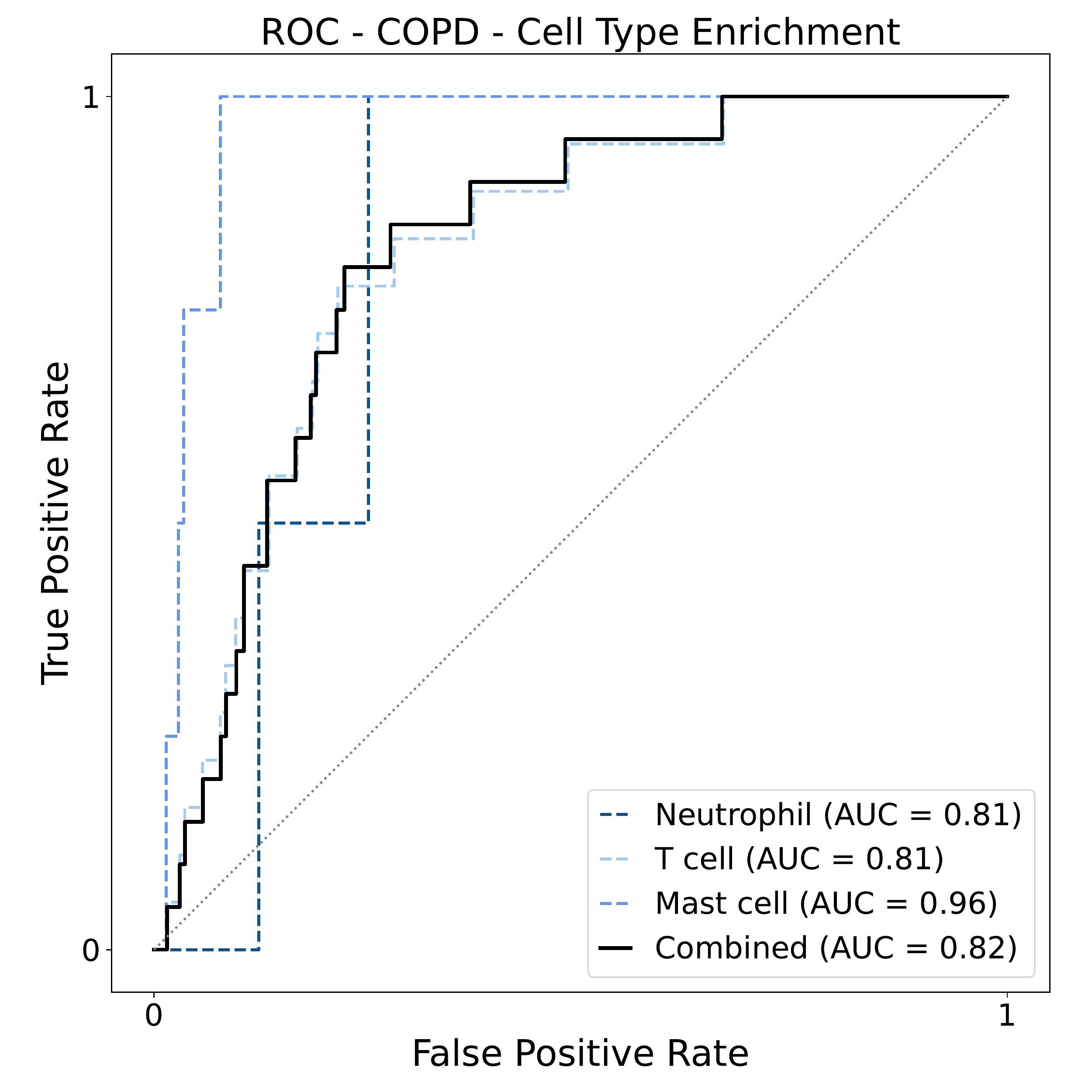}
\includegraphics[width=0.35\textwidth]{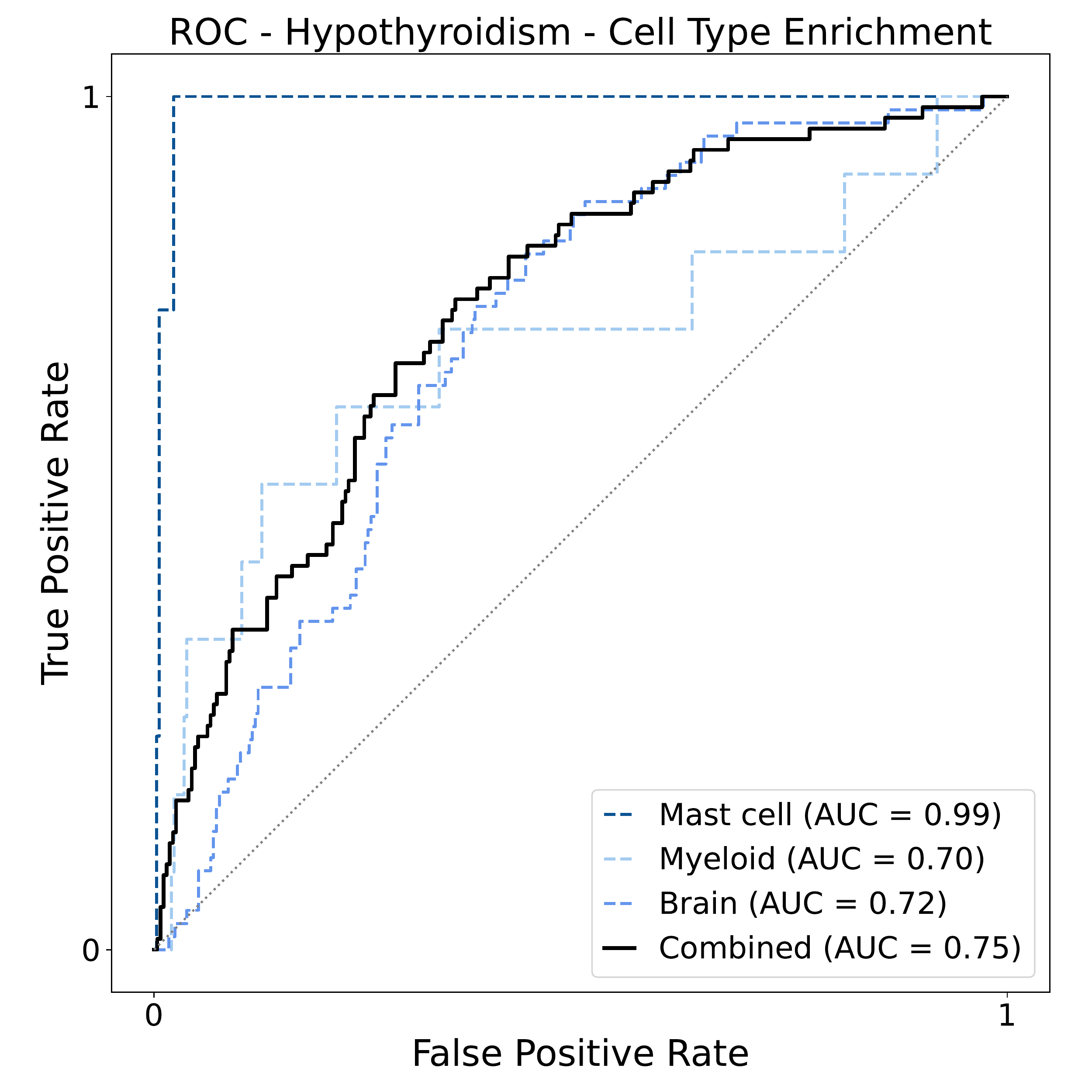}
	
\caption{\textbf{Cell type enrichment analysis shows that \themethod{} attributions emphasise disease-associated cell types.} Selected ROC curves (bottom) for cell type-specific gene enrichment across major diseases. We selected putatively disease-associated categories of cell types for each disease based on established associations reported in literature. For example for psoriasis, ROC curves include enrichment of fibroblasts\cite{guban2016abnormal}, skin\cite{albanesi2007resident}, smooth muscle cells (SMCs)\cite{armstrong2011angiogenesis}, and T cells\cite{albanesi2007resident} in the cell-type ranking for psoriasis. For type 1 diabetes, ROC curves include enrichment of T cells\cite{roep2003role}, B cells\cite{silveira2006b}, lymph node\cite{hoglund1999initiation} and spleen\cite{saito2012diabetes}. Each curve illustrates the true and false positive rates associated with each cell type walking along the cell type ranking from top to bottom - demonstrating the ability of \themethod{} to attribute disease-relevant cell types. The 'Combined' curve (black) represents the predictive accuracy when considering any of the putatively disease-associated cell types. AUC values above the 0.5 reference line show that \themethod{} effectively identifies and prioritises cell types putatively pertinent to the respective disease. Please note that cell types were assigned to the most specific category, i.e. mast cells were not also included in the myeloid cells category.
}
\label{fig:interpretation_roc}
\end{figure*}

\paragraph{Cell type enrichment.} To identify cell types that are enriched in \themethod{} cell type rankings, we performed a receiver operator curve (ROC) analysis that walks through the aggregated disease cell type rankings for each disease and assesses the relative ranking of specific categories of cell types (\Cref{fig:interpretation_roc}). We selected putatively disease-associated cell types based on established associations reported in literature. Area under the curves (AUCs) higher than 0.5 for all diseases and relevant cell types demonstrate that \themethod{} rankings prioritise putatively disease-associated cell types. To generate the cell type-disease association overview plot (\Cref{fig:interpretation_quantification}) for all considered diseases, we converted the enrichment AUCs into a linear \% enrichment score where 0.50 and 1.00 AUC enrichment correspond to 0\% and 100\% enrichment, respectively. We note that some Enformer per-cell type output tracks map ambiguously to the broader categories of cell types analysed (e.g. mast cells and myeloid cells) and we resolved such ambiguities by mapping each cell type to the most specific category.

\paragraph{Cell type-disease associations supported by literature.}\label{par:methods_literature} To identify cell types putatively involved in disease according to scientific literature, we searched PubMed\footnote{\url{https://pubmed.ncbi.nlm.nih.gov/} (accessed 1 April 2024)} for the top 100 abstracts involving the respective disease and cell type for each of the 6 diseases and 21 cell types/tissues studied in this work (query: "(disease) AND (cell\_type)"). This yielded a total of \numprint{10694} abstracts (for some combinations, fewer than 100 abstracts existed). We then used Claude Sonnet (Anthropic Ireland, Ltd., accessed 1st April 2024) to automatically score each abstract where integer scores (-5 to 5) ranged from the strongest possible evidence against (-5) over no evidence (0) to the strongest possible evidence for (5) a cell type/tissue being involved in a particular disease. We used the resulting score distributions to derive enrichment scores indicating the frequency and magnitude of published evidence for an association between each cell type and disease by counting the fraction of abstracts scored with at least a score of 1 out of the top 100 abstracts. Because scientific abstracts that provide very strong evidence can be an indicator of a potential association even if few in relative number, we also counted the absolute number of abstracts with a score greater or equal to 4 and indicated the cell type and disease combinations with at least 5 such abstracts reporting strong evidence in \Cref{fig:interpretation_quantification}.

\paragraph{Baseline methods for cell type identification.} We compared \themethod{} predicted cell-disease associations with those provided by state-of-the-art cell type identification methods, including \citet{ongen2017estimating,finucane2018heritability,watanabe2019genetic,jagadeesh2022identifying} and \citet{amariuta2023modeling}. We used published cell-disease associations where available, and standardised tissue and cell identifiers across the baseline methods to enable comparison. We limited comparisons to the overlap in tissues and cell types between each baseline and \themethod{} to avoid penalising methods that produce associations for fewer cell types. For \citet{jagadeesh2022identifying}, we followed the authors instructions (using E-value threshold of 5) to generate cell type and disease associations using scRNAseq datasets from T1D\cite{honardoost2024systematic}, psoriasis\cite{reynolds2021developmental}, and COPD\cite{salcher2022high} to increase the number of overlapping diseases available for comparison. 

\paragraph{Clustering and subtyping on genetic predisposition.}  We utilised the individual-level mechanisms attributed by \themethod{} to perform a cluster analysis based on predicted genetic predisposition (\Cref{fig:clusters}). Model attributions for test set individuals were generated with a modified attribution algorithm to address the lack of reference individuals to get samples throughout the decision manifold in the individual interpretation setting, and to obtain the direction of change (i.e. whether an increase in expression correlates or anti-correlates with disease prediction). To increase the number of samples we used the integrated gradients (IG\cite{sundararajan2017axiomatic}) method from the \texttt{captum}\cite{kokhlikyan2020captum} package. We randomly selected 100 undiagnosed individuals to use as reference baselines for IG. For each diagnosed-undiagnosed  pairing, the IG was calculated with 20 steps, then translated to CAGE track gradients. To prevent the Enformer head's nonlinear activation from amplifying sampling error, an average CAGE track gradient was calculated by applying the gradient descent step at 20 values linearly-interpolated between embeddings of individuals and the negative baseline reference.
To elucidate subtypes, the individual attributions for diagnosed individuals were reduced to a two dimensional embedding with UMAP and clustered using HDBSCAN from \texttt{scikit-learn} using a minimum cluster size of 5\% of the number of diagnosed individuals. For reference, an equal number of undiagnosed individuals were embedded with the pre-fitted UMAP, and included in the HDBSCAN clustering.

\section*{Code availability}
Source code will be made available on Github upon publication.

\section*{Data availability}
The genome sequencing and phenotypic annotation data used in this study is available to researchers through the UK Biobank\cite{sudlow2015uk}. This study was completed under UK Biobank application No. 20361. Attributions (mechanistic hypotheses) derived from \themethod{} will be made available upon publication. 

\section*{Acknowledgements}
LS, AG, AM, RS, MC, KB, and PS are employees and shareholders of GSK plc. FT is a former employee of GSK plc. PN received funding from GSK plc.

\bibliographystyle{unsrtnat} 
\bibliography{references} 

\begin{figure*}[pt!] 
\vspace{-1em}
\centering
\figtitle{Individual Disease Risk Prediction Performance}
 
 \begin{subfigure}[b]{\rowheadersize\textwidth}
    \hspace{0.5em}
  \end{subfigure}
\begin{subfigure}[t]{\resultsubplotsize\textwidth}\centering
    \textsf{AUROC}
  \end{subfigure}
  \begin{subfigure}[t]{\resultsubplotsize\textwidth}\centering
    \textsf{AUPRC}
  \end{subfigure} 
  \begin{subfigure}[t]{\resultsubplotsize\textwidth}\centering
    \textsf{PPV}
  \end{subfigure}\hfill 
  
  \begin{subfigure}[b]{\rowheadersize\textwidth}
    \rotatebox{90}{\hspace{1.5em}\textsf{Psoriasis}}
  \end{subfigure}
  \begin{subfigure}[t]{\resultsubplotsize\textwidth}
  \centering
	\includegraphics[width=1.0\textwidth, valign=b]{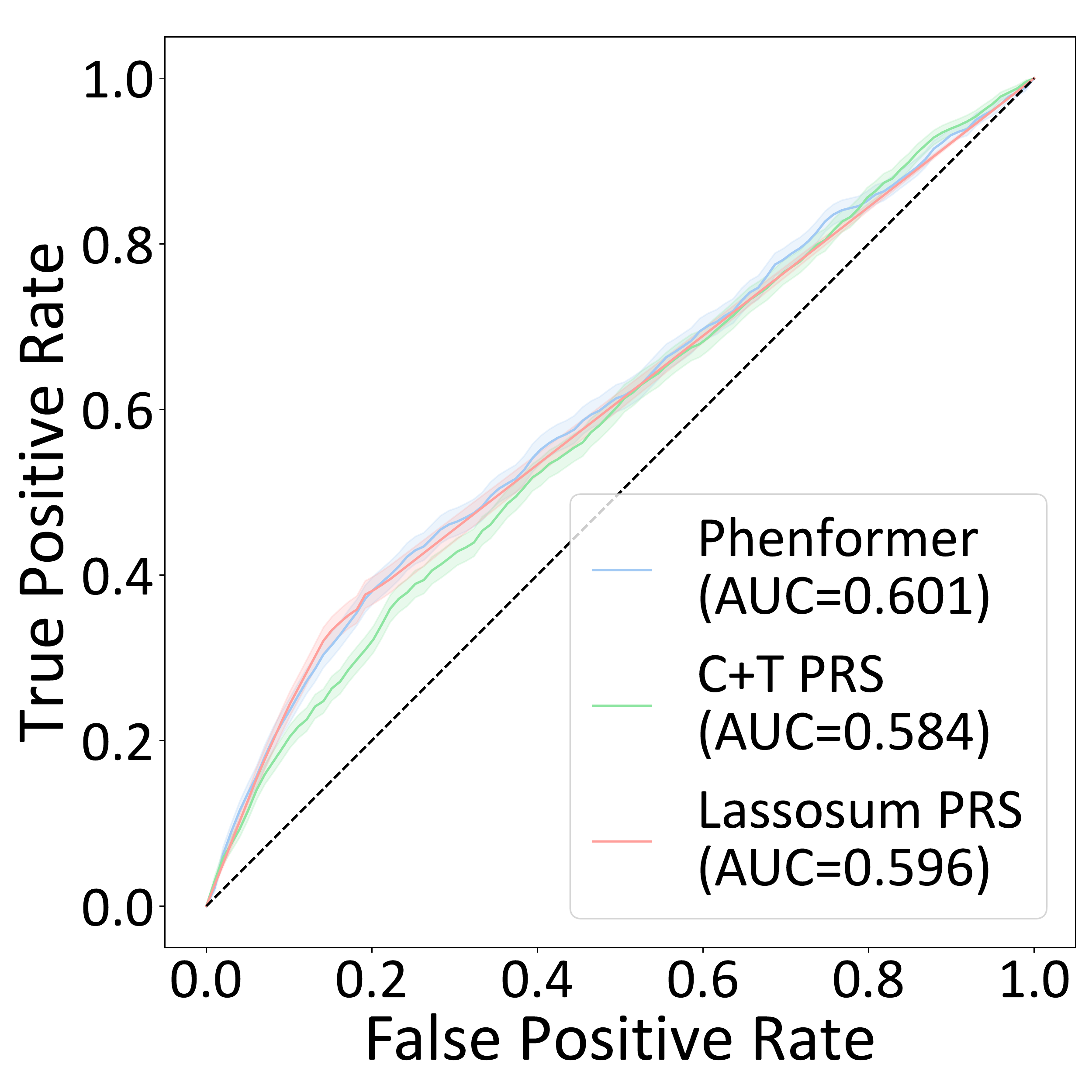}
  \end{subfigure}
  \begin{subfigure}[t]{\resultsubplotsize\textwidth}
\centering	
	\includegraphics[width=1.0\textwidth, valign=b]{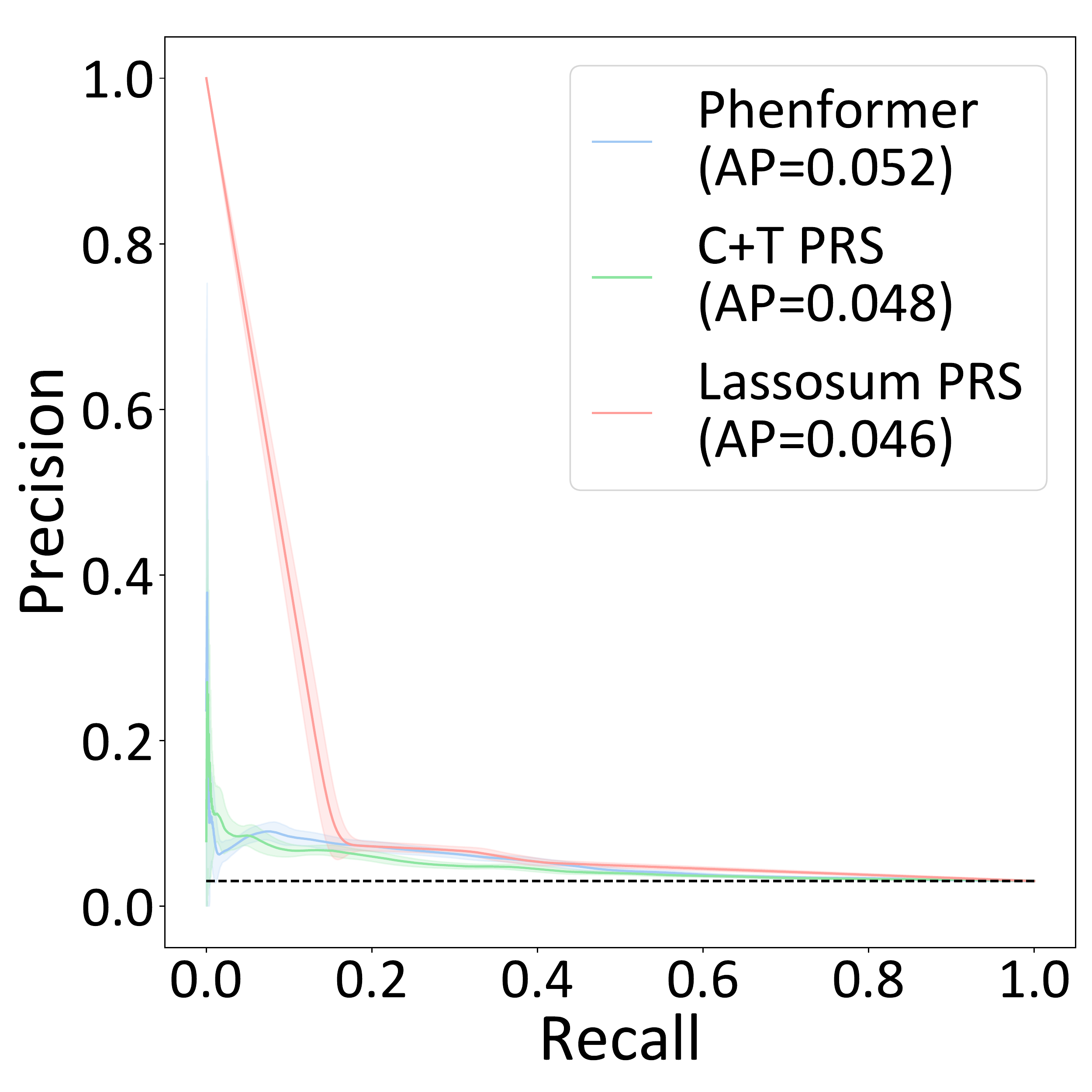}
  \end{subfigure}
  \begin{subfigure}[t]{\resultsubplotsize\textwidth}
\centering	
	\includegraphics[width=1.0\textwidth, valign=b]{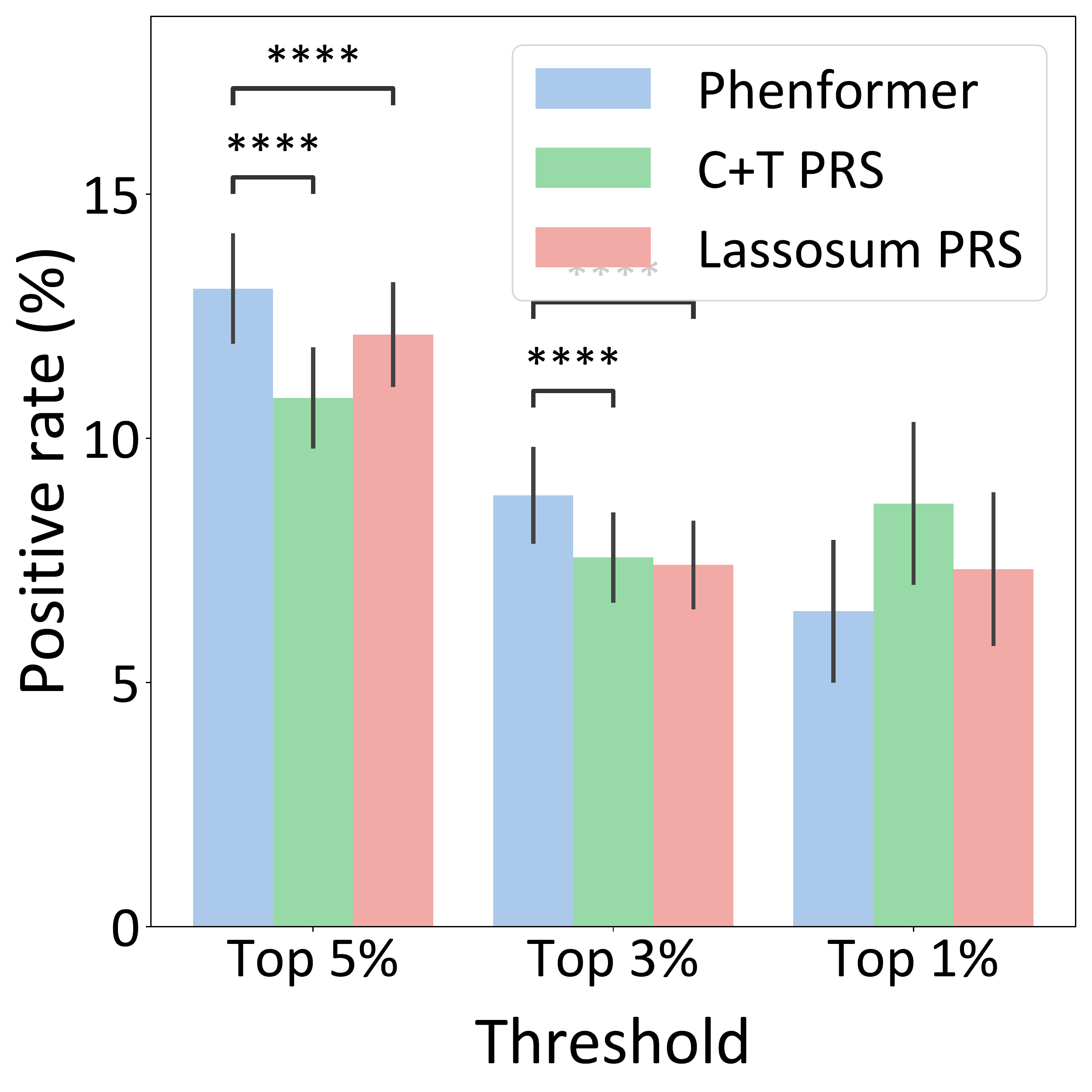}
  \end{subfigure}\hfill 
 
  \begin{subfigure}[b]{\rowheadersize\textwidth}
    \rotatebox{90}{\hspace{0.1em}\textsf{Type 1 Diabetes}}
  \end{subfigure}
  \begin{subfigure}[t]{\resultsubplotsize\textwidth}
  \centering
	\includegraphics[width=1.0\textwidth, valign=b]{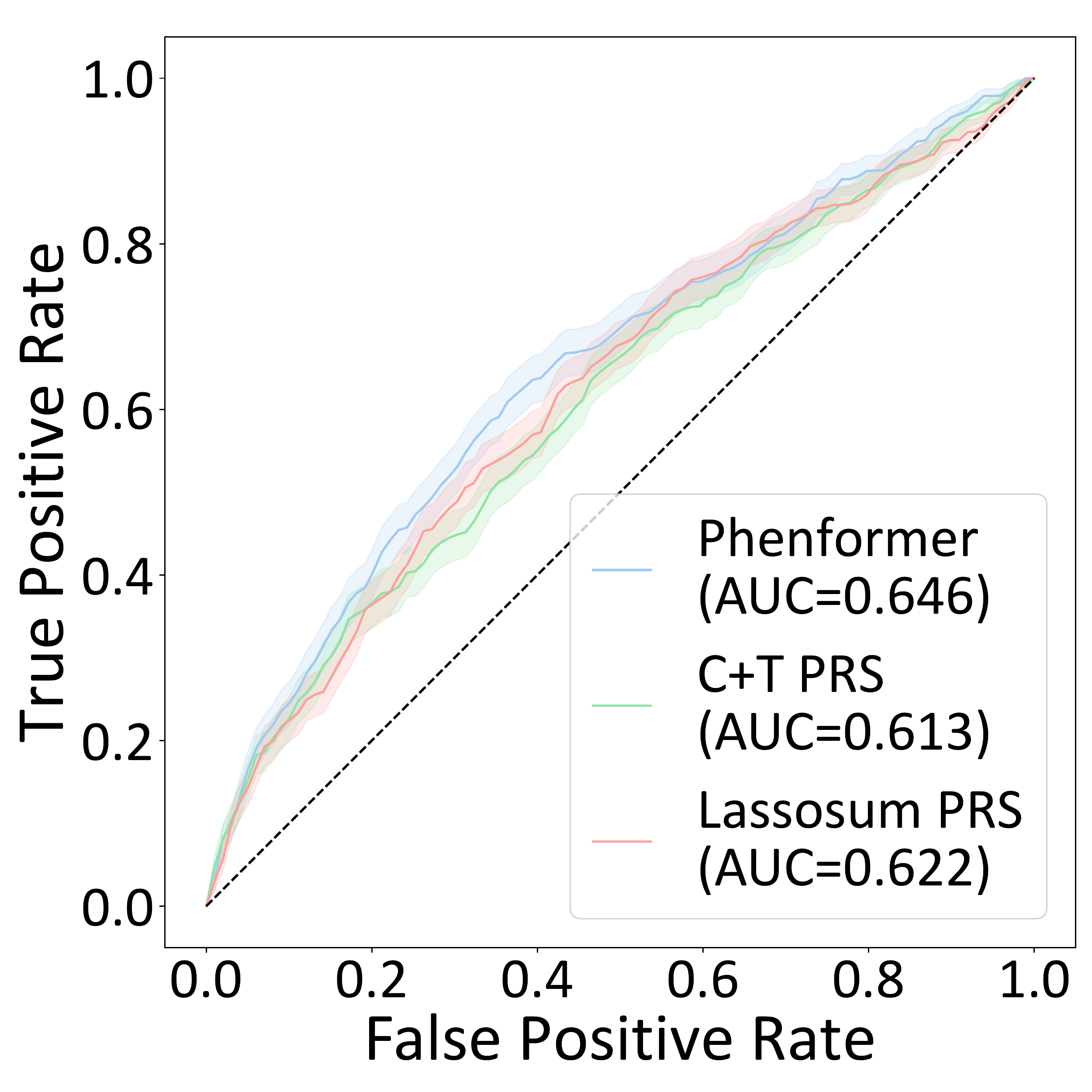}
  \end{subfigure}
  \begin{subfigure}[t]{\resultsubplotsize\textwidth}
\centering	
	\includegraphics[width=1.0\textwidth, valign=b]{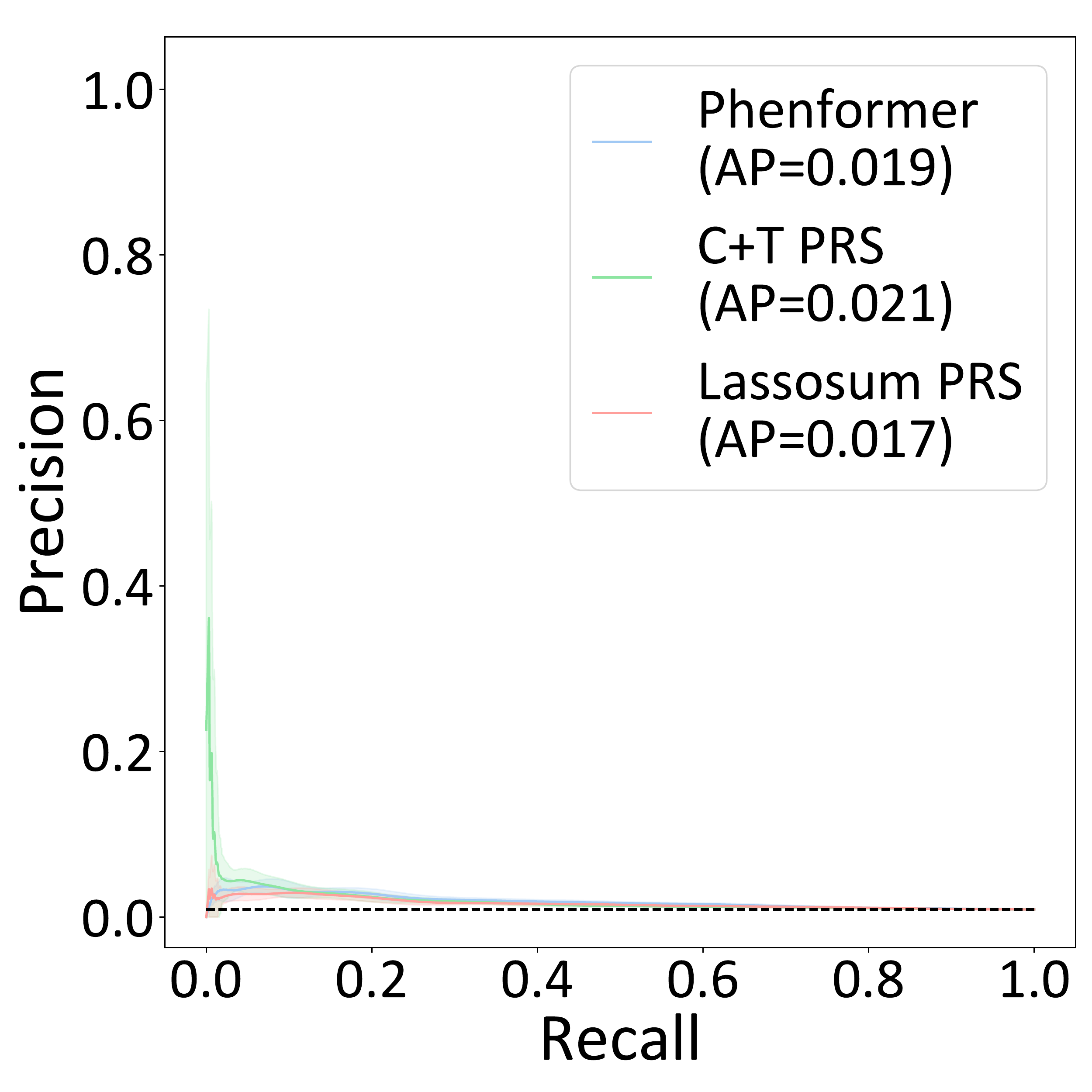}
  \end{subfigure}
  \begin{subfigure}[t]{\resultsubplotsize\textwidth}
\centering	
	\includegraphics[width=1.0\textwidth, valign=b]{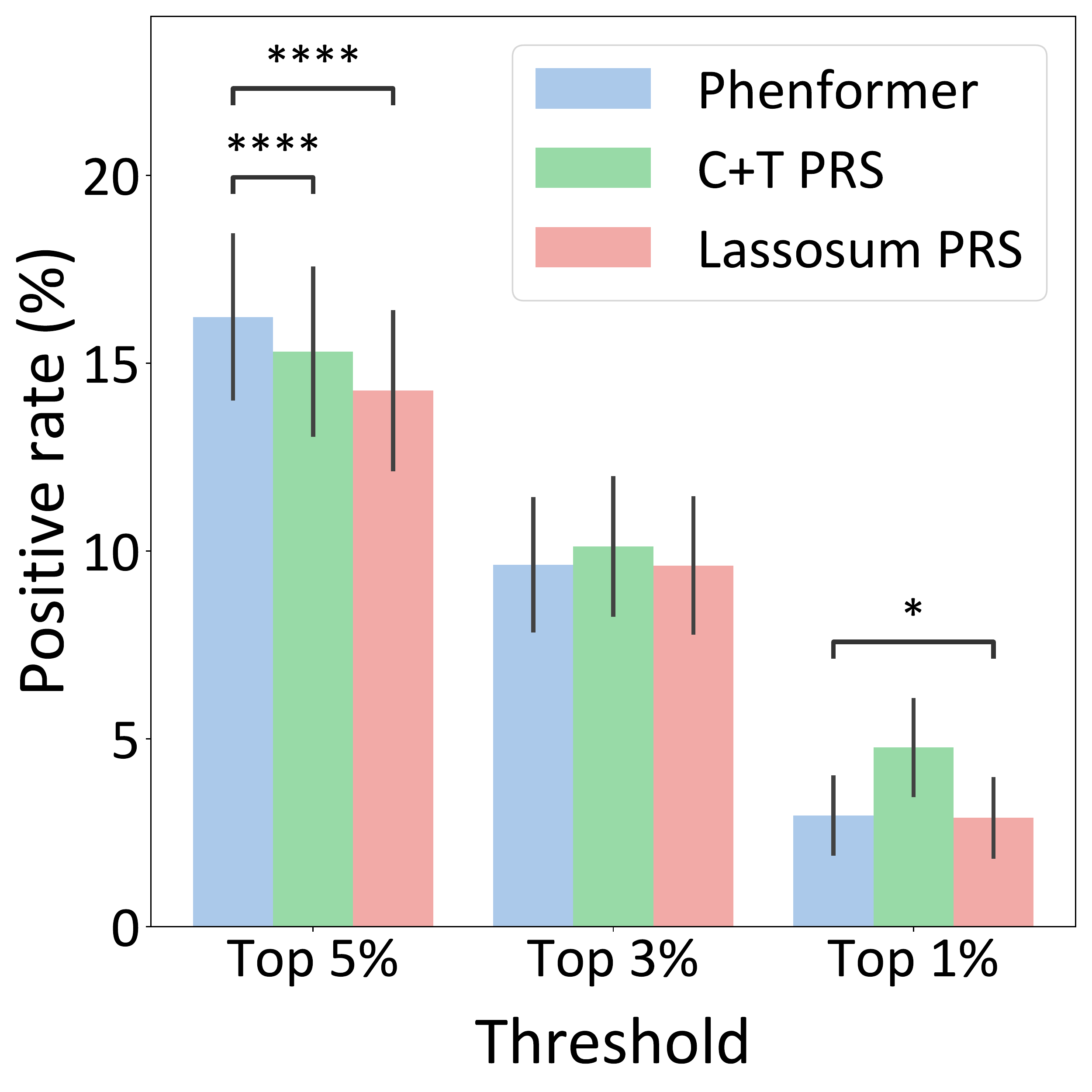}
  \end{subfigure}\hfill  
  
  \begin{subfigure}[b]{\rowheadersize\textwidth}
    \rotatebox{90}{\hspace{0.1em}\textsf{Type 2 Diabetes}}
  \end{subfigure}
  \begin{subfigure}[t]{\resultsubplotsize\textwidth}
  \centering
	\includegraphics[width=1.0\textwidth, valign=b]{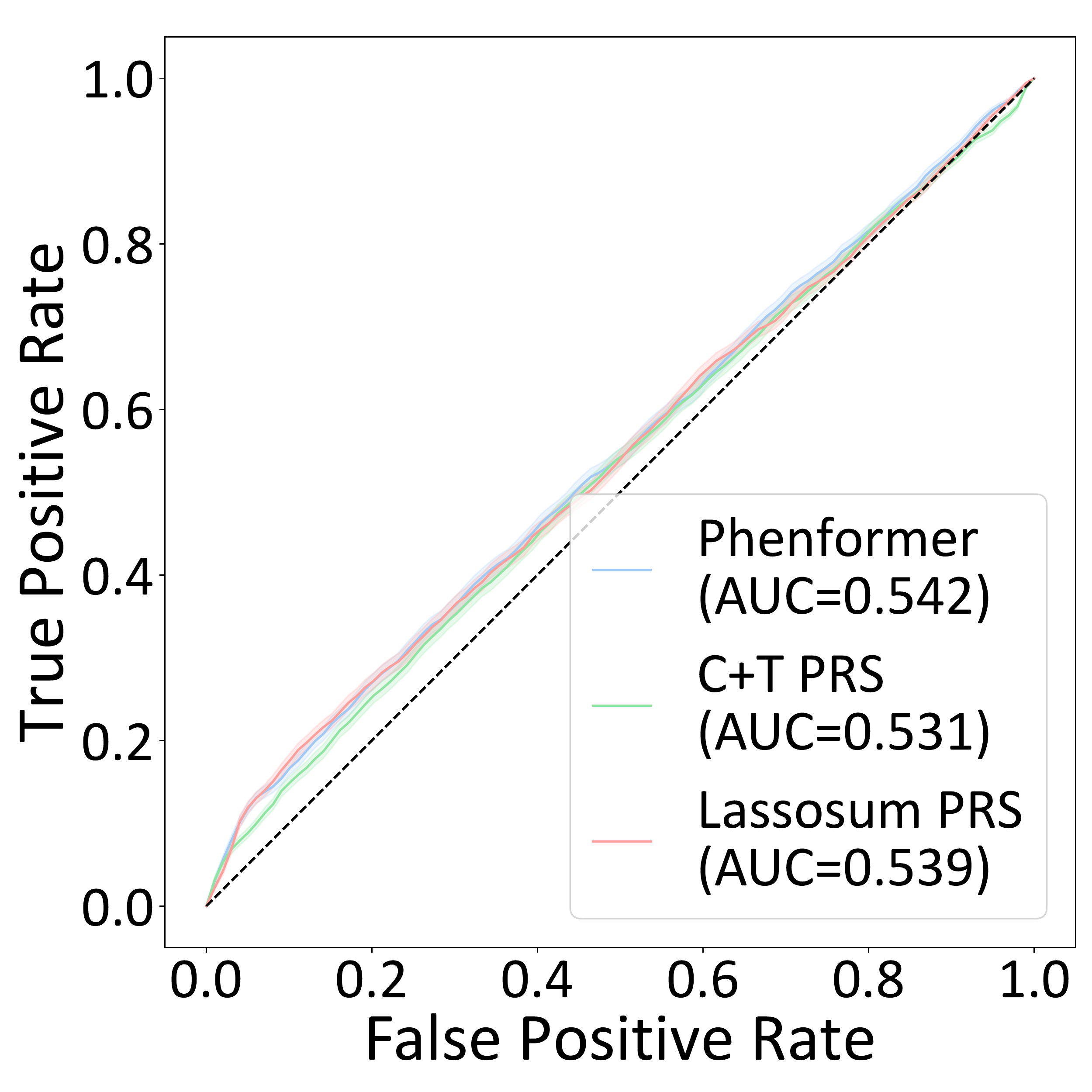}
  \end{subfigure}
  \begin{subfigure}[t]{\resultsubplotsize\textwidth}
\centering	
	\includegraphics[width=1.0\textwidth, valign=b]{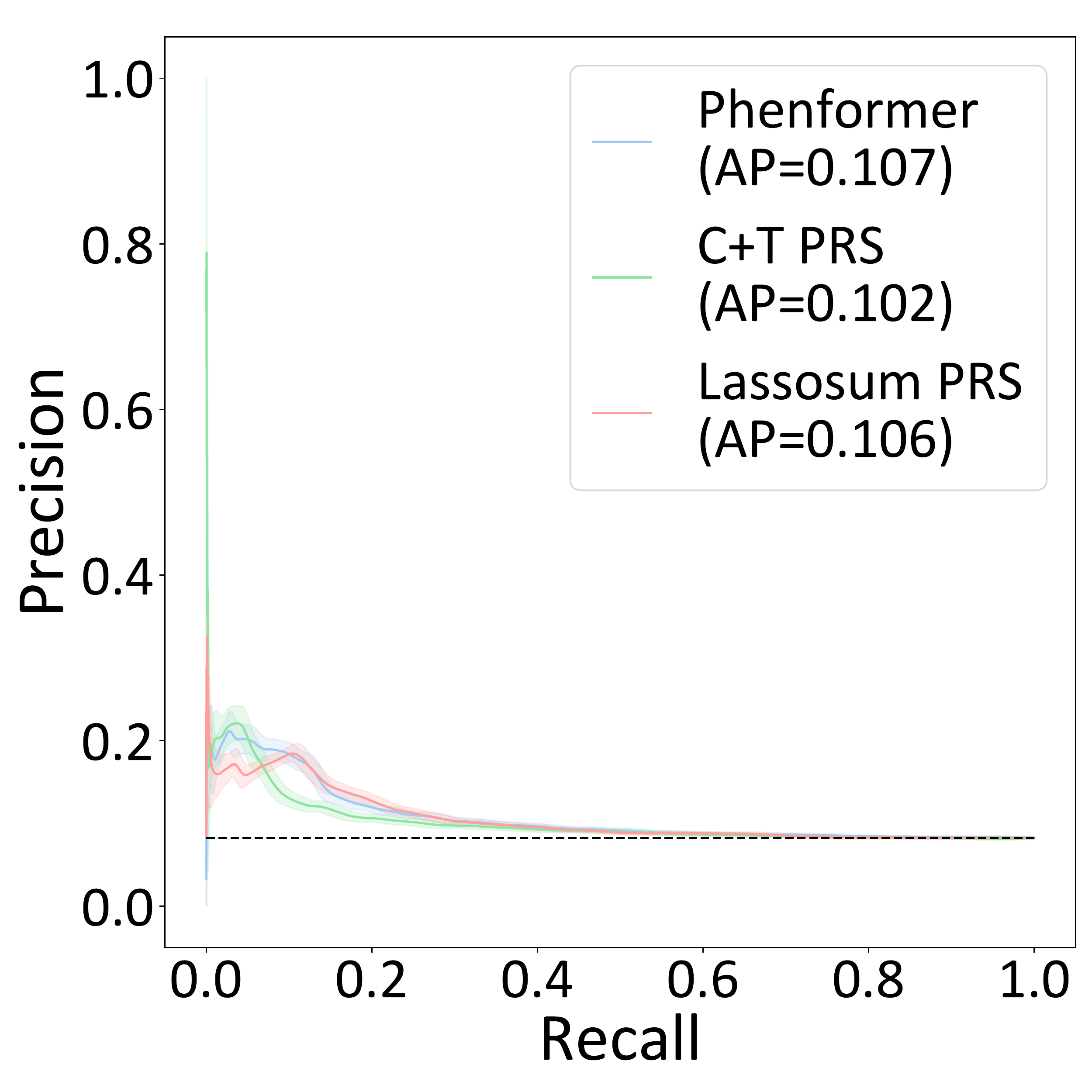}
  \end{subfigure}
  \begin{subfigure}[t]{\resultsubplotsize\textwidth}
\centering	
	\includegraphics[width=1.0\textwidth, valign=b]{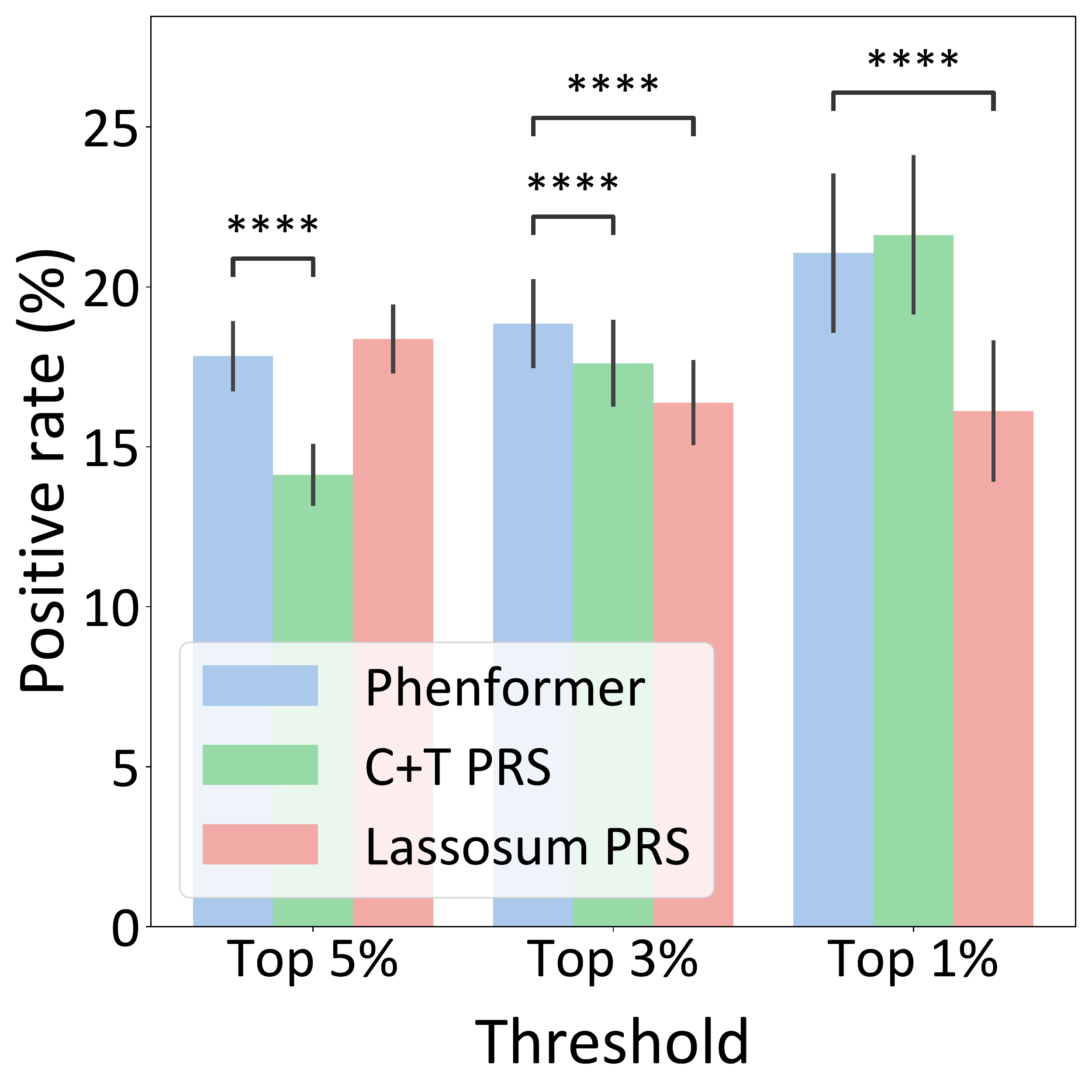}
  \end{subfigure}\hfill  
  
  \begin{subfigure}[b]{\rowheadersize\textwidth}
    \rotatebox{90}{\hspace{0em}\textsf{Diabetic Retino.}}
  \end{subfigure}
  \begin{subfigure}[t]{\resultsubplotsize\textwidth}
  \centering
	\includegraphics[width=1.0\textwidth, valign=b]{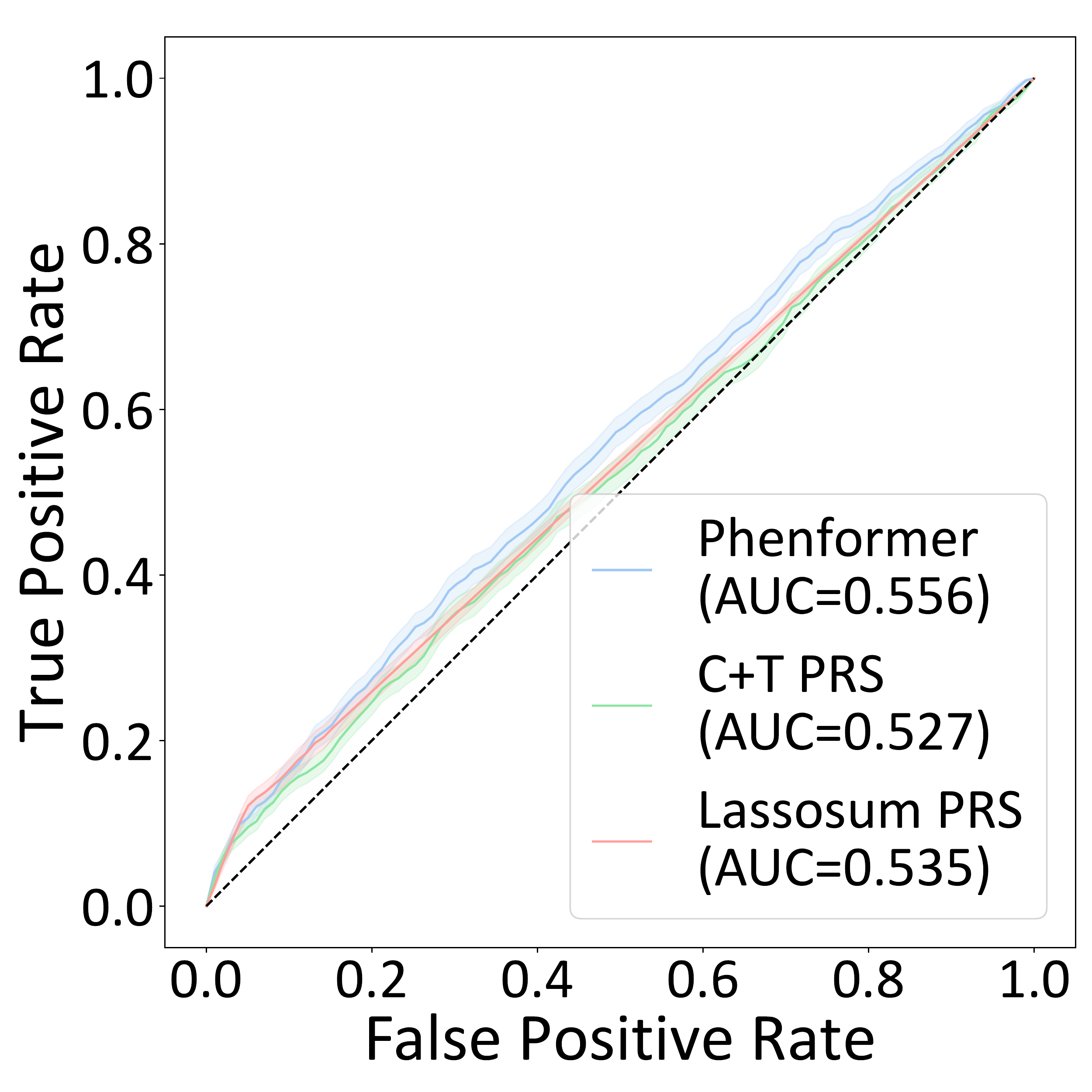}
  \end{subfigure}
  \begin{subfigure}[t]{\resultsubplotsize\textwidth}
\centering	
	\includegraphics[width=1.0\textwidth, valign=b]{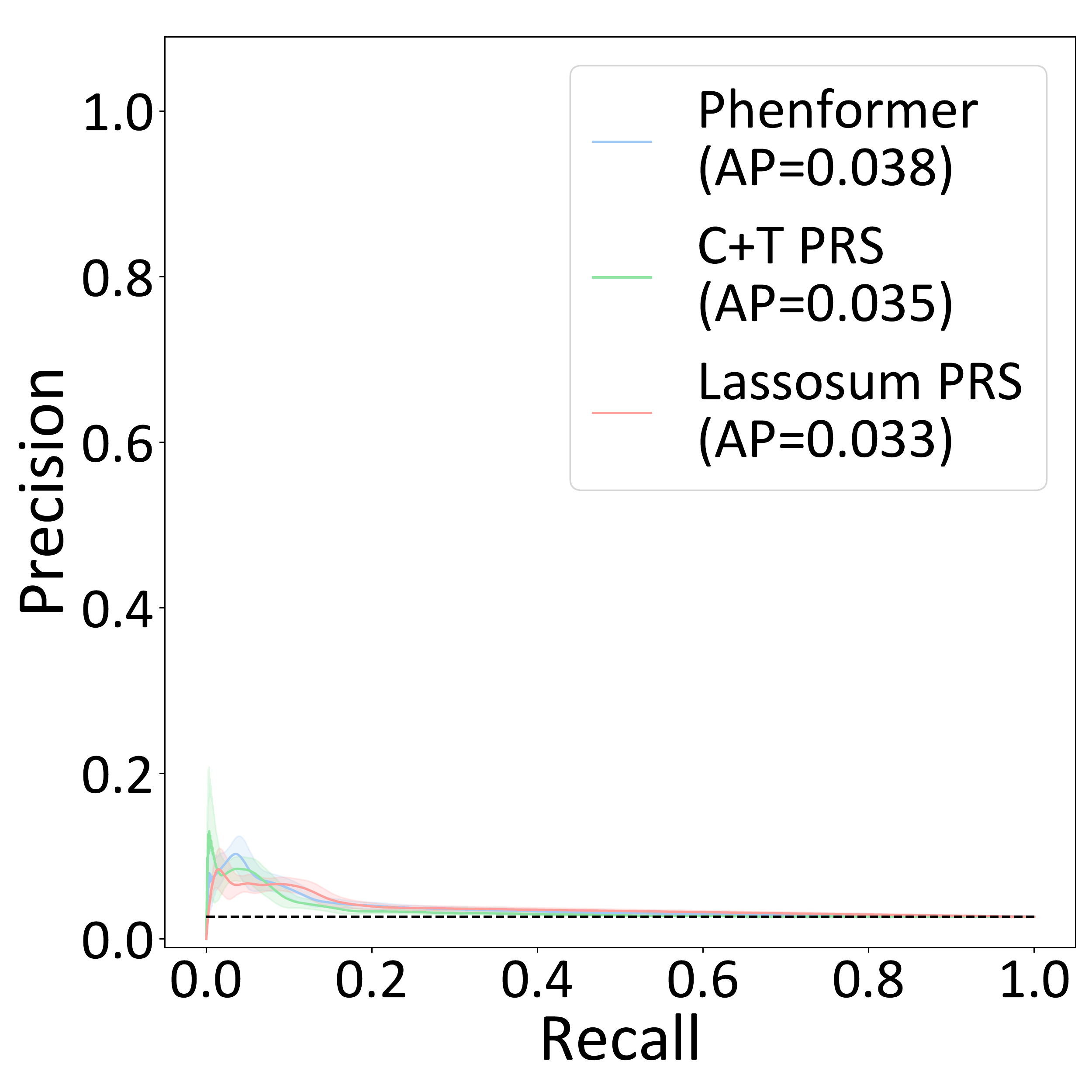}
  \end{subfigure}
  \begin{subfigure}[t]{\resultsubplotsize\textwidth}
\centering	
	\includegraphics[width=1.0\textwidth, valign=b]{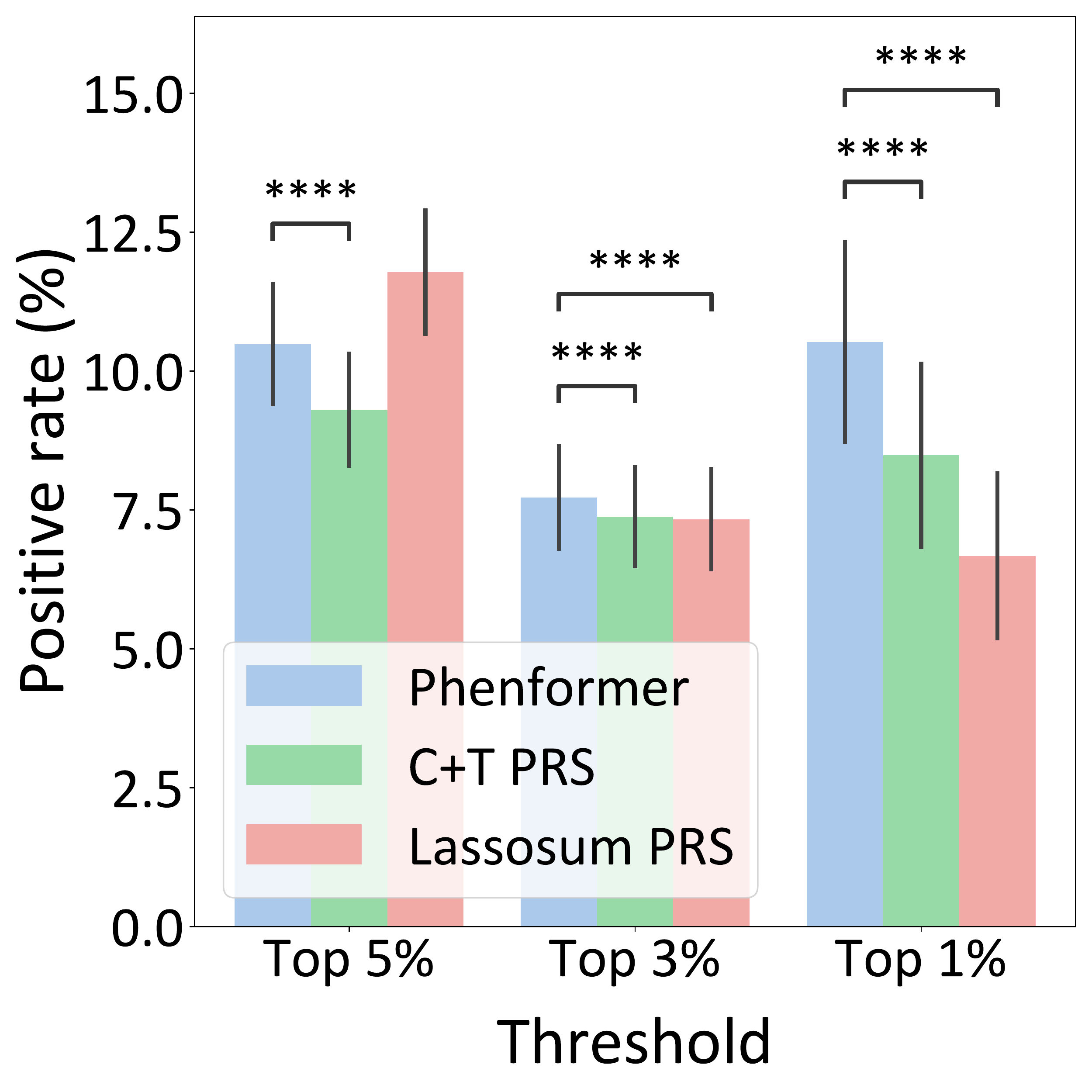}
  \end{subfigure}\hfill
  
  \begin{subfigure}[b]{\rowheadersize\textwidth}
    \rotatebox{90}{\hspace{2em}\textsf{COPD}}
  \end{subfigure}
  \begin{subfigure}[t]{\resultsubplotsize\textwidth}
  \centering
	\includegraphics[width=1.0\textwidth, valign=b]{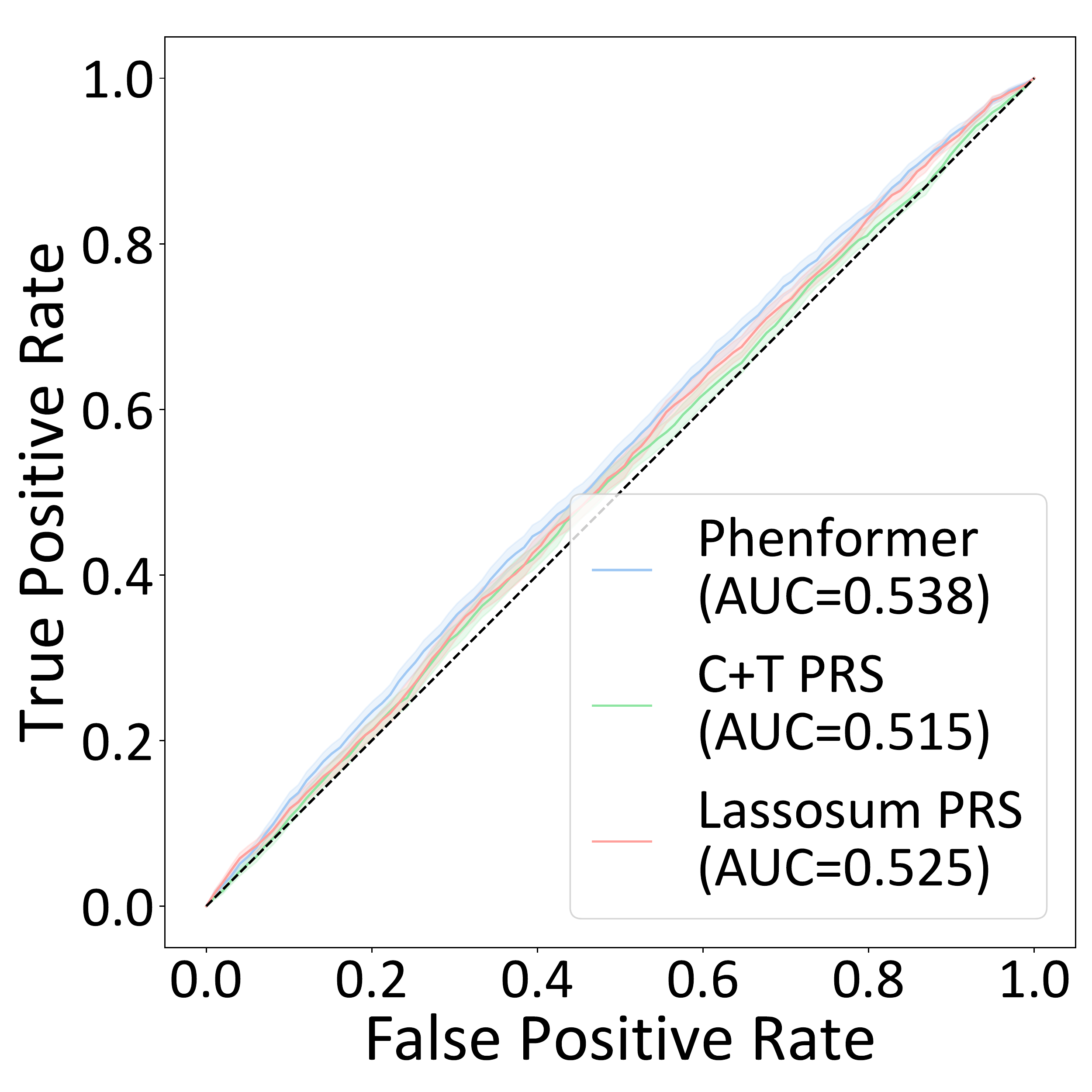}
  \end{subfigure}
  \begin{subfigure}[t]{\resultsubplotsize\textwidth}
\centering	
	\includegraphics[width=1.0\textwidth, valign=b]{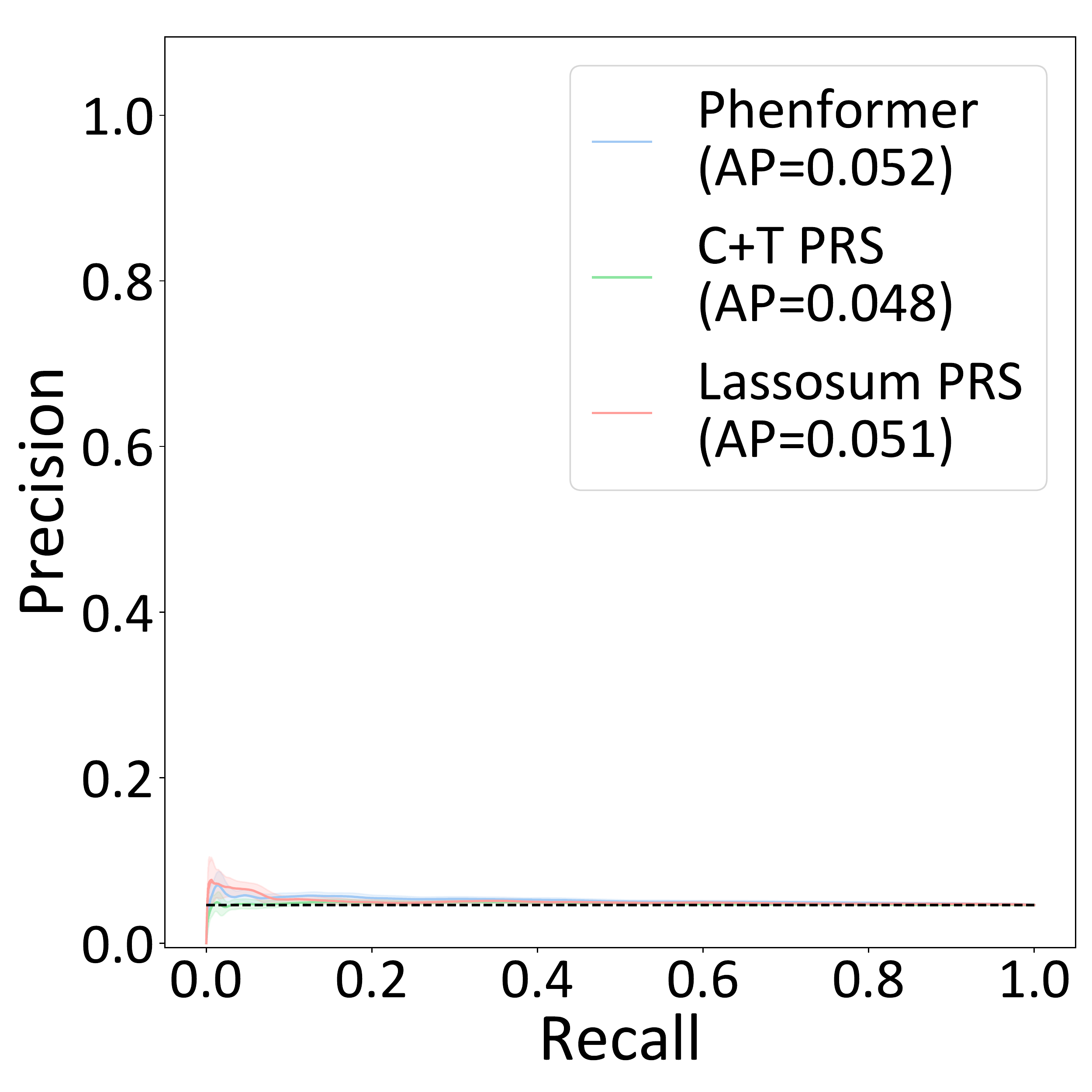}
  \end{subfigure}
  \begin{subfigure}[t]{\resultsubplotsize\textwidth}
\centering	
	\includegraphics[width=1.0\textwidth, valign=b]{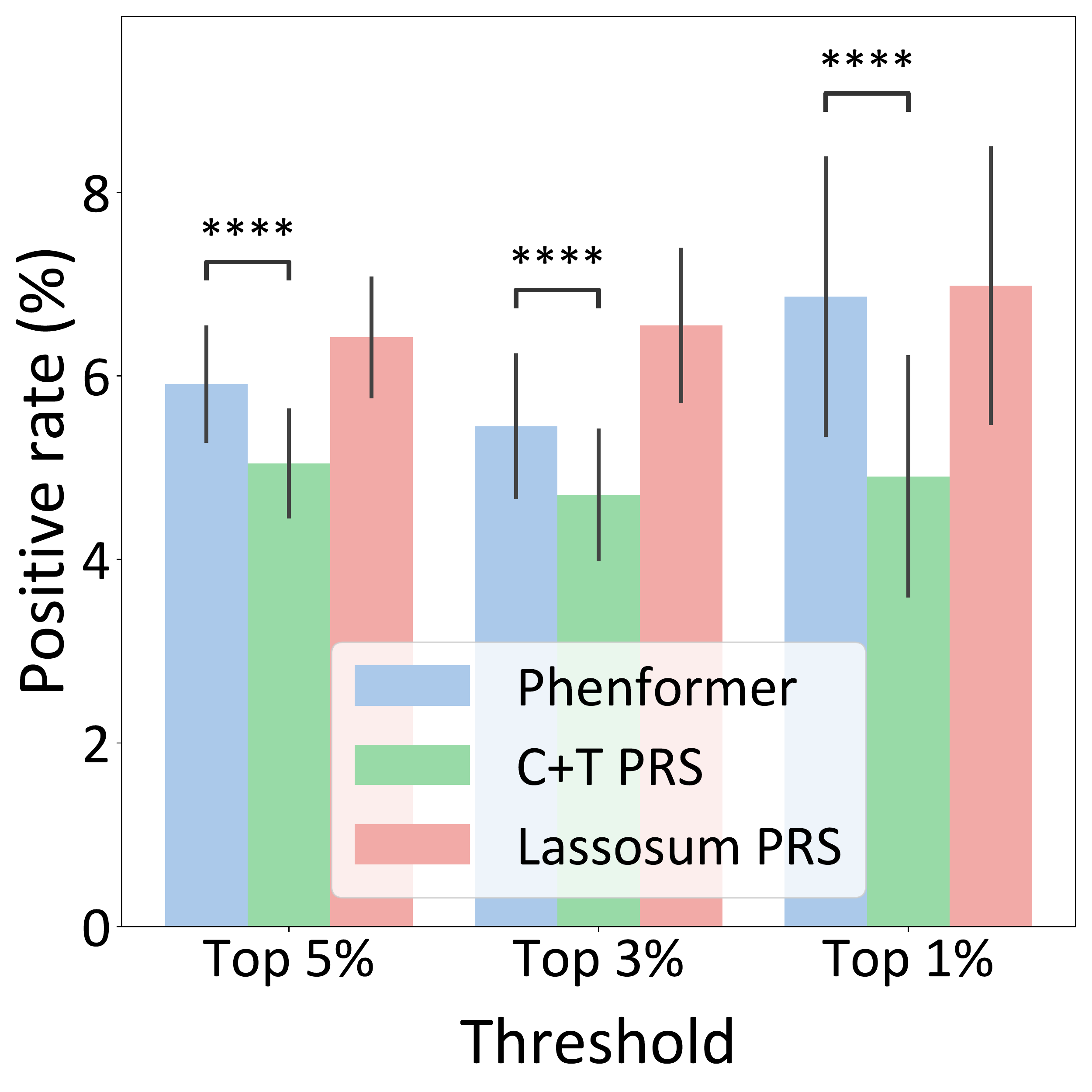}
  \end{subfigure}\hfill
    
  \begin{subfigure}[b]{\rowheadersize\textwidth}
    \rotatebox{90}{\hspace{0.5em}\textsf{Hypothyroidism}}
  \end{subfigure}
  \begin{subfigure}[t]{\resultsubplotsize\textwidth}
  \centering
	\includegraphics[width=1.0\textwidth, valign=b]{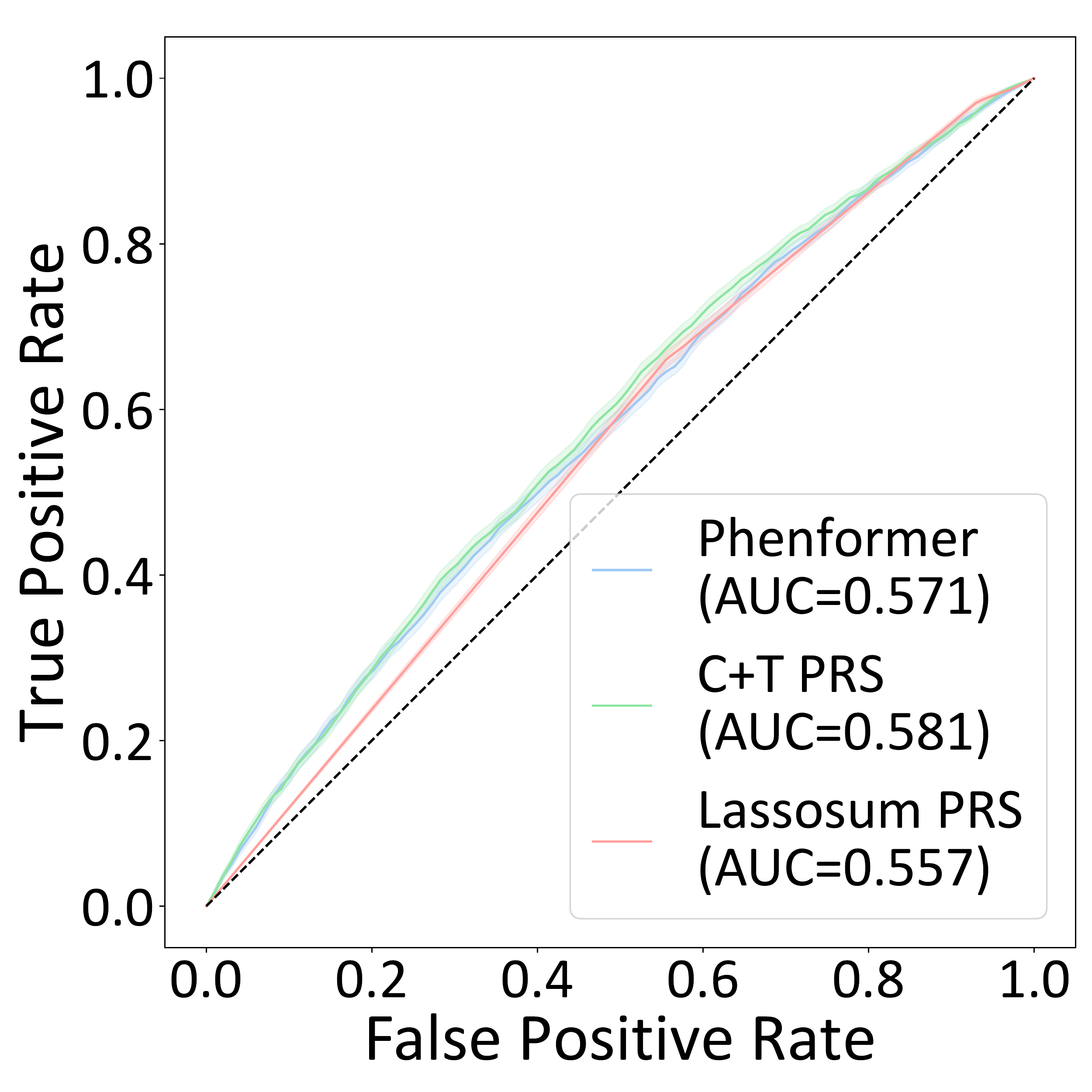}
  \end{subfigure}
  \begin{subfigure}[t]{\resultsubplotsize\textwidth}
\centering	
	\includegraphics[width=1.0\textwidth, valign=b]{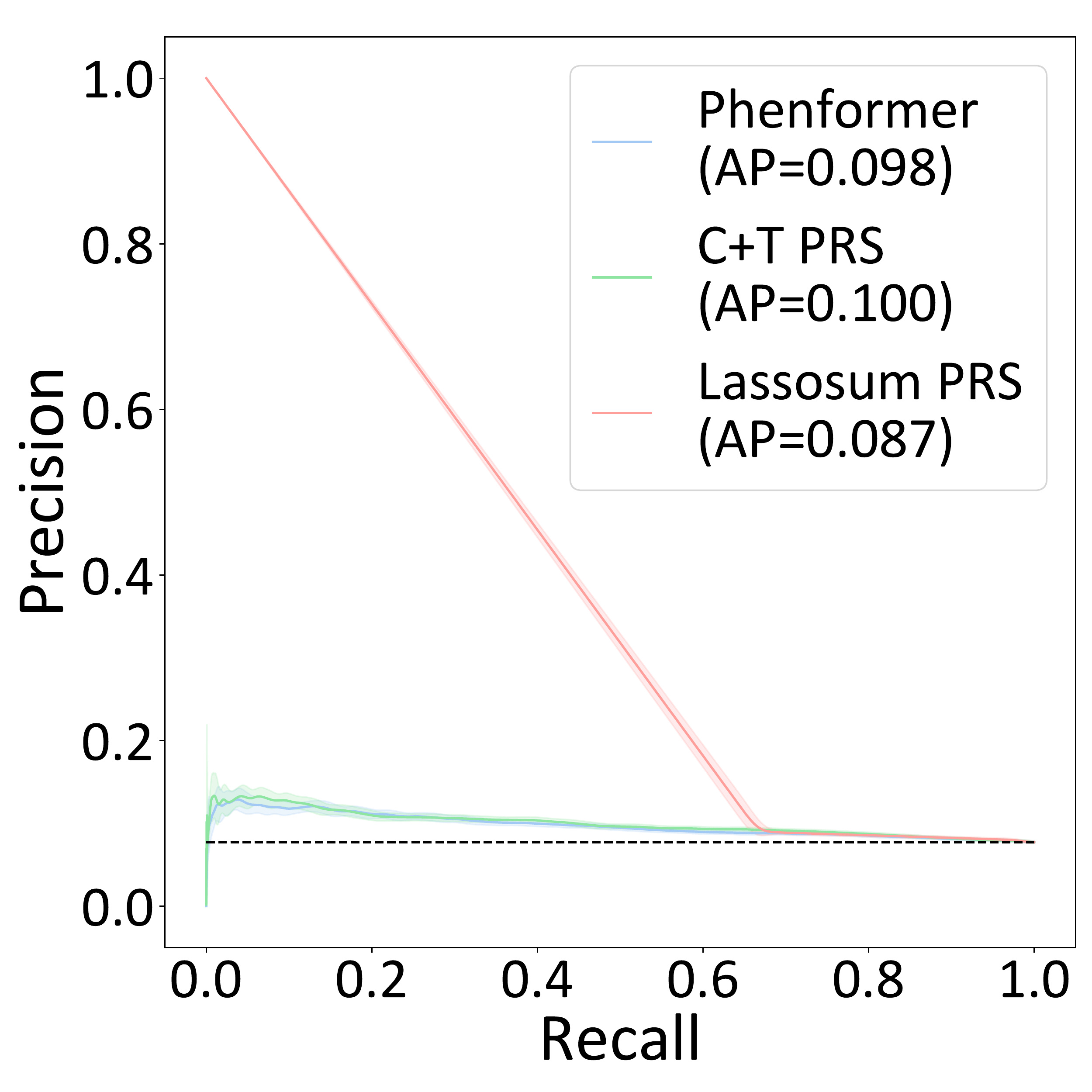}
  \end{subfigure}
  \begin{subfigure}[t]{\resultsubplotsize\textwidth}
\centering	
	\includegraphics[width=1.0\textwidth, valign=b]{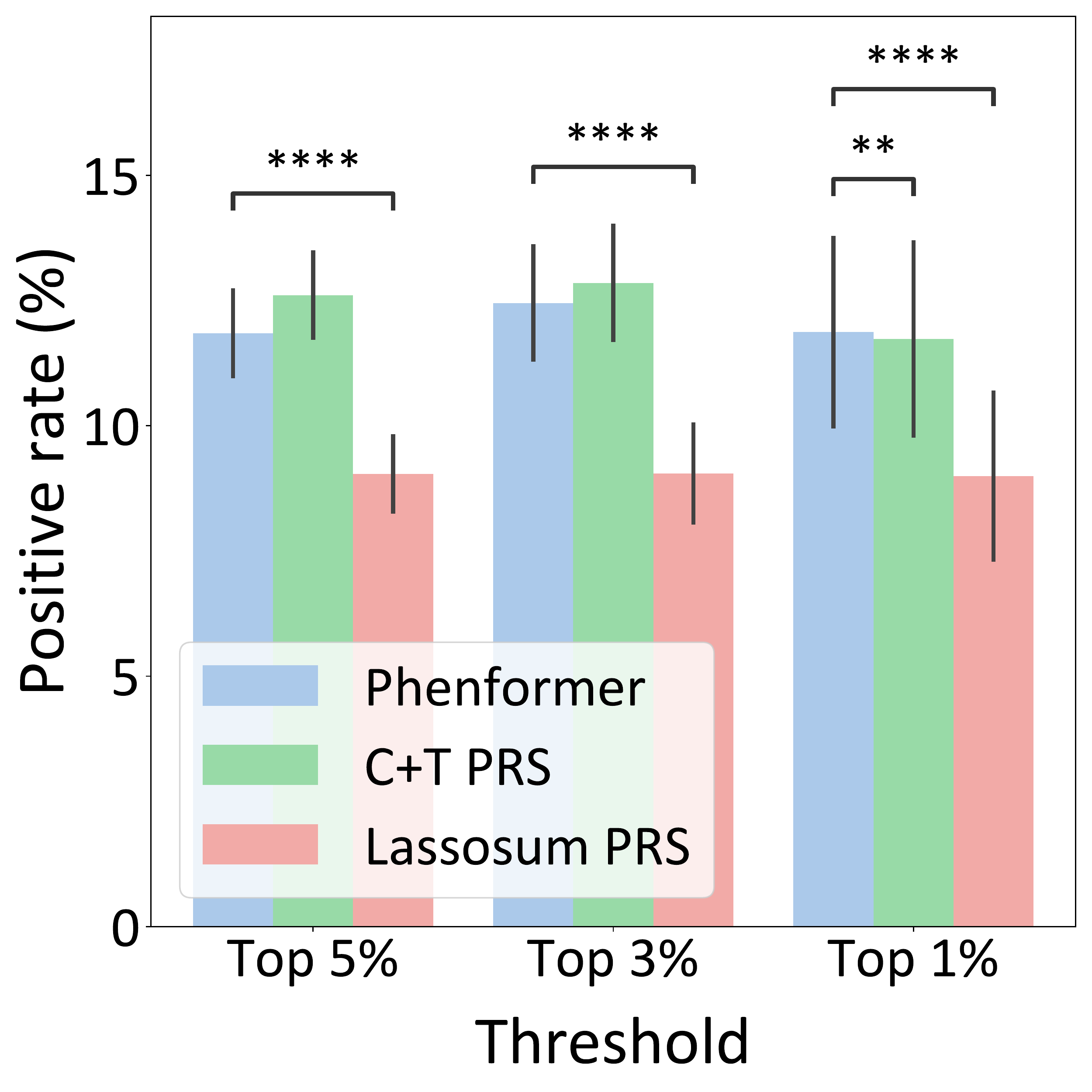}
  \end{subfigure}\hfill
 
\caption{\textbf{\themethod{} outperforms polygenic risk score (PRS) methods on several major diseases across ancestries.} The performance of \themethod{} compared to state-of-the-art polygenic risk scores (PRS) methods in terms of Area under the Receiver Operator Curve (AUROC; leftmost column), Area under the Precision Recall Curve (AUPRC; center column) and positive predictive value among the top 3\% highest predictions stratified by age group (top 3\% PPV; rightmost column) on the same held-out test set of individuals, variants and diseases (psoriasis, type 1 diabetes, type 2 diabetes, diabetic retinopathy, chronic obstructive pulmonary disease [COPD], hypothyroidism). \themethod{} outperforms PRS methods significantly (p $\leq 0.05$) on all diseases except C+T PRS on Hypothyroidism. Stars (****) indicate statistical significance (p $\leq 0.001$, Mann-Whitney Wilcoxon test for superiority, \numprint{2000} bootstrap samples).}
\label{fig:performance}
\end{figure*} %
  
  \begin{figure*}[pt!] 
  \centering
  
\figtitle{Risk Prediction for Individuals of Non-European Ancestry}
 
 \begin{subfigure}[b]{\rowheadersize\textwidth}
    \hspace{0.5em}
  \end{subfigure}
\begin{subfigure}[t]{\resultsubplotsize\textwidth}\centering
    \textsf{AUROC}
  \end{subfigure}
  \begin{subfigure}[t]{\resultsubplotsize\textwidth}\centering
   \textsf {AUPRC}
  \end{subfigure} 
  \begin{subfigure}[t]{\resultsubplotsize\textwidth}\centering
    \textsf{PPV}
  \end{subfigure}\hfill 
  
  \begin{subfigure}[b]{\rowheadersize\textwidth}
    \rotatebox{90}{\hspace{1.5em}\textsf{Psoriasis}}
  \end{subfigure}
  \begin{subfigure}[t]{\resultsubplotsize\textwidth}
  \centering
	\includegraphics[width=1.0\textwidth, valign=b]{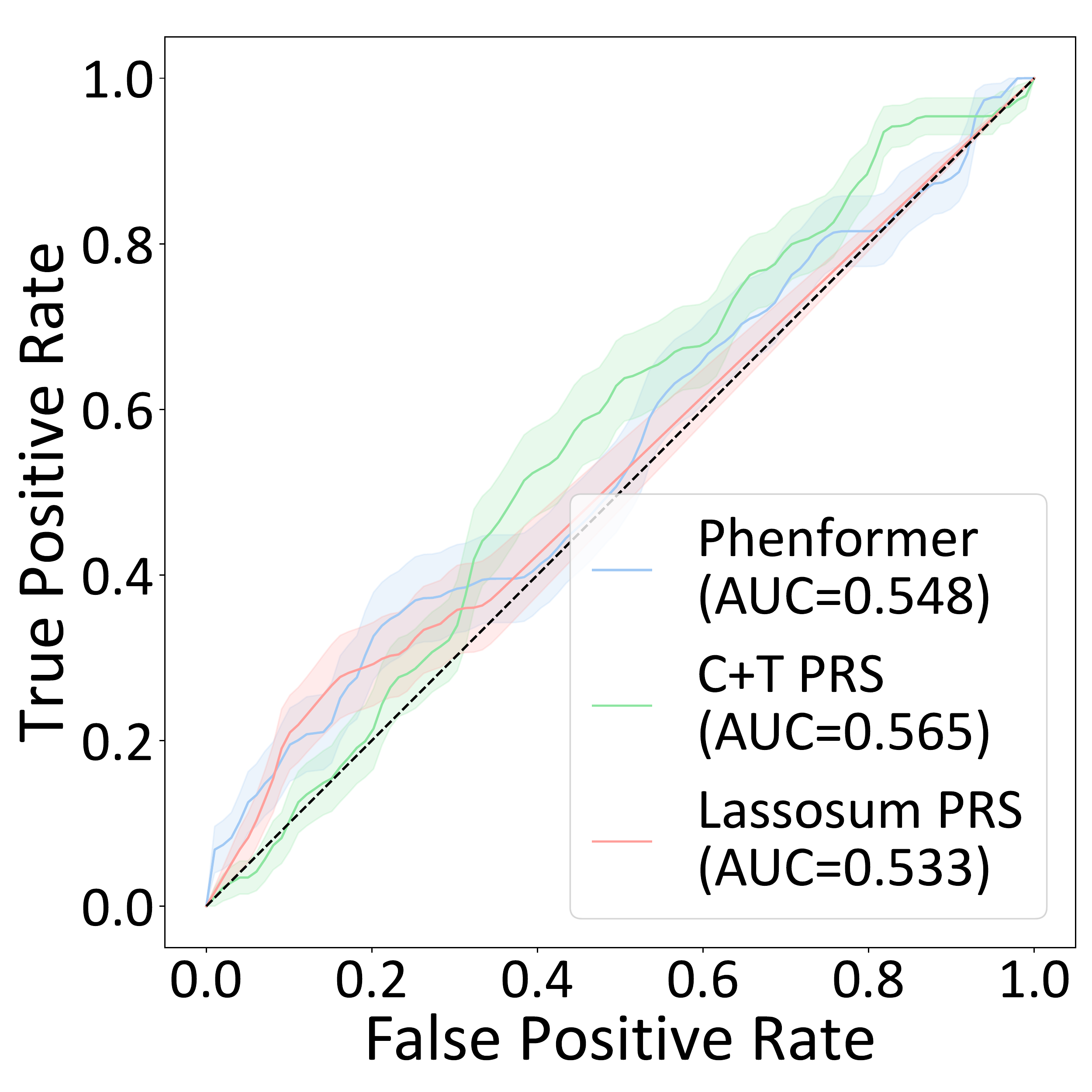}
  \end{subfigure}
  \begin{subfigure}[t]{\resultsubplotsize\textwidth}
\centering	
	\includegraphics[width=1.0\textwidth, valign=b]{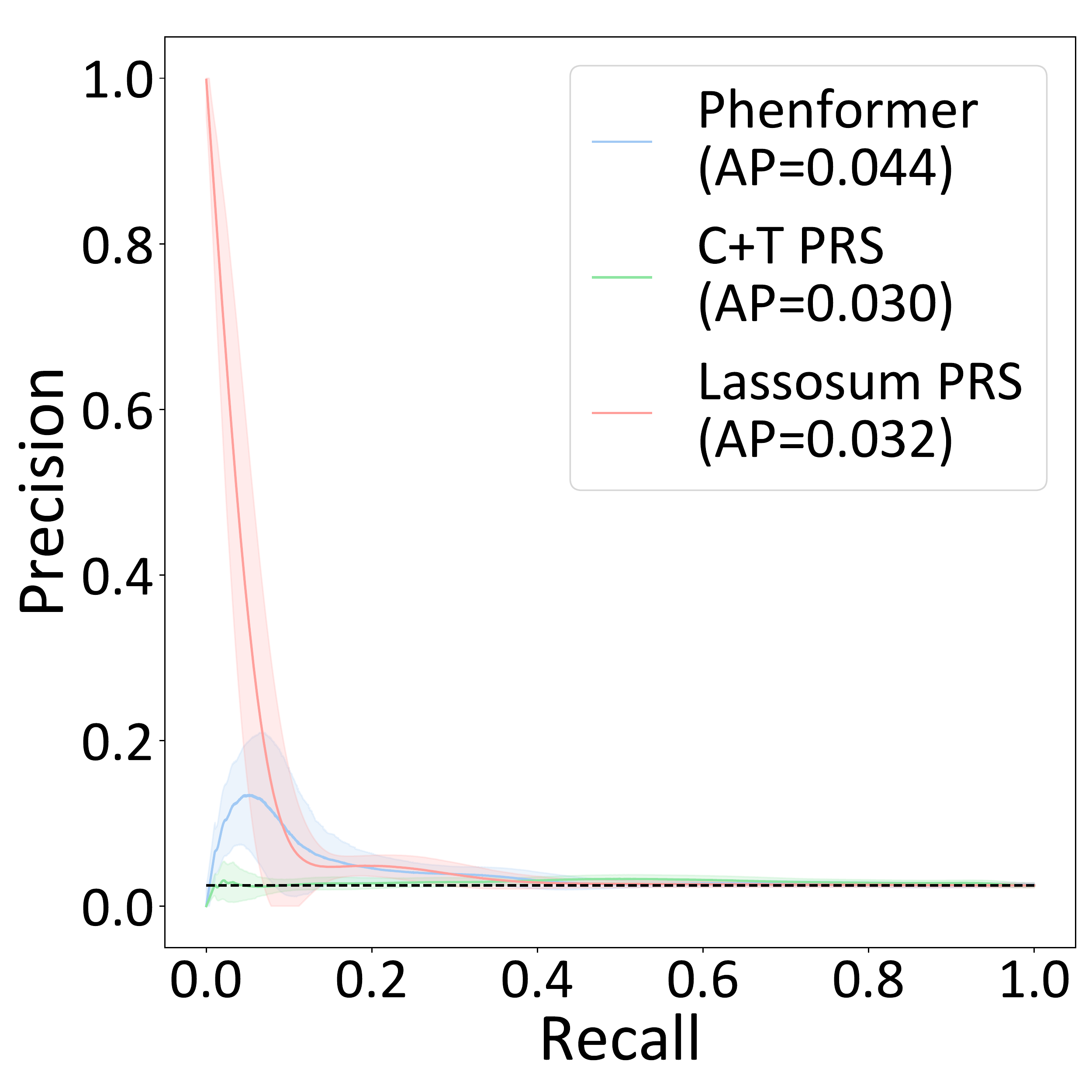}
  \end{subfigure}
  \begin{subfigure}[t]{\resultsubplotsize\textwidth}
\centering	
	\includegraphics[width=1.0\textwidth, valign=b]{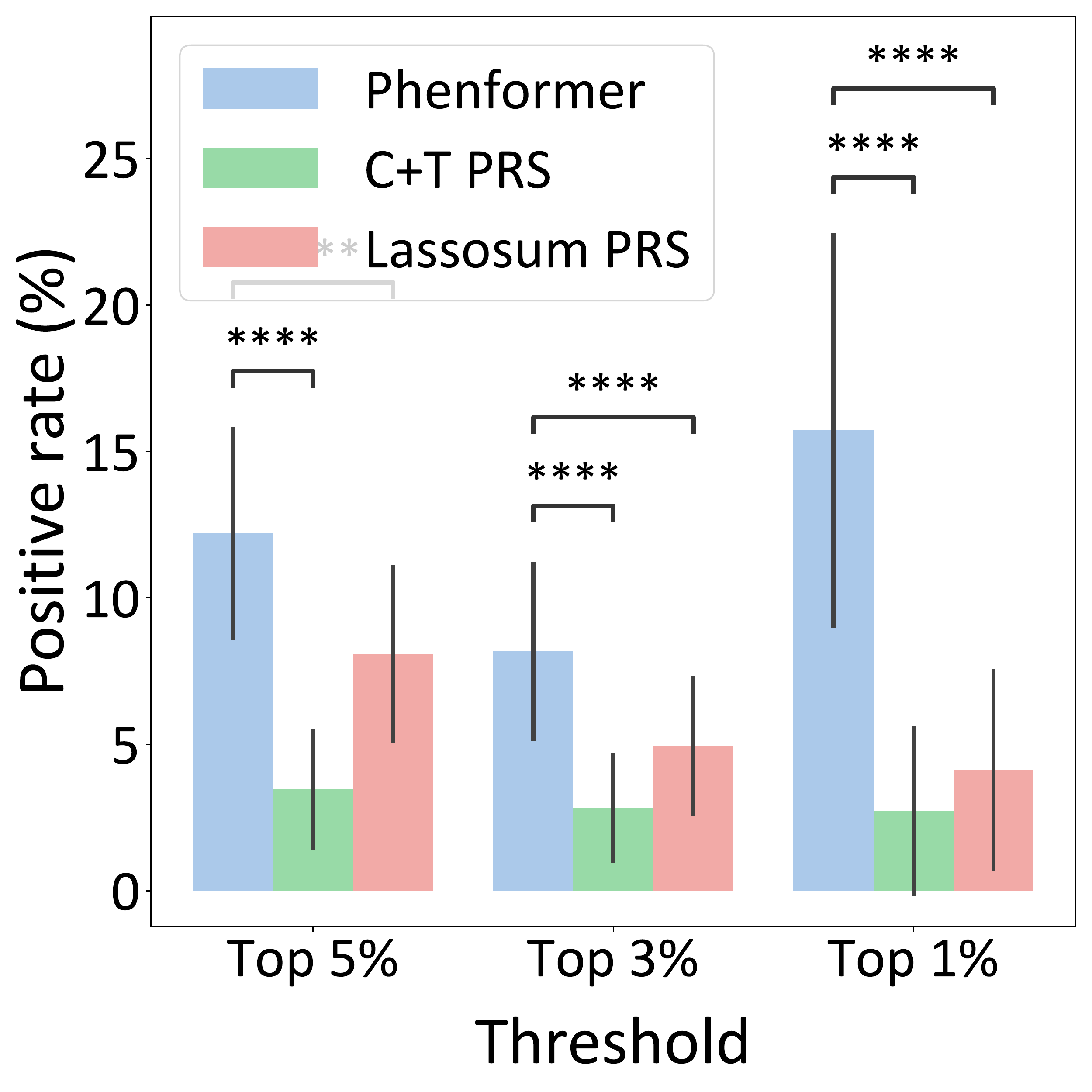}
  \end{subfigure}\hfill 
 
  \begin{subfigure}[b]{\rowheadersize\textwidth}
    \rotatebox{90}{\hspace{0.1em}\textsf{Type 1 Diabetes}}
  \end{subfigure}
  \begin{subfigure}[t]{\resultsubplotsize\textwidth}
  \centering
	\includegraphics[width=1.0\textwidth, valign=b]{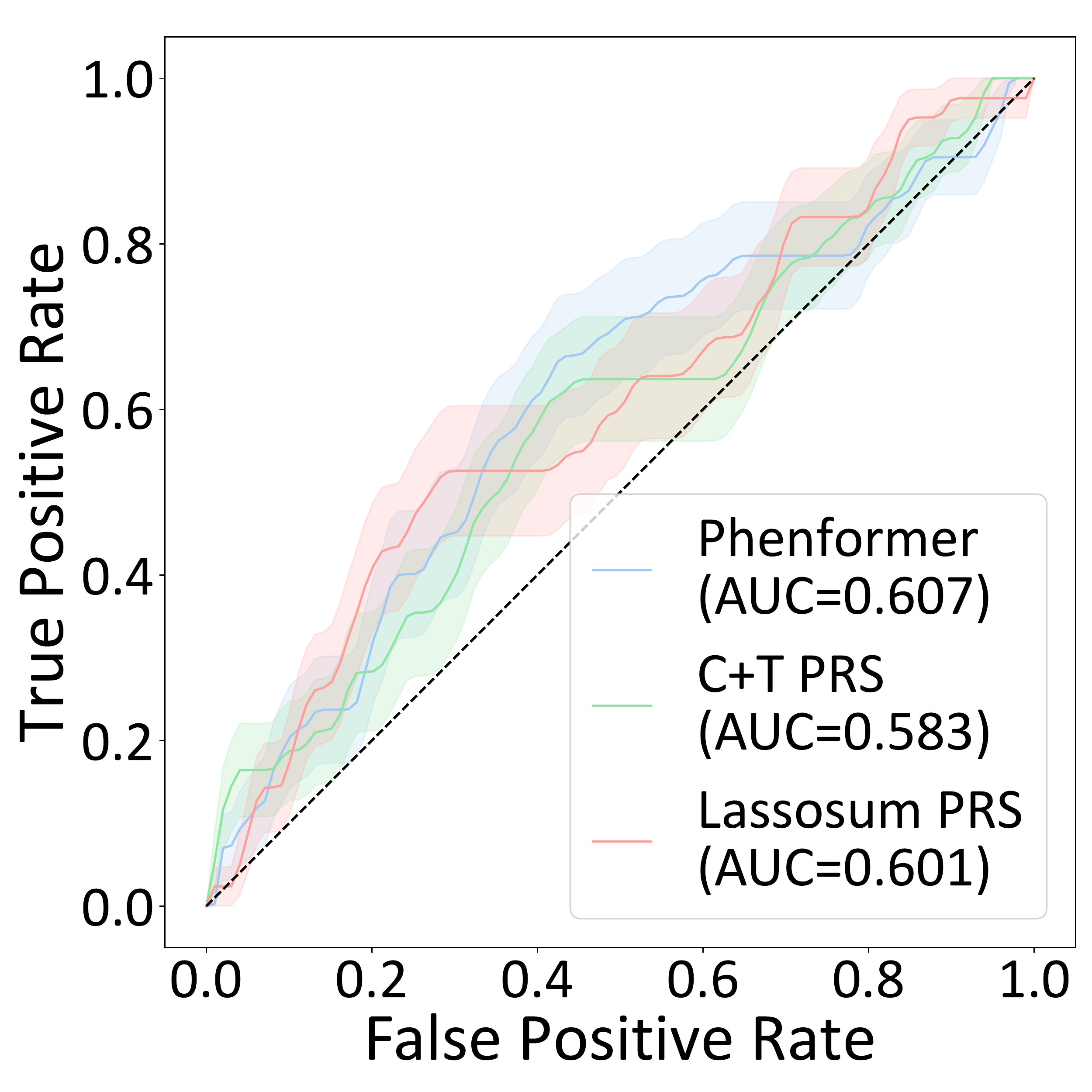}
  \end{subfigure}
  \begin{subfigure}[t]{\resultsubplotsize\textwidth}
\centering	
	\includegraphics[width=1.0\textwidth, valign=b]{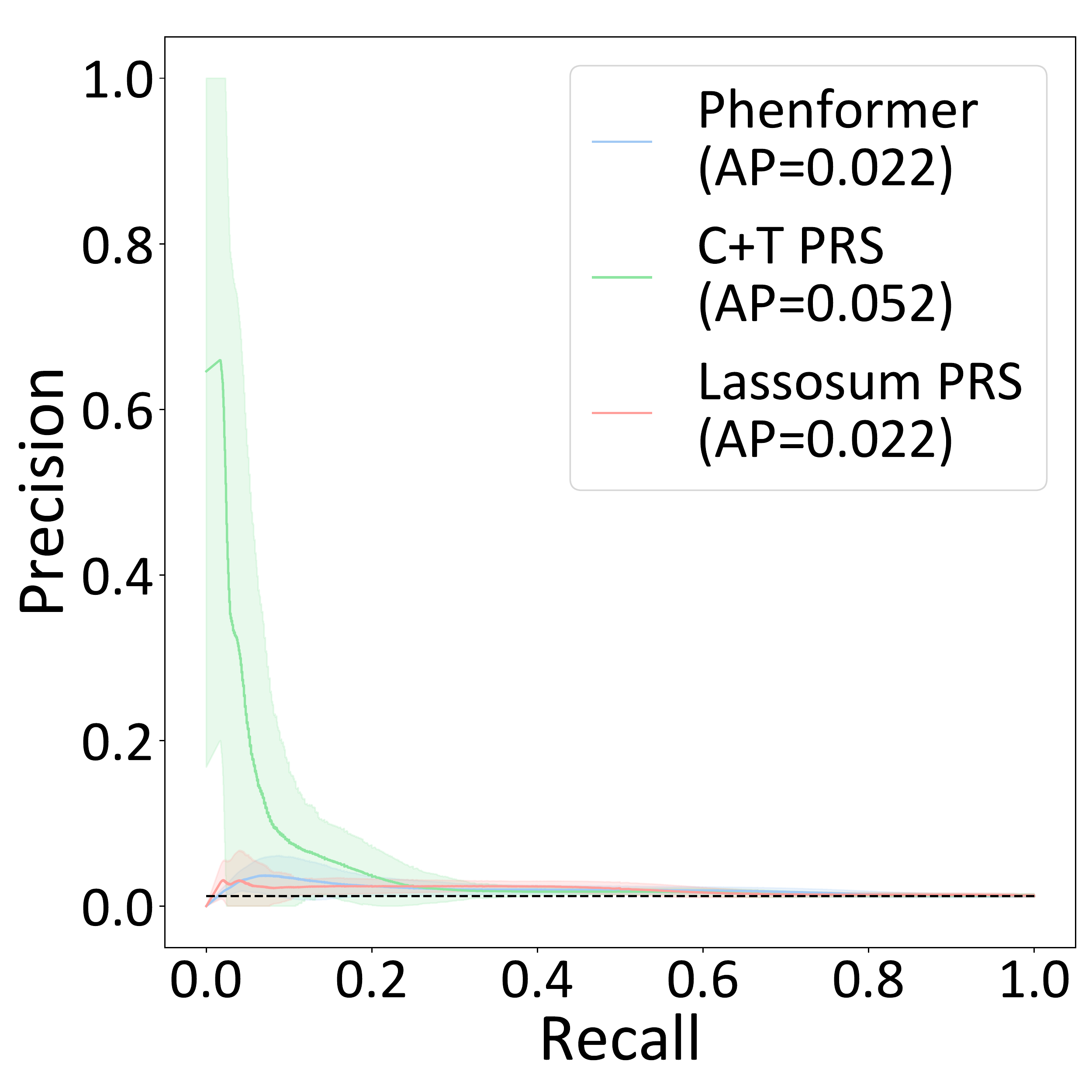}
  \end{subfigure}
  \begin{subfigure}[t]{\resultsubplotsize\textwidth}
\centering	
	\includegraphics[width=1.0\textwidth, valign=b]{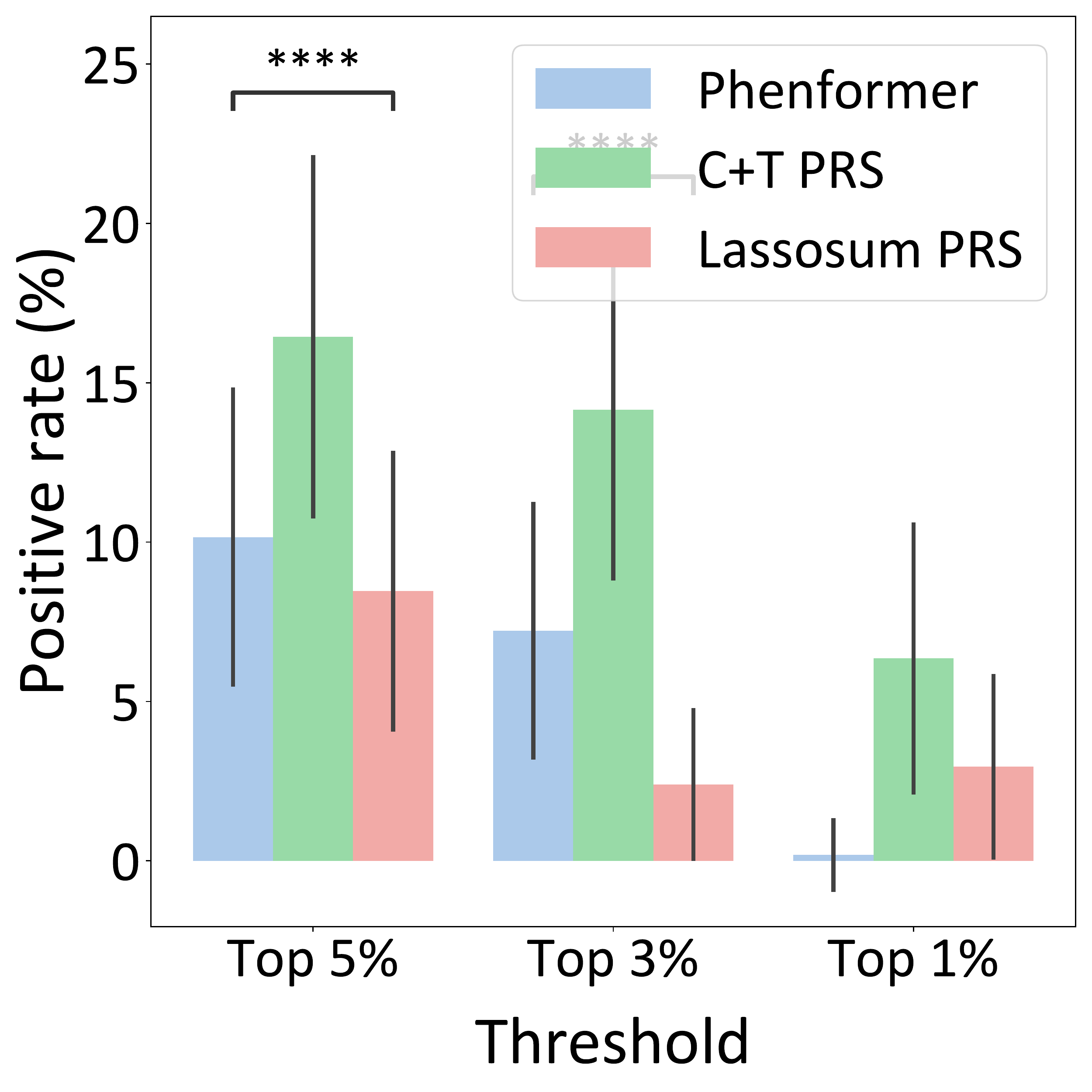}
  \end{subfigure}\hfill  
  
  \begin{subfigure}[b]{\rowheadersize\textwidth}
    \rotatebox{90}{\hspace{0.1em}\textsf{Type 2 Diabetes}}
  \end{subfigure}
  \begin{subfigure}[t]{\resultsubplotsize\textwidth}
  \centering
	\includegraphics[width=1.0\textwidth, valign=b]{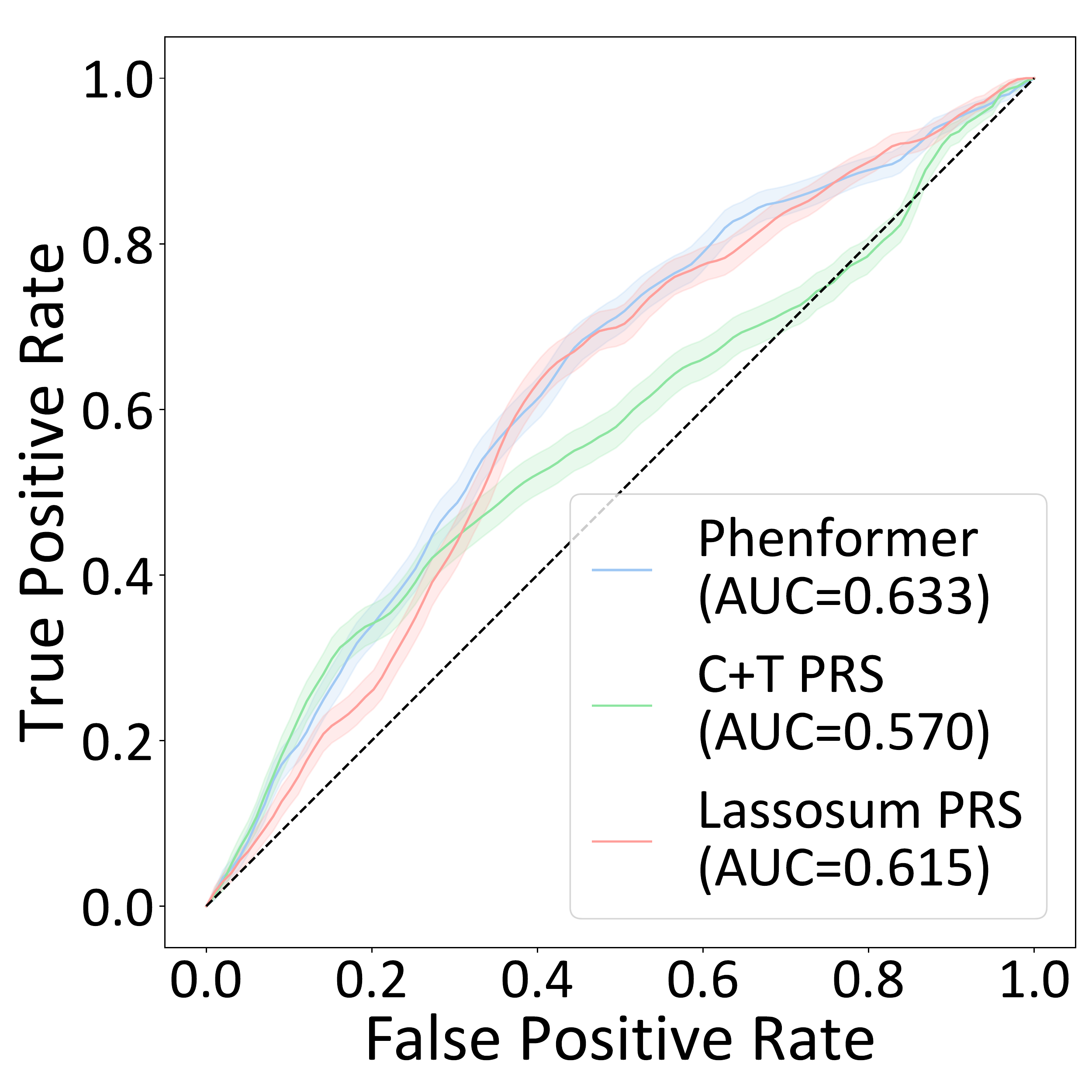}
  \end{subfigure}
  \begin{subfigure}[t]{\resultsubplotsize\textwidth}
\centering	
	\includegraphics[width=1.0\textwidth, valign=b]{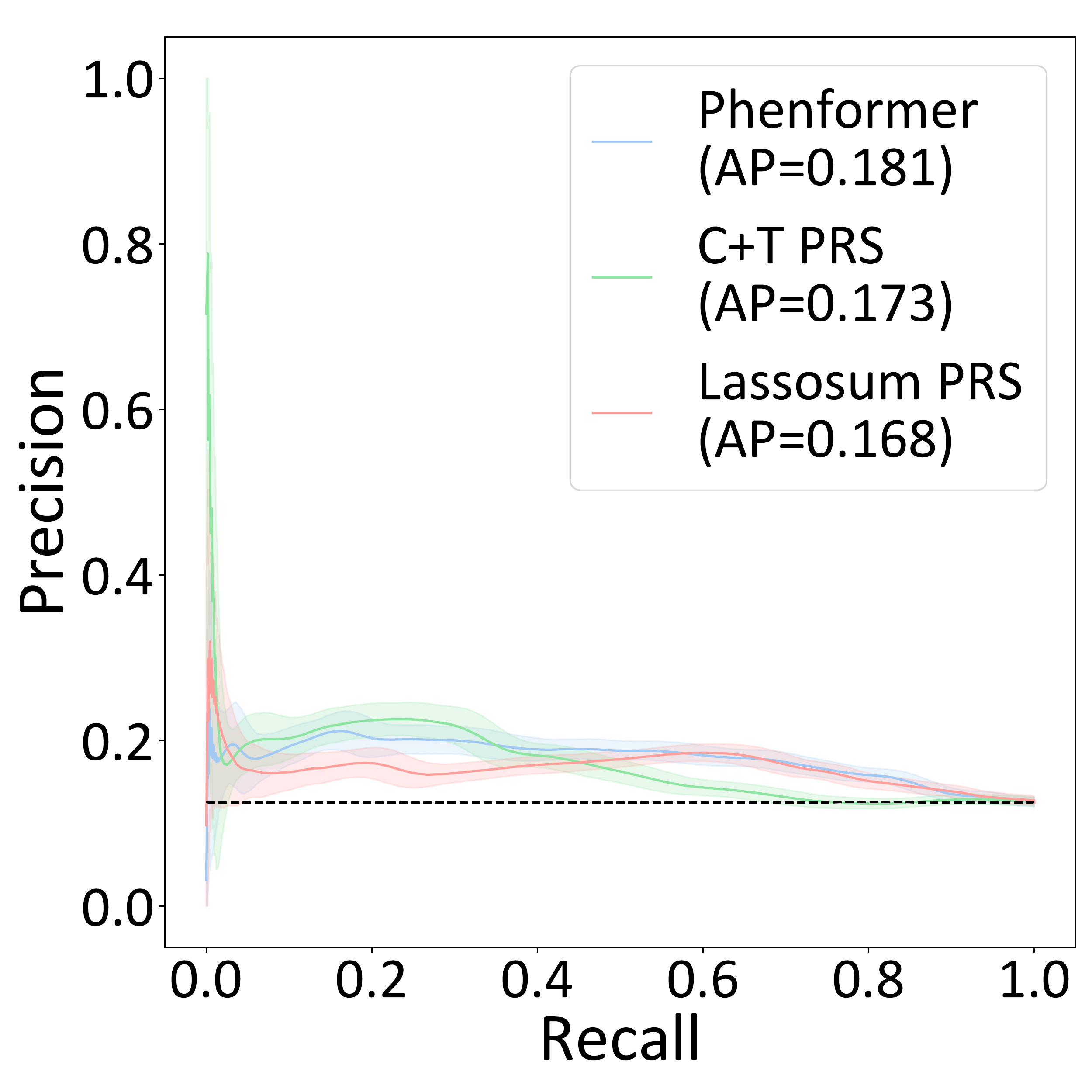}
  \end{subfigure}
  \begin{subfigure}[t]{\resultsubplotsize\textwidth}
\centering	
	\includegraphics[width=1.0\textwidth, valign=b]{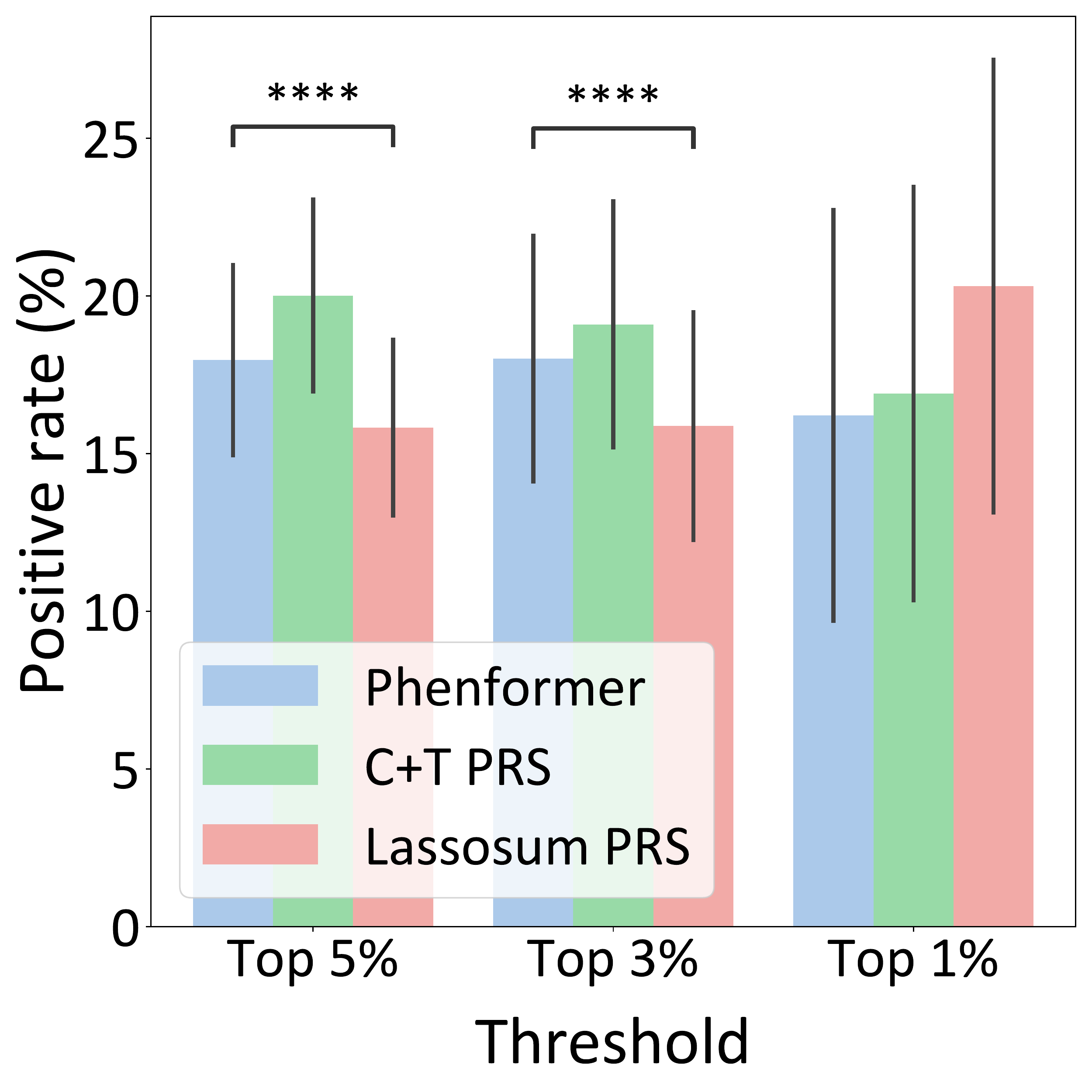}
  \end{subfigure}\hfill  
  
  \begin{subfigure}[b]{\rowheadersize\textwidth}
    \rotatebox{90}{\hspace{0em}\textsf{Diabetic Retino.}}
  \end{subfigure}
  \begin{subfigure}[t]{\resultsubplotsize\textwidth}
  \centering
	\includegraphics[width=1.0\textwidth, valign=b]{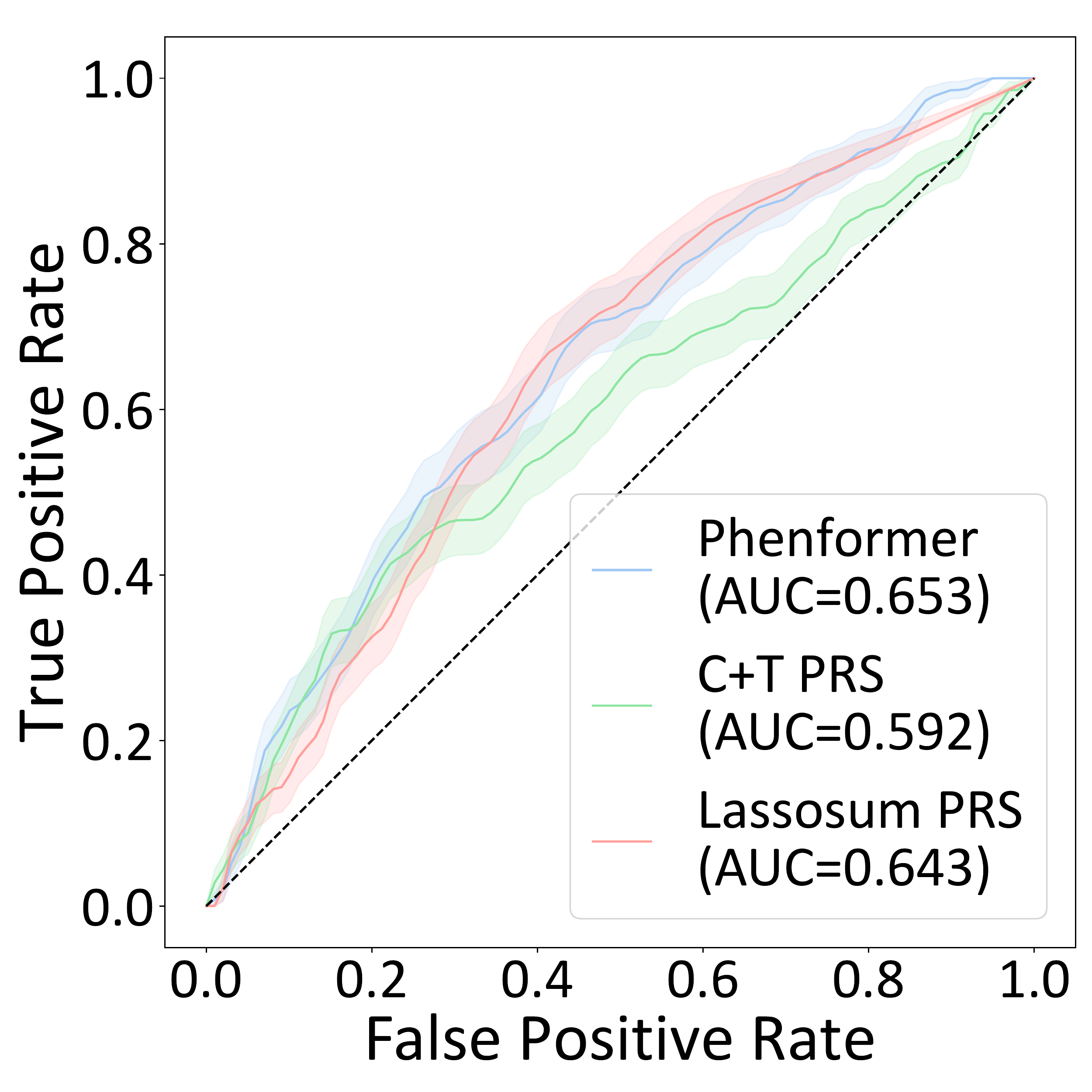}
  \end{subfigure}
  \begin{subfigure}[t]{\resultsubplotsize\textwidth}
\centering	
	\includegraphics[width=1.0\textwidth, valign=b]{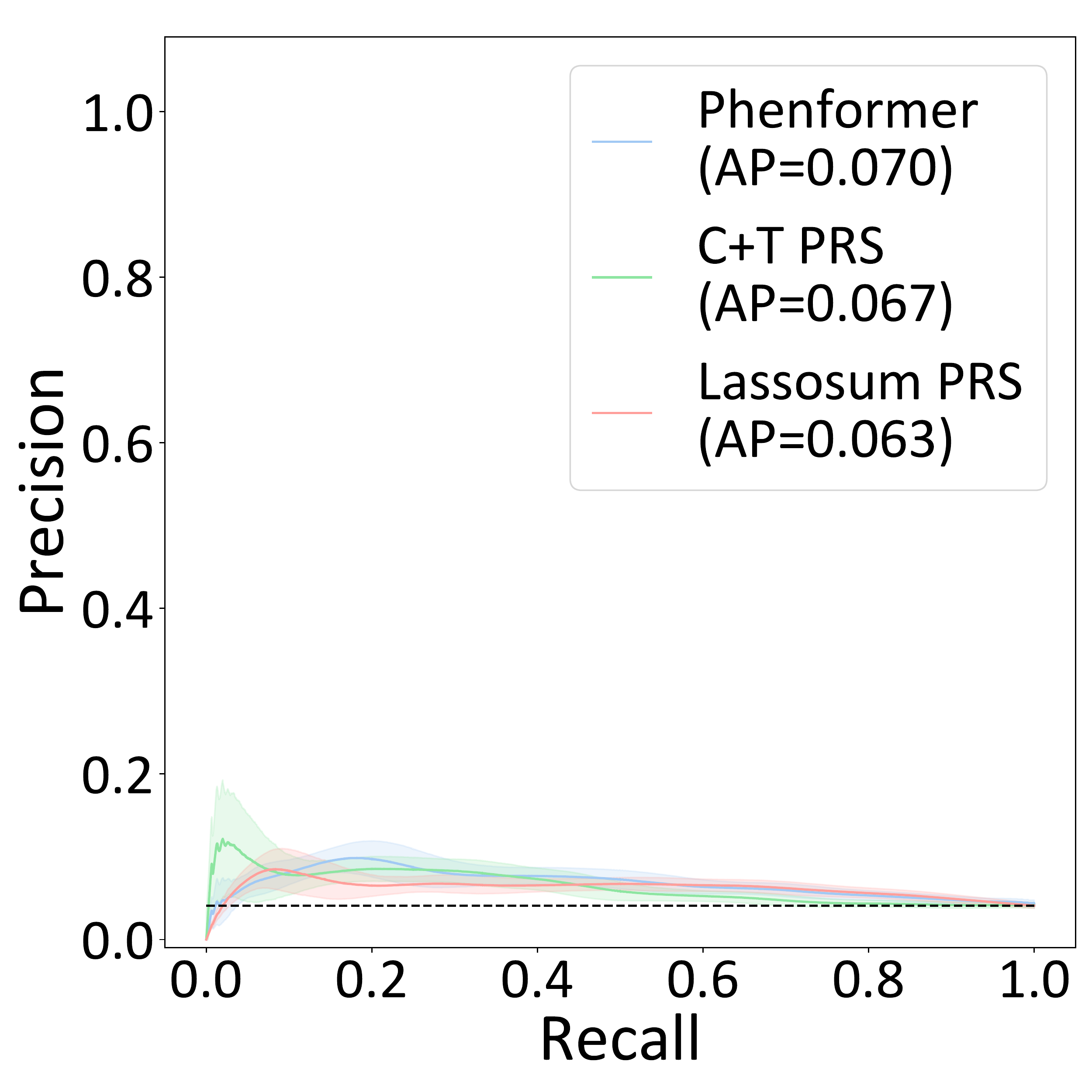}
  \end{subfigure}
  \begin{subfigure}[t]{\resultsubplotsize\textwidth}
\centering	
	\includegraphics[width=1.0\textwidth, valign=b]{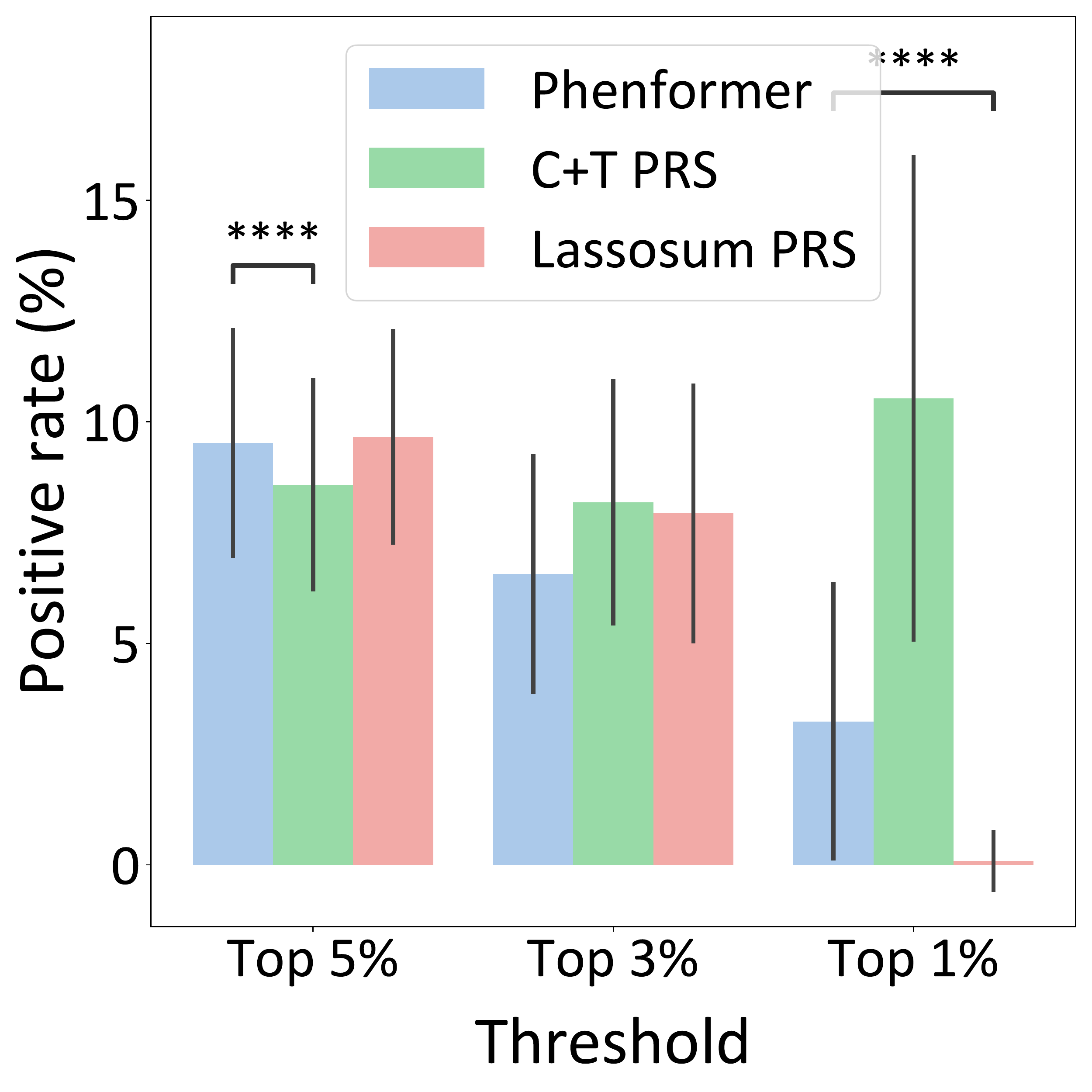}
  \end{subfigure}\hfill
  
  \begin{subfigure}[b]{\rowheadersize\textwidth}
    \rotatebox{90}{\hspace{2em}\textsf{COPD}}
  \end{subfigure}
  \begin{subfigure}[t]{\resultsubplotsize\textwidth}
  \centering
	\includegraphics[width=1.0\textwidth, valign=b]{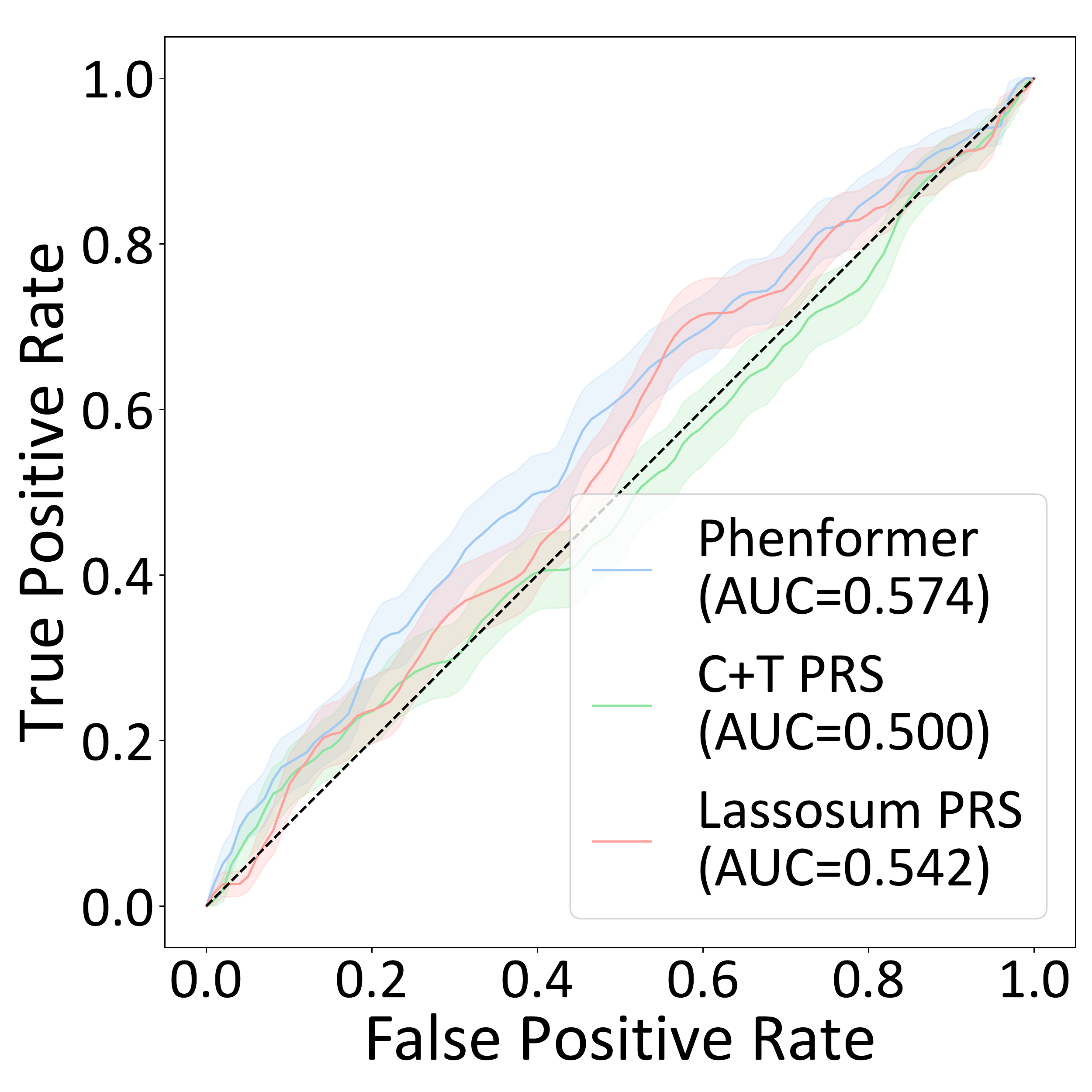}
  \end{subfigure}
  \begin{subfigure}[t]{\resultsubplotsize\textwidth}
\centering	
	\includegraphics[width=1.0\textwidth, valign=b]{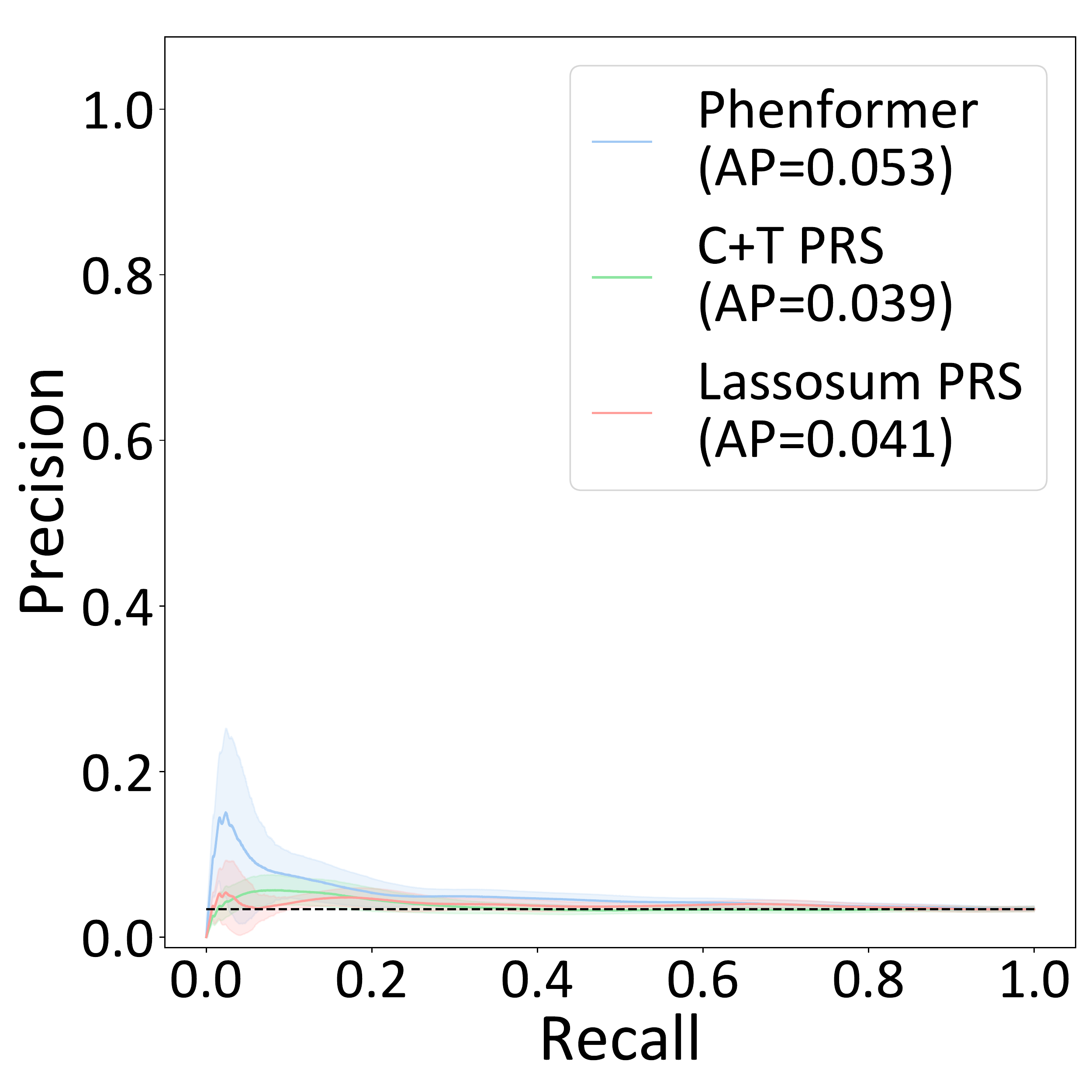}
  \end{subfigure}
  \begin{subfigure}[t]{\resultsubplotsize\textwidth}
\centering	
	\includegraphics[width=1.0\textwidth, valign=b]{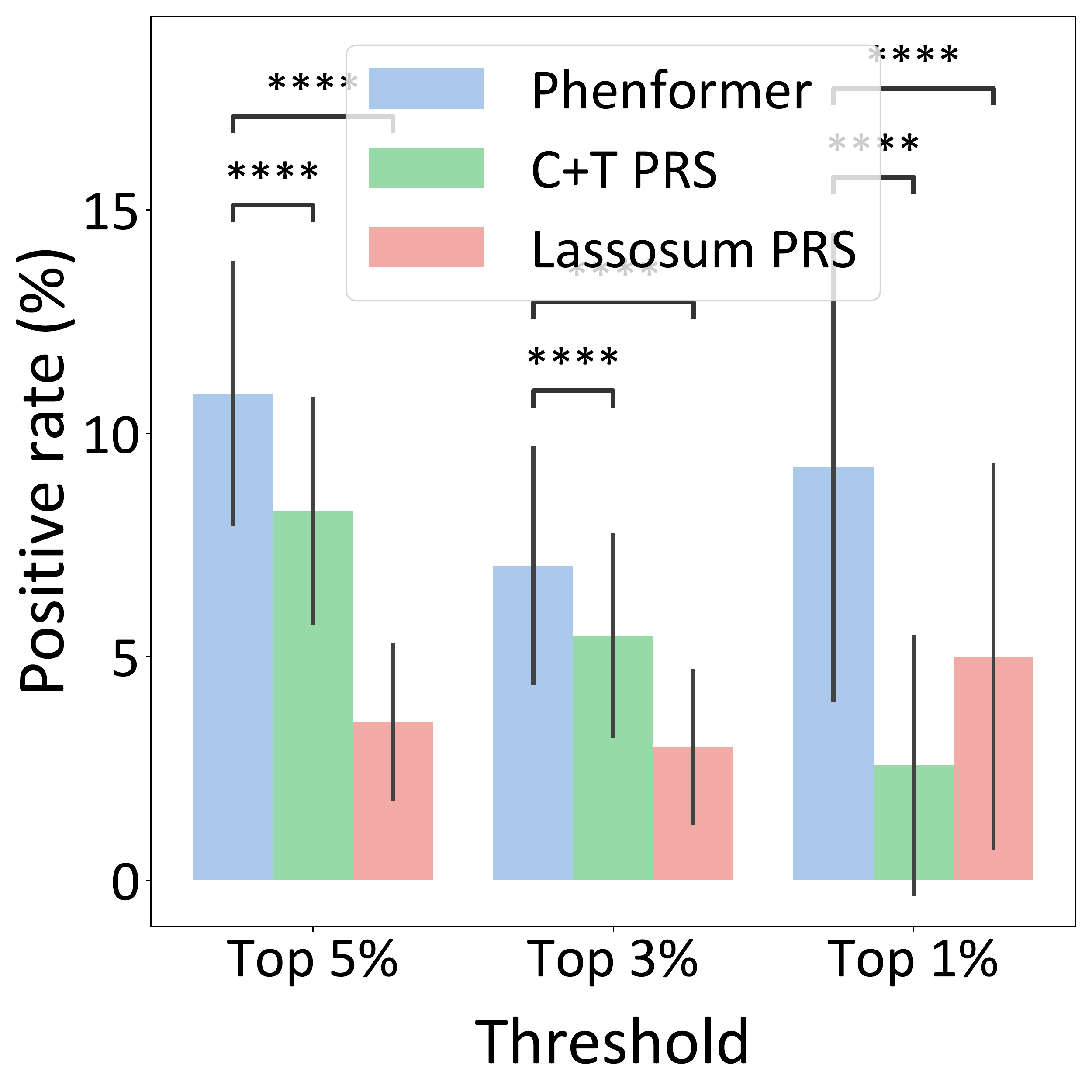}
  \end{subfigure}\hfill
 
  \begin{subfigure}[b]{\rowheadersize\textwidth}
    \rotatebox{90}{\hspace{0.5em}\textsf{Hypothyroidism}}
  \end{subfigure}
  \begin{subfigure}[t]{\resultsubplotsize\textwidth}
  \centering
	\includegraphics[width=1.0\textwidth, valign=b]{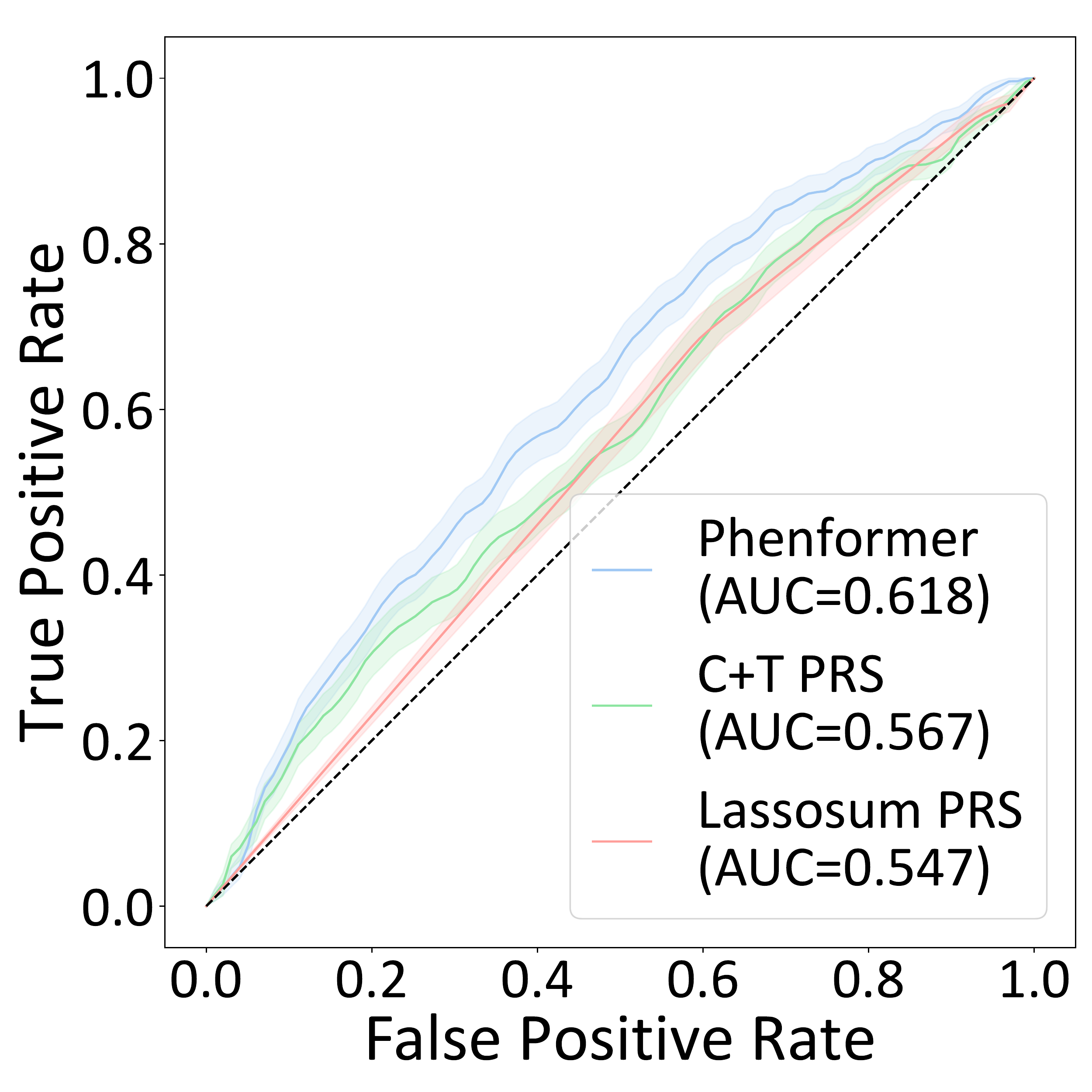}
  \end{subfigure}
  \begin{subfigure}[t]{\resultsubplotsize\textwidth}
\centering	
	\includegraphics[width=1.0\textwidth, valign=b]{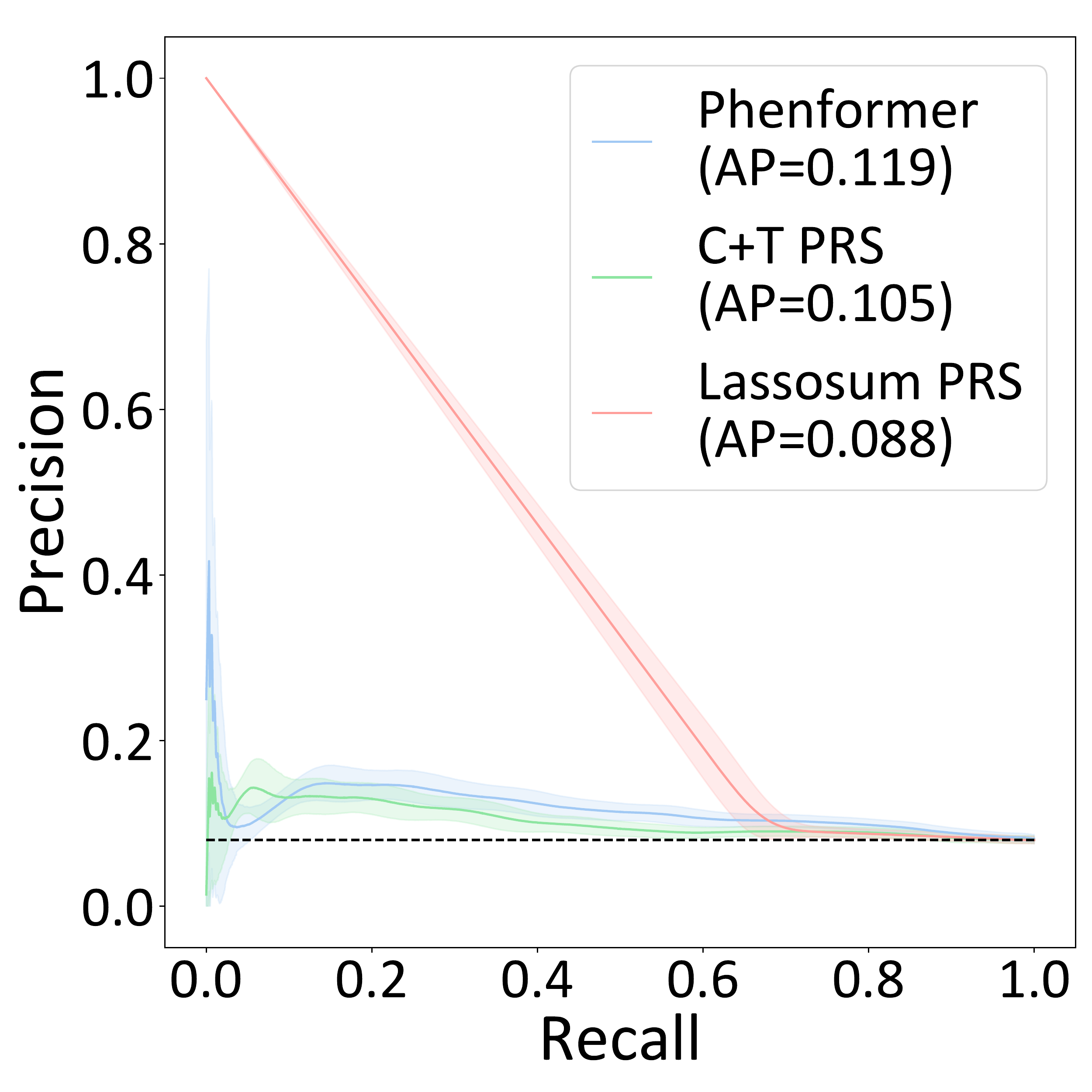}
  \end{subfigure}
  \begin{subfigure}[t]{\resultsubplotsize\textwidth}
\centering	
	\includegraphics[width=1.0\textwidth, valign=b]{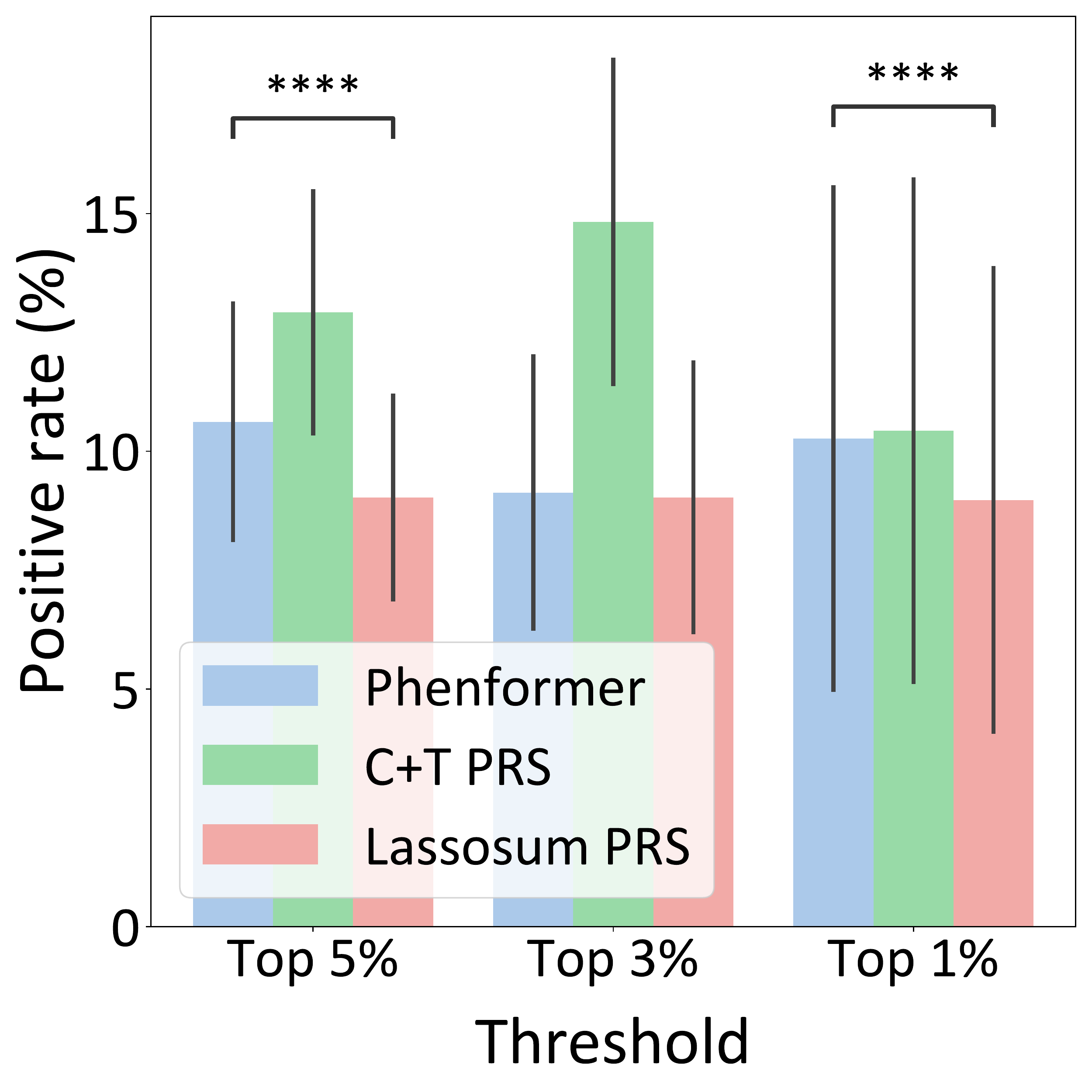}
  \end{subfigure}\hfill
  
\vspace{1pt}
\caption{\textbf{\themethod{} maintains better performance in predicting individual disease risk in individuals of diverse, non-European backgrounds than PRS methods.} The performance of \themethod{} compared to PRS methods in terms of Area under the Receiver Operator Curve (AUROC; leftmost column), Area under the Precision Recall Curve (AUPRC; center column) and positive predictive value among the top 3\% highest predictions (top 3\% PPV; rightmost column) on a subset of individuals of diverse, non-European ancestry. We find that \themethod{} is more transportable than PRS methods with relatively greater performance in diverse ancestries and significantly better performance also in Hypothyroidism. Stars (****) indicate statistical significance (p $\leq 0.001$, Mann-Whitney Wilcoxon test for superiority, \numprint{2000} bootstrap samples).}
\label{fig:performance_noneur}
\end{figure*}

\begin{figure*}[pt!] 
\centering	

\figtitle{Sequence region and cell type attributions (type 2 diabetes and diabetic retinopathy)}

  \begin{subfigure}[b]{\rowheadersize\textwidth}
    \rotatebox{90}{\hspace{5.0em}\textsf{Type 2 Diabetes}}
  \end{subfigure}
  \begin{subfigure}[t]{0.47\textwidth}\centering
	\includegraphics[width=1.0\textwidth, valign=b]{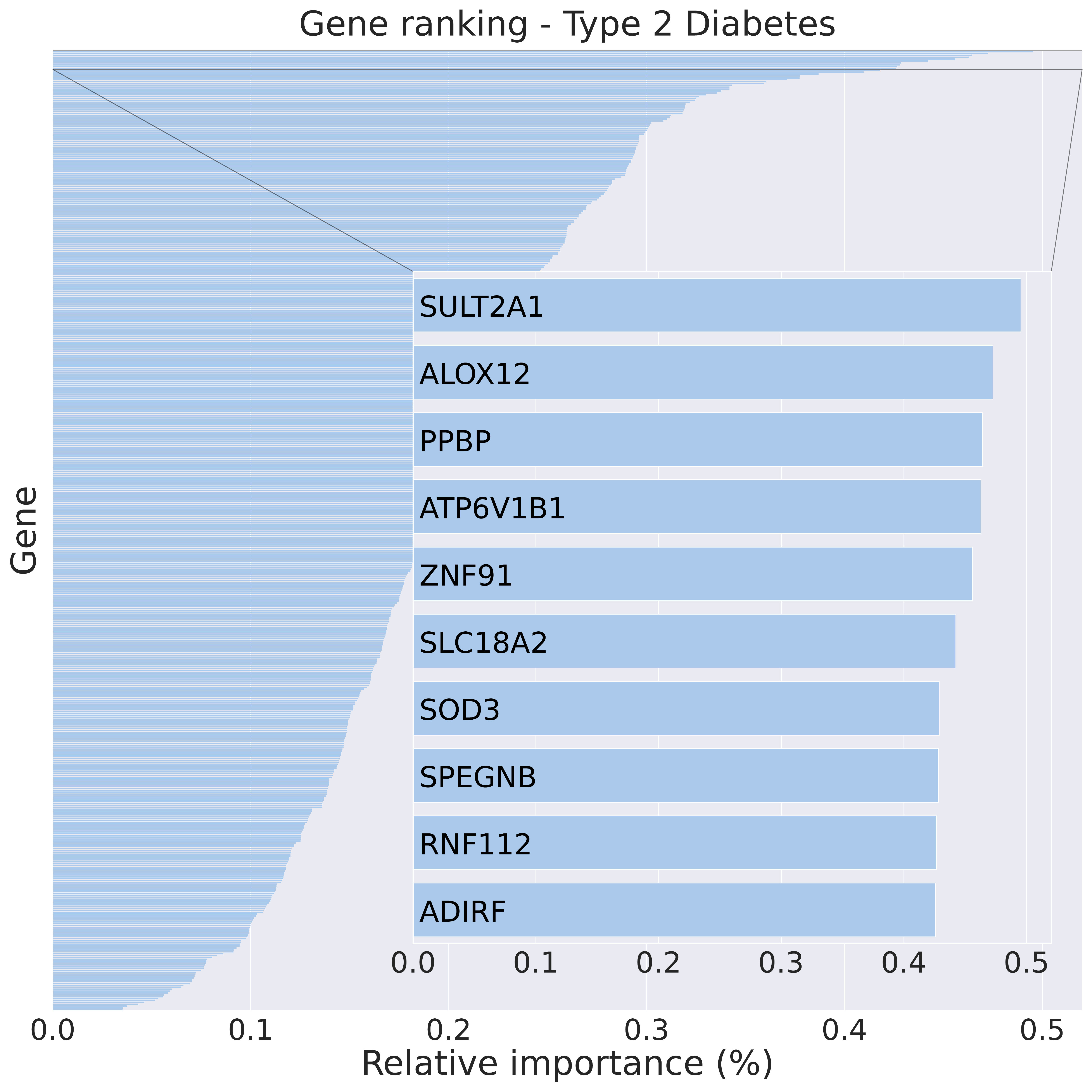}
  \end{subfigure}
  \begin{subfigure}[t]{0.47\textwidth}
\centering	
	\includegraphics[width=1.0\textwidth, valign=b]{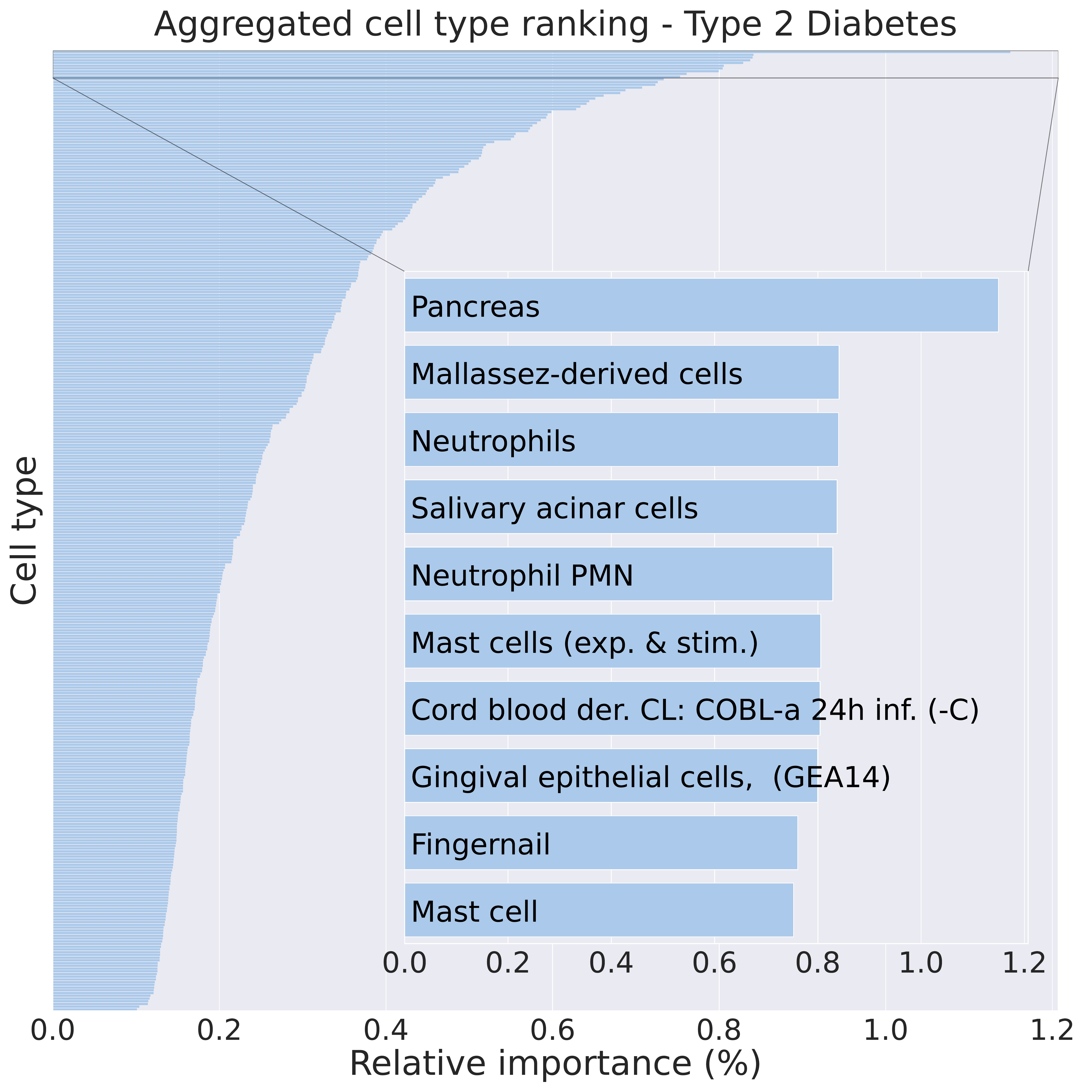}
  \end{subfigure}	
  \begin{subfigure}[b]{\rowheadersize\textwidth}
    \rotatebox{90}{\hspace{5.0em}\textsf{Diabetic Retinopathy}}
  \end{subfigure}
  \begin{subfigure}[t]{0.47\textwidth}\centering
	\includegraphics[width=1.0\textwidth, valign=b]{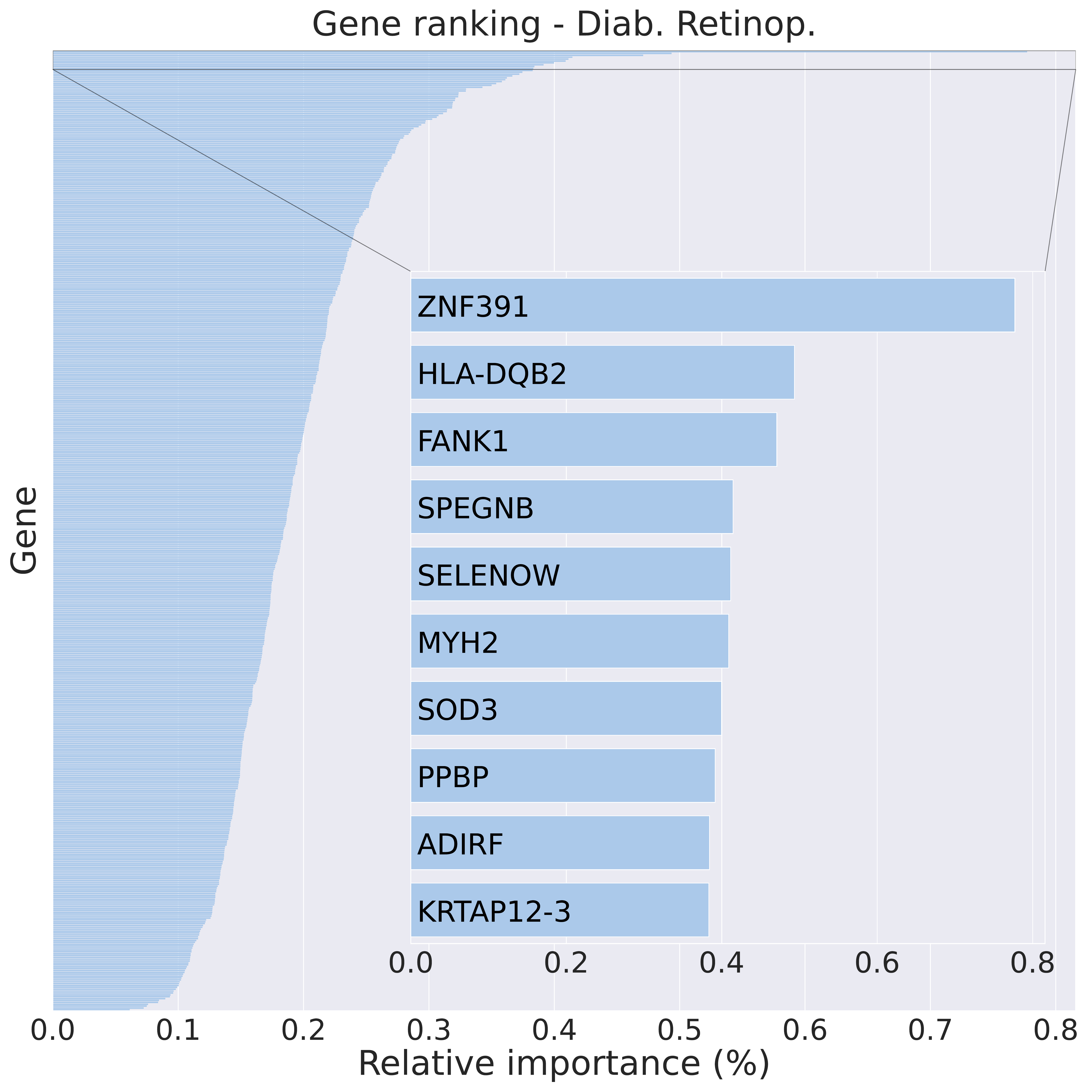}
  \end{subfigure}
  \begin{subfigure}[t]{0.47\textwidth}
\centering	
	\includegraphics[width=1.0\textwidth, valign=b]{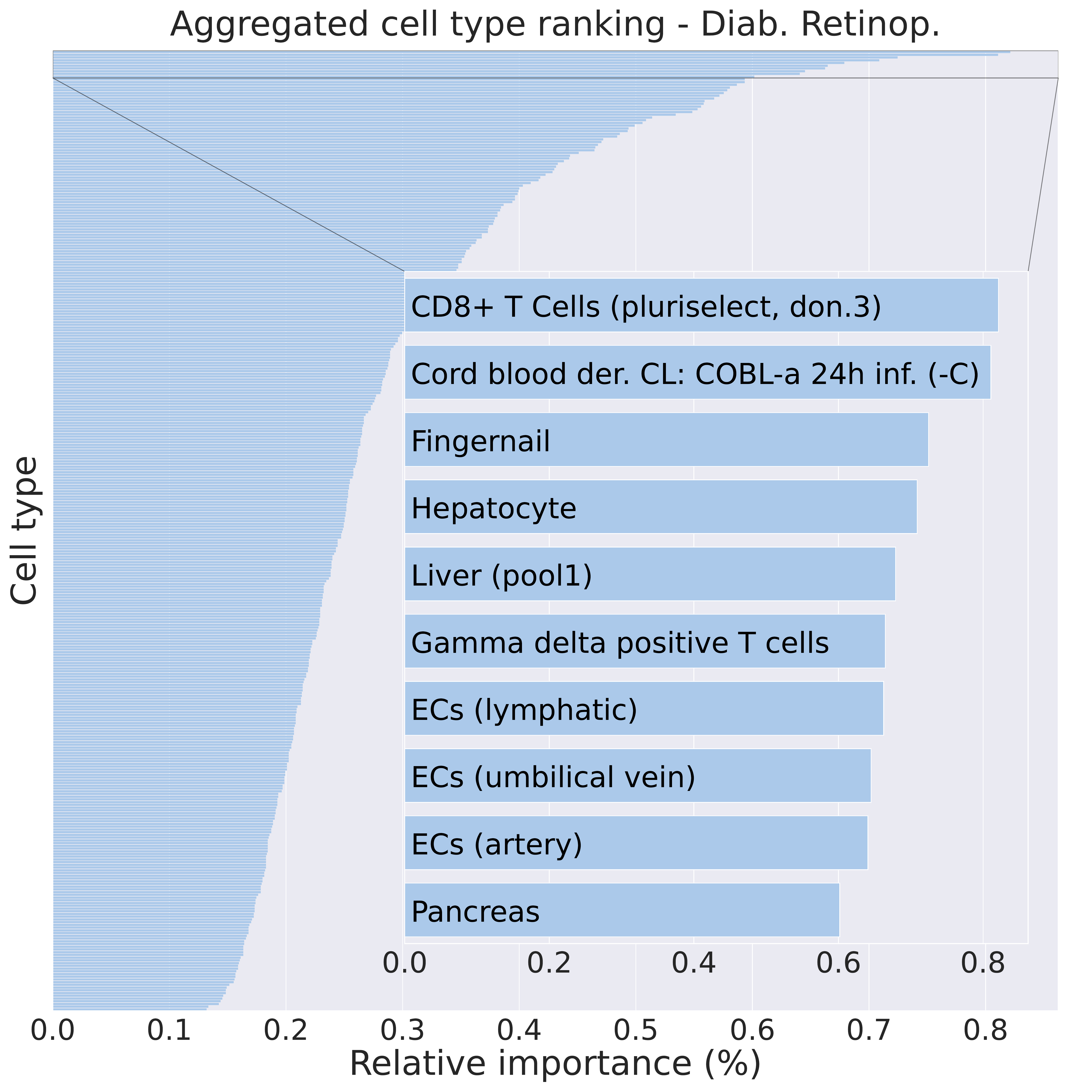}
  \end{subfigure}	
  
\caption{\textbf{Cell types and sequence windows associated with predicted risk in type 2 diabetes and diabetic retinopathy.} \themethod{} attributions highlight what changes in the transcripts of which sequence window (left column; referred to by the TSS gene) and cell and tissue types (right column) differentiate individuals that are predicted to go on to develop a disease (top row: type 2 diabetes, bottom: diabetic retinopathy) compared to those that are not.
}
\label{fig:interpretation_cont}
\end{figure*}

\begin{figure*}[pt!] 
\centering	

\figtitle{Sequence region and cell type attributions (COPD and hypothyroidism)}

  \begin{subfigure}[b]{\rowheadersize\textwidth}
    \rotatebox{90}{\hspace{8.0em}\textsf{COPD}}
  \end{subfigure}
  \begin{subfigure}[t]{0.47\textwidth}\centering
	\includegraphics[width=1.0\textwidth, valign=b]{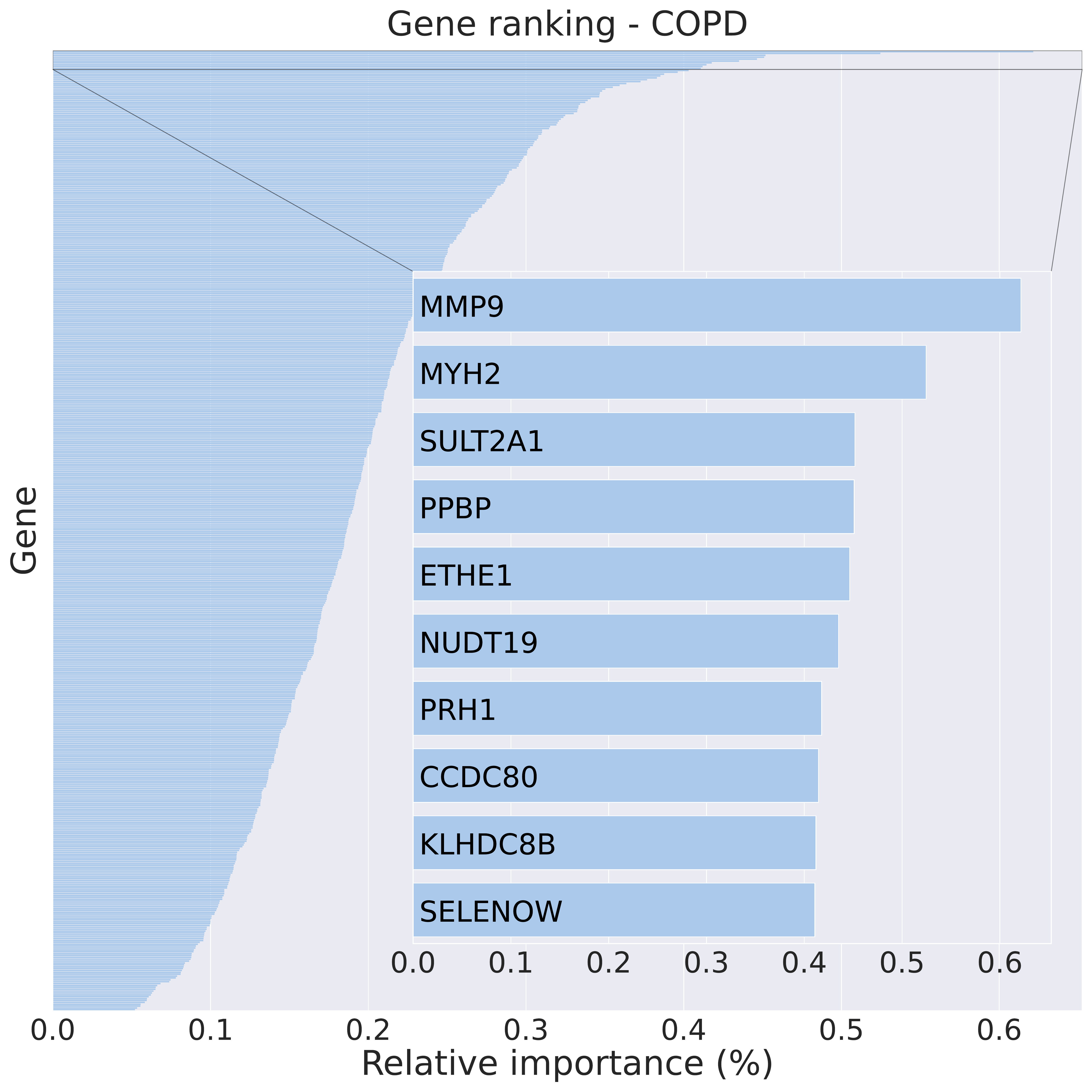}
  \end{subfigure}
  \begin{subfigure}[t]{0.47\textwidth}
\centering	
	\includegraphics[width=1.0\textwidth, valign=b]{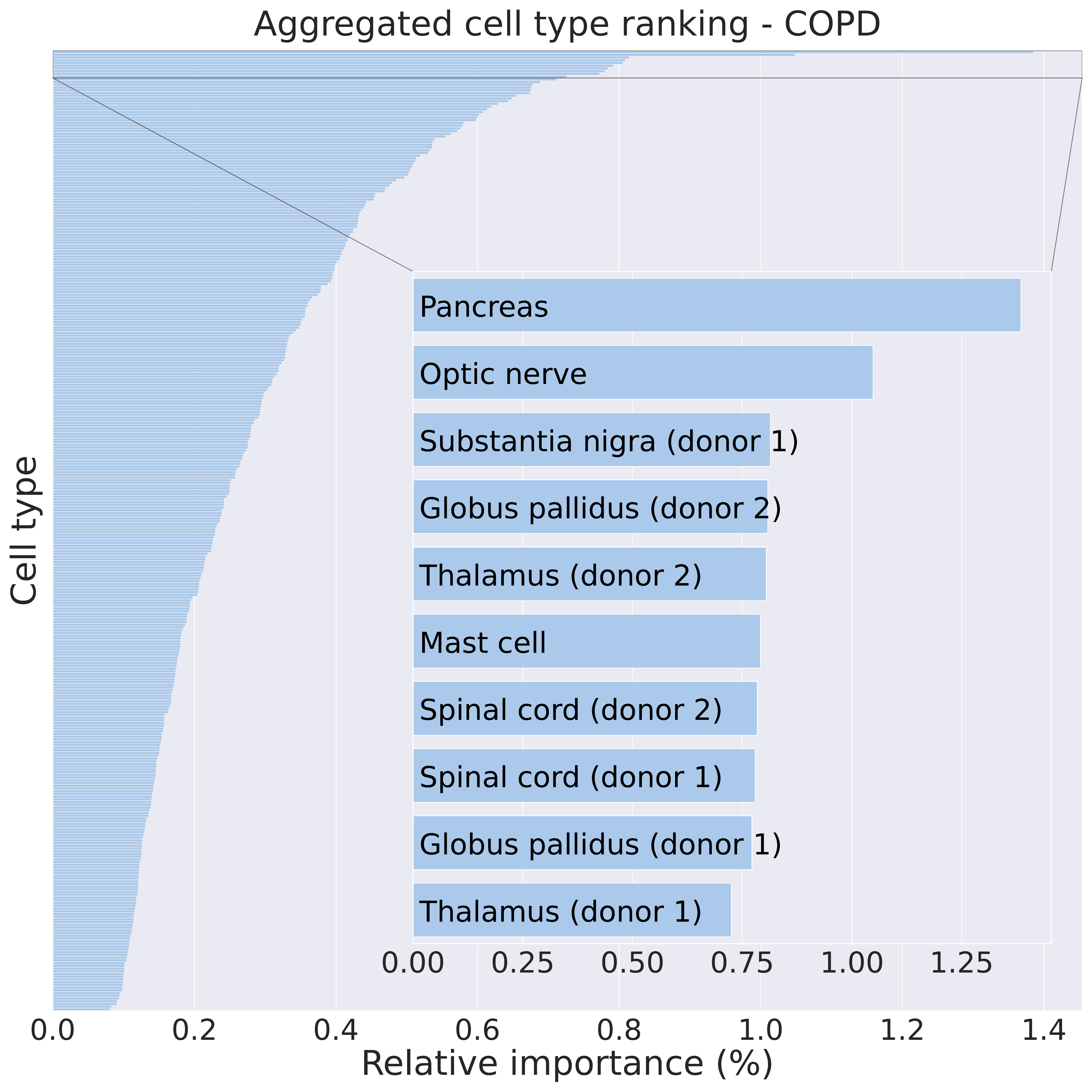}
  \end{subfigure}
  \begin{subfigure}[b]{\rowheadersize\textwidth}
    \rotatebox{90}{\hspace{6.0em}\textsf{Hypothyroidism}}
  \end{subfigure}
  \begin{subfigure}[t]{0.47\textwidth}\centering
	\includegraphics[width=1.0\textwidth, valign=b]{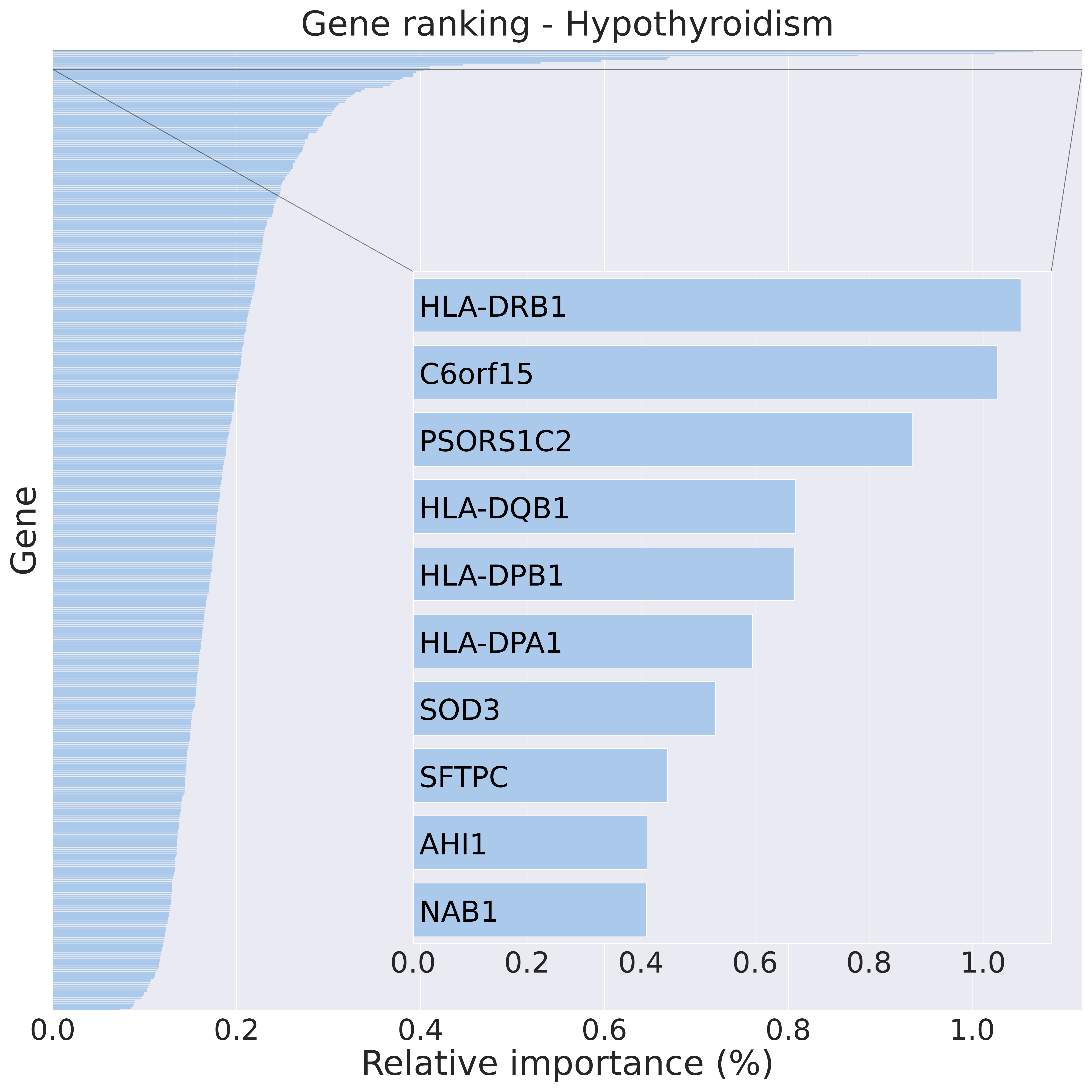}
  \end{subfigure}
  \begin{subfigure}[t]{0.47\textwidth}
\centering	
	\includegraphics[width=1.0\textwidth, valign=b]{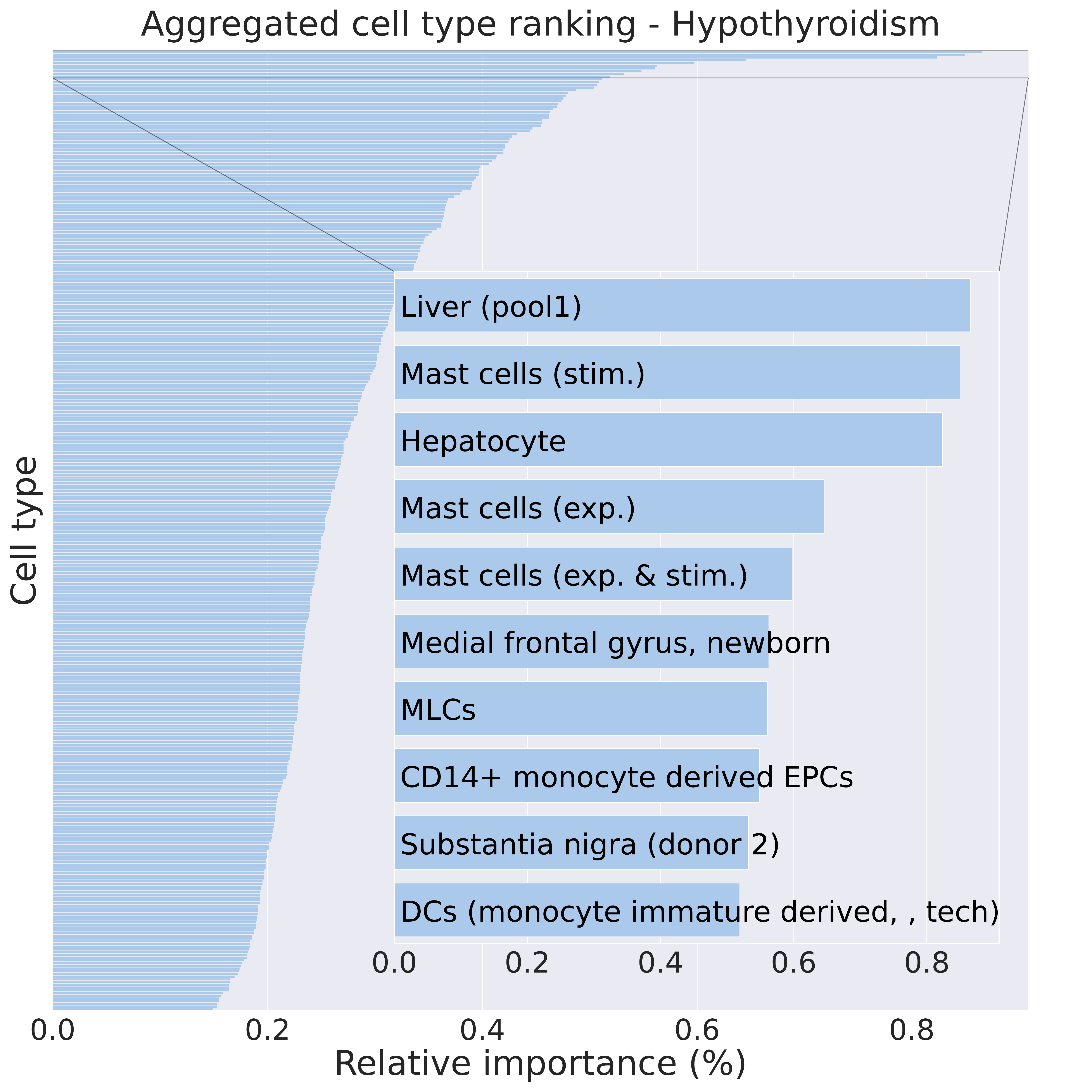}
  \end{subfigure}
  
\caption{\textbf{Cell types and sequence windows associated with predicted risk in COPD and hypothyroidism.} \themethod{} attributions highlight what changes in the transcripts of which sequence window (left column; referred to by the TSS gene) and cell and tissue types (right column) differentiate individuals that are predicted to go on to develop a disease (top row: COPD, bottom: hypothyroidism) compared to those that are not.
}
\label{fig:interpretation_cont2}
\end{figure*}

\begin{figure*}[pt!] 
\centering	

\figtitle{Subtyping by molecular mechanisms (additional diseases)}

\begin{subfigure}[b]{\rowheadersize\textwidth}
    \rotatebox{90}{\hspace{2.25em}\textsf{Type 1 Diabetes}}
  \end{subfigure}
  \begin{subfigure}[t]{0.43\textwidth}\centering
\includegraphics[width=\textwidth]{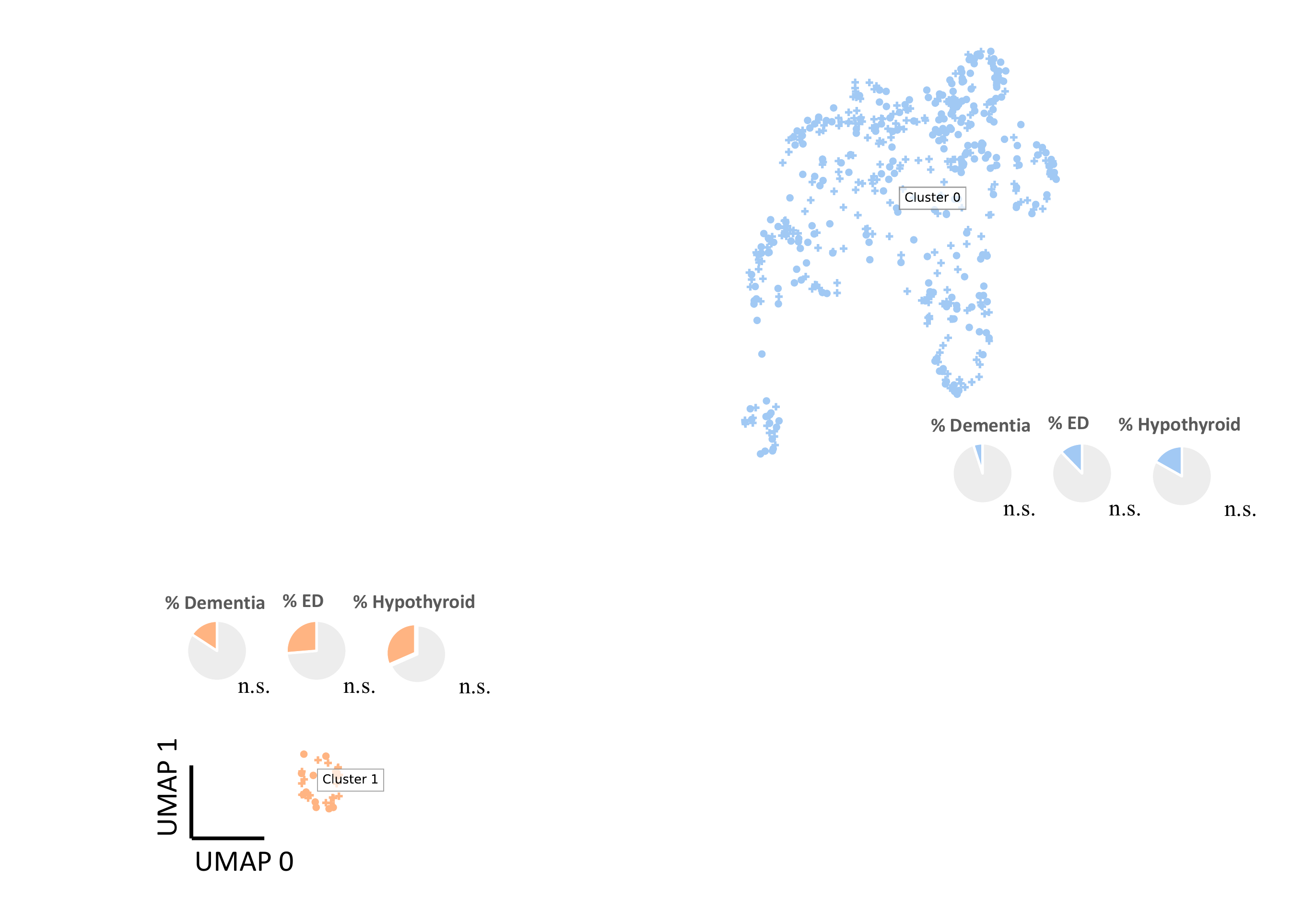} 
\end{subfigure}
  \begin{subfigure}[b]{\rowheadersize\textwidth}
    \rotatebox{90}{\hspace{2.25em}\textsf{Type 2 Diabetes}}
  \end{subfigure}
  \begin{subfigure}[t]{0.43\textwidth}\centering
\includegraphics[width=\textwidth]{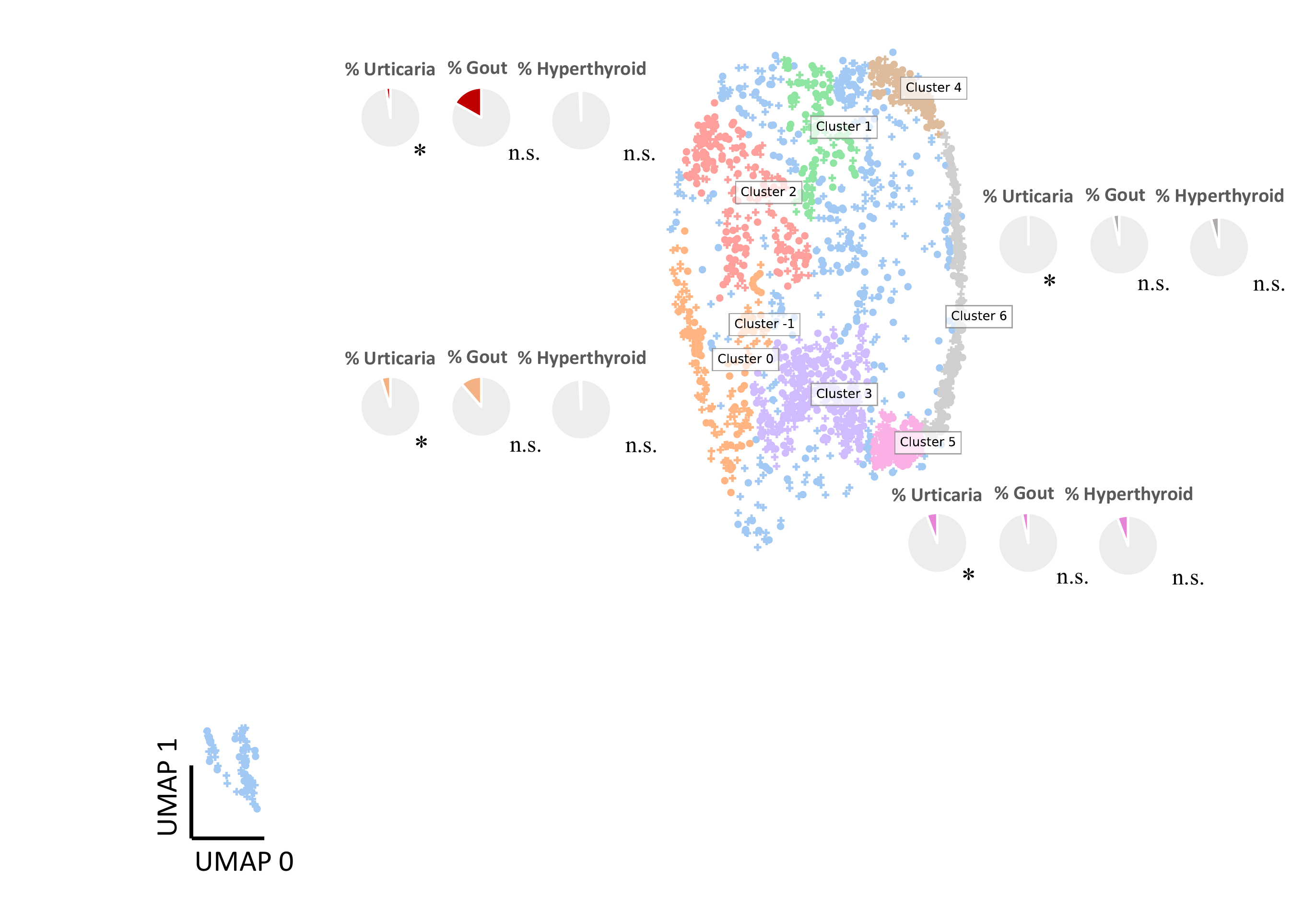} 
\end{subfigure}\quad

  \begin{subfigure}[b]{\rowheadersize\textwidth}
    \rotatebox{90}{\hspace{4.5em}\textsf{COPD}}
  \end{subfigure}
  \begin{subfigure}[t]{0.43\textwidth}\centering
\includegraphics[width=\textwidth]{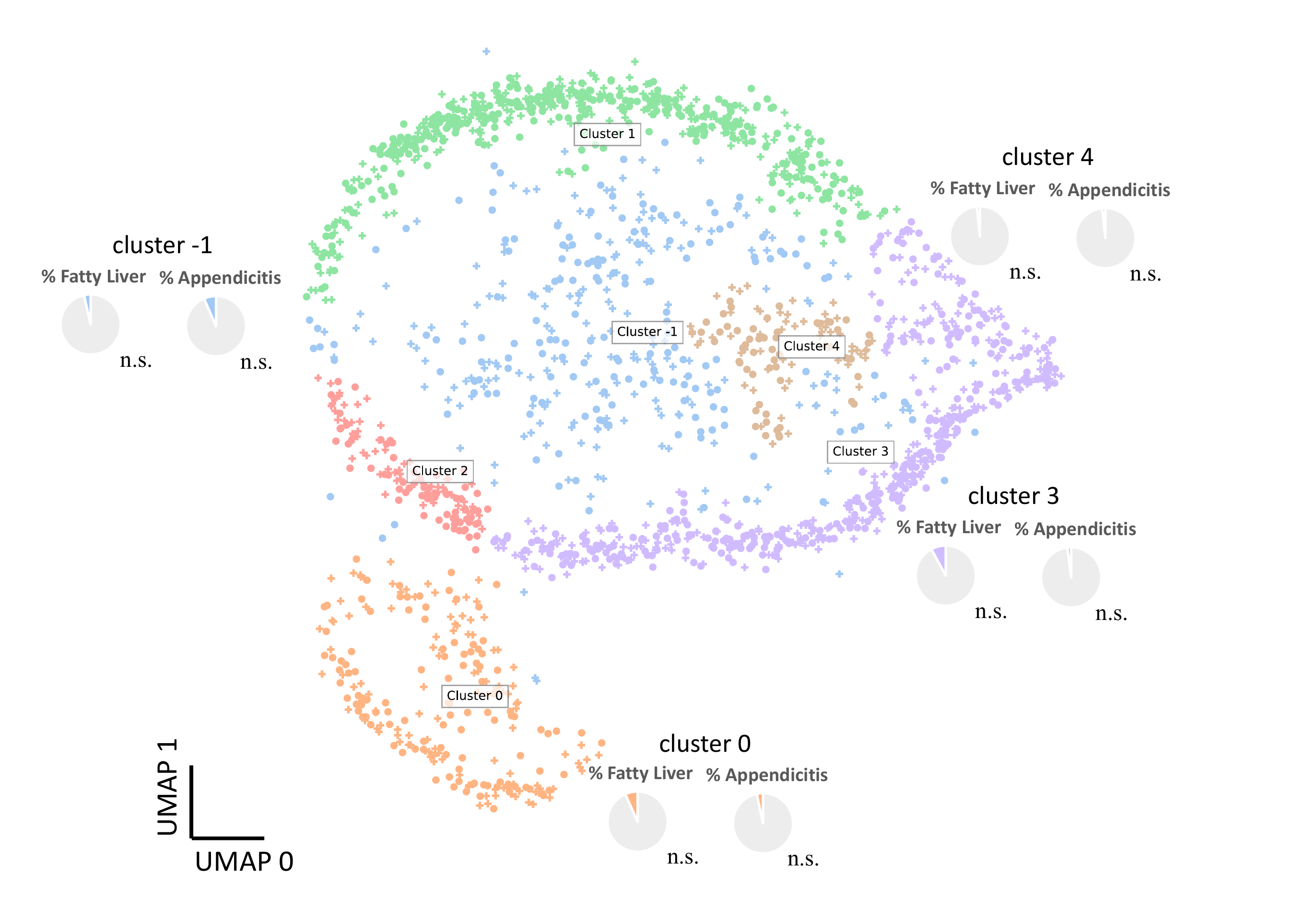} 
\end{subfigure}
  \begin{subfigure}[b]{\rowheadersize\textwidth}
    \rotatebox{90}{\hspace{3em}\textsf{Hypothyroidism}}
  \end{subfigure}
  \begin{subfigure}[t]{0.43\textwidth}\centering
\includegraphics[width=\textwidth]{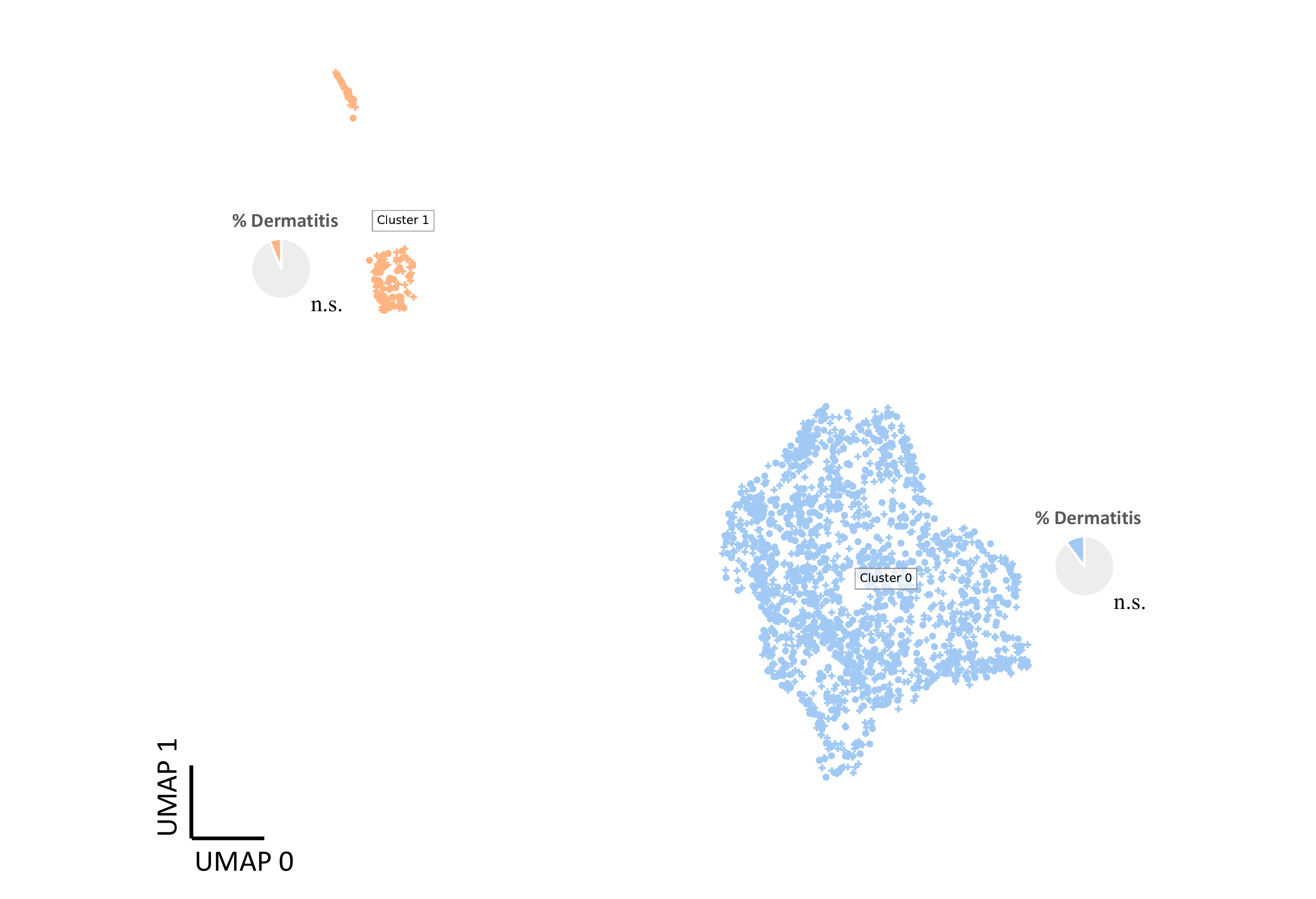} 
\end{subfigure}\quad

\caption{\textbf{\themethod{} groups individual genomes by underlying differences in molecular mechanisms.} Latent space embeddings of \themethod{} can be used to subtype individuals according to their differences in molecular processes induced by genetic variation, enabling a fine-grained understanding of molecular subtypes in broader disease categories. Circles and plus (+) symbols represent diagnosed and an equal amount of reference undiagnosed individuals (not used for clustering), respectively. We identified molecular subtypes (colors with associated cluster labels) using \themethod{} trained to predict T1D (top left), T2D (top right), COPD (bottom left) and hypothyroidism (bottom right; visualised using UMAP \cite{mcinnes2018umap}). Subtypes were associated with differences in terms of co-morbidity rates (pie chart insets) among diagnosed cluster members (highlighted for clusters with the largest differences). We find statistically significant (* = p $\leq 0.05$; $\chi^2$ test) differences in predisposition for urticaria in T2D subtypes, and several additional appreciable differences that do not reach significance (n.s.) in T2D and other diseases.}
\label{fig:clusters_more}
\end{figure*}

\end{document}